\documentclass[prd,superscriptaddress,nofootinbib,floatfix,notitlepage]{revtex4-1}

\usepackage[utf8]{inputenc}

\usepackage{amsmath,amsbsy}    
\usepackage{graphicx}   
\usepackage[lofdepth,lotdepth,caption=false]{subfig}  
\usepackage{hyperref}   
\usepackage[export]{adjustbox}
\usepackage{natbib,ifthen}
\usepackage[usenames, dvipsnames]{color}
\usepackage{dsfont}
\usepackage{csquotes}

\newcommand{\jcap}{JCAP}
\newcommand{\apjl}{Astrophys.\ J.}
\newcommand{\physrep}{Phys.\ Rep.}
\newcommand{\aap}{Astron.\ Astrophys.}

\newcommand{\mnras}{Mon.\ Not.\ R.\ Astron.\ Soc.}
\newcommand{\apjs}{Astrophys.\ J.\ Suppl.\ Ser.}
\newcommand{\araa}{Ann.\ Rev.\ Astron.\ Astrophys.}

\newif\ifincludefigs
\includefigsfalse 

\newcommand{\beq}{\begin{equation}}
\newcommand{\eeq}{\end{equation}}

\newcommand{\vk}{\mathbf{k}}

\newcommand{\ve}{\ensuremath{\boldsymbol\epsilon}}
\newcommand{\vtheta}{\ensuremath{\boldsymbol\theta}}
\newcommand{\vx}{\mathbf{x}}
\newcommand{\vy}{\mathbf{y}}
\newcommand{\vz}{\mathbf{z}}
\newcommand{\vq}{\mathbf{q}}

\renewcommand{\vr}{\mathbf{r}}

\renewcommand{\(}{\left(}
\renewcommand{\)}{\right)}
\renewcommand{\[}{\left[}
\renewcommand{\]}{\right]}

\newcommand{\dd}{\mathrm{d}}
\newcommand{\ii}{\mathrm{i}}

\newcommand{\DiracDelta}{\delta_\mathrm{D}}

\newcommand{\clight}{\ensuremath{\mathrm{c}}}

\newcommand{\OmegaMatter}{\Omega_{\mathrm{m}}}

\newcommand{\kms}{\ensuremath{\mathrm{km\,s}^{-1}}}

\newcommand{\Mpc}{\ensuremath{\mathrm{Mpc}}}
\newcommand{\Gpc}{\ensuremath{\mathrm{Gpc}}}

\begin{document}

\title{Bayesian weak lensing tomography: Reconstructing the 3D large-scale distribution of matter with a lognormal prior}
\author{Vanessa B\"ohm}
\affiliation{Max-Planck-Institut f\"ur Astrophysik, Karl-Schwarzschild-Str. 1, 85748 Garching, Germany}
\author{Stefan Hilbert}
\affiliation{Exzellenzcluster Universe, Boltzmannstr. 2, 85748 Garching, Germany}
\affiliation{Ludwig-Maximilians-Universit\"at, Universit\"ats-Sternwarte, Scheinerstr. 1, 81679 M\"unchen, Germany}
\author{Maksim Greiner}
\affiliation{Max-Planck-Institut f\"ur Astrophysik, Karl-Schwarzschild-Str. 1, 85748 Garching, Germany}
\affiliation{Exzellenzcluster Universe, Boltzmannstr. 2, 85748 Garching, Germany}
\author{Torsten A. En{\ss}lin}
\affiliation{Max-Planck-Institut f\"ur Astrophysik, Karl-Schwarzschild-Str. 1, 85748 Garching, Germany}
\affiliation{Ludwig-Maximilians-Universit\"at, Universit\"ats-Sternwarte, Scheinerstr. 1, 81679 M\"unchen, Germany}
\affiliation{Exzellenzcluster Universe, Boltzmannstr. 2, 85748 Garching, Germany}
\date{\today}

\begin{abstract}
We present a Bayesian reconstruction algorithm that infers the three-dimensional large-scale matter distribution from the weak gravitational lensing effects measured in the image shapes of galaxies. The algorithm is designed to also work with non-Gaussian posterior distributions which arise, for example, from a non-Gaussian prior distribution. In this work, we use a lognormal prior and compare the reconstruction results to a Gaussian prior in a suite of increasingly realistic tests on mock data. We find that in cases of high noise levels (i.e. for low source galaxy densities and/or high shape measurement uncertainties), both normal and lognormal priors lead to reconstructions of comparable quality, but with the lognormal reconstruction being prone to mass-sheet degeneracy. In the low-noise regime and on small scales, the lognormal model produces better reconstructions than the normal model: The lognormal model 1) enforces non-negative densities, while negative densities are present when a normal prior is employed, 2) better traces the extremal values and the skewness of the true underlying distribution, and 3) yields a higher pixel-wise correlation between the reconstruction and the true density. 
\end{abstract}

\maketitle

\section{Introduction}
\label{sec:Intro}

Weak gravitational lensing of galaxies offers a unique way to study the distribution of matter in the Universe (see ~\citep{BartelmannSchneiderRev} for a review on weak gravitational lensing and~\citep{KilbingerReview,HoekstraReview,RefregierReview} for recent reviews on weak galaxy lensing). Lensing by structures along the line of sight causes distortions in the images of distant galaxies (which in this context are often referred to as sources), which leads to correlations between the apparent shapes of these galaxies. The dominant and most easily detectable image distortion that lensing induces is a shearing of the galaxy images. Because of this, the effect is often referred to as cosmic shear.

Galaxy shape measurements allow to constrain the clustering of matter at different scales and redshifts, which can then be translated into constraints on cosmological models and their parameters. The integrated lensing signal is mostly sensitive to a combination of the cosmic mean matter density $\OmegaMatter$ and the matter power spectrum amplitude $\sigma_8$~\citep{1997Bernardeau}. Tomographic methods can yield additional constraints on all parameters guiding the growth rate of structure~\citep{Hu1999ApJ,Simon2004}, notably also on the properties of dark energy~\citep{Hu2002,Huterer2002,Song2004}, but rely on low photometric redshift uncertainties~\citep{Ma_RedUncer}. Since lensing is a direct probe of the total matter, luminous and dark, it can be combined with measurements of the luminous matter distribution in order to learn about the relationship between baryons and dark matter. Another important feature of weak galaxy lensing lies in its ability to probe the matter distribution over a wide range of scales, from many tens of Mega parsec, where structure formation is still linear today and comparably easy to model, down to non-linear sub-Mega parsec scales. Due to its sensitivity to such a wide range of scales, lensing can provide a large amount of information to constrain models of non-linear structure formation and cosmology, in particular if also higher-order statistics are considered.

First firm statistical detections of cosmic shear were reported in 2000 by four different groups~\citep{Bacon2000,VanWaerbeke2000,Kaiser2000,Wittman2000}. Since then, the field has seen a tremendous increase in the amount and quality of lensing data~\citep[e.g.]{COSMOS_Survey,CFHTLenS_Survey,KiDS_Survey, DLens_Survey, DES_sv_shear_catalogue} as well as a notable improvement in analysis techniques~\citep[e.g.][]{2007MNRASMiller1,2008MNRASMiller2,2013MNRASMiller3,im3shape,gfit,Schneider2015,CFHTLens_2Tomo,2003Heymans,2003King,2012MNRASHilde}. Estimates of cosmological parameters have been inferred from these data by comparing the power spectrum of the fully projected 2D shear field (or related quantities) as well as auto- and cross-spectra in a number of redshift bins to theoretical predictions~\citep{COSMOS_Res3D, DLens_Res2D,100deg_Res2D,CFHTLens_2D,COSMOS_ResTomo,CFHTLens_2Tomo,KiDSLens_CONS,DLens_Tomo,CFHTLens_Tomo+IA,DES_sv_shear_Tomo}. Several authors have also investigated the additional constraining power that can be achieved by incorporating third-order statistics and/or shear peak counts and correlations \citep{1997Bernardeau,2004MNRASTakadaJain,SemboloniEtal2011,DietrichHartlap2010,MarianEtal2011,MarianEtal2012,MarianEtal2013,HilbertEtal2012,LiuEtal2015,KacprzakEtal2016}, which helps to break parameter degeneracies. 

The full information content, however, lies in the three-dimensional non-linear shear field.
3D weak shear analysis methods have been proposed by a number of authors~\citep{2003MNRASHeavens,2005PhRvDCastro,2011MNRASKitching} and were recently applied to data from the Canada France Hawaii Telescope Lensing Survey (CFHTLenS) ~\citep{2014MNRASKitching}. Furthermore, the measured shear field can be used for 3D reconstructions of the underlying density field. This can then be directly compared to models of structure formation and simplifies the cross correlation with other tracers of matter. Algorithms that invert the lens equation to obtain the density have been worked out by a number of authors~\citep{1993ApJKaiserSquires,2002PhRvDHuKeeton,2001Taylor}. Since this inversion is under-constrained, it requires some regularization method or choice of prior on the density field. Most of the algorithms employ a Wiener filter, which corresponds to a normal (Gaussian) prior, that can be complemented with information about galaxy clustering~\citep{2013A&ASimon,2014MNRASSzepietowski}.

We aim to extend the work on tomographic reconstruction of the 3D matter distribution by designing a fully Bayesian reconstruction algorithm that uses a lognormal prior on the density field.
The algorithm is designed to reconstruct the 3D cosmic density fluctuation field $\delta(\vx,\tau)$ from weak galaxy lensing data, i.e. a measurement of galaxy ellipticities at different (photometrically measured) redshifts. Its derivation is based on the language of information field theory~\citep{2009PhRvDEnsslin}, which has already been used to address similar tomography problems~\citep{2016MaksimTomo}. We do not make use of the flat-sky approximation i.e. lines of sight are allowed to be non-perpendicular to a fixed 3D coordinate grid. Further, we do not bin the data into pixels but take each galaxy into account as an individual contribution to the likelihood. This allows us to incorporate distance uncertainties of individual galaxies instead of sample redshift distributions.

In contrast to a normal prior for the density field, as it has often been assumed before, a lognormal prior  automatically enforces the strict positivity of the field and allows to capture some of the non-Gaussian features that are imprinted on the density distribution by non-linear structure formation. 
Hubble was the first to notice that galaxy number counts could be well approximated by a lognormal distribution~\citep{1934ApJHubble}. Characterizing the matter overdensities in the Universe as a lognormal field was first assessed by Coles \& Jones in 1991~\citep{1991MNRASCJ}. Subsequent studies showed that a logarithmic mapping of the nonlinear matter distribution can partly re-gaussianize the field, and that non-linear features in the matter power spectrum can be reproduced by a lognormal transformation of the linear matter power spectrum~\citep{2009ApJNeyrinck,2015A&AGreiner}.
A lognormal prior has already been used and shown to be superior to a Gaussian one in Bayesian algorithms that reconstruct the large-scale matter distribution from the observed galaxy distribution~\citep{2009PhRvDEnsslin,2010MNRASKitaura,2010MNRASJasche}.
Lognormal distributions have also already been considered in the context of weak lensing:  Analyses of ray-tracing simulations and the Dark Energy Survey (DES) Science Verification data showed that the 1-point distribution function of the lensing convergence is better described by a lognormal than a Gaussian model~\citep{2002ApJTaruya,2016MNRASClerkin}. Also the cosmic shear covariance can be modeled to better accuracy under the assumption that the underlying convergence field follows a lognormal distribution instead of a Gaussian one~\citep{2011A&AHilbert}.

Bayesian inference methods are widely used in weak shear analyses, most prominently in the context of shear measurements from galaxy images~\citep{2007MNRASMiller1,2014MNRASBernstein,Schneider2015}. Recently, notable effort has been put into developing a fully Bayesian analysis pipeline that propagates all uncertainties consistently from the raw image to the inferred cosmological parameters~\citep{2016MNRASAlsing,2016Heavens}.

This paper is organized as follows: This introduction is followed by a short section, Sec.~\ref{sec:Notations}, in which we briefly introduce the notations and coordinate systems that will be used in the derivation of the formalism. Our lognormal prior model for the density is described in detail in Sec.~\ref{sec:PriorModel}. In Sec.~\ref{sec:datamodel}, we present the data model, i.e. the lensing formalism that connects the data from a cosmic shear measurement to the underlying density field and give a brief overview over its implementation in Sec.~\ref{sec:impldetails}. The maximum a posteriori estimator that is used to infer the matter distribution is introduced in Sec.~\ref{Sec:MAPesti} and extended to include redshift uncertainties of individual sources in Sec.~\ref{sec:z-marg-Likelihood}. In Sec.~\ref{sec:validation}, we show  results of the density reconstruction on increasingly realistic mock data. We conclude this work with a summary and discussion in Sec.~\ref{sec:conclusions}.

\section{Coordinate systems and notational conventions}
\label{sec:Notations}

In the derivation of the formalism we work with three different types of coordinate systems. First, we use three-dimensional purely spatial comoving coordinates $\vx=(x_0,x_1,x_2)$ at fixed comoving lookback time $\tau$, that, combined with the time coordinate, form the four-dimensional coordinate system $(\vx,\tau)$.

Second, we use a 3D comoving coordinate system on the light cone of an observer at the origin. Vectors on the light cone are marked by a prime, e.g. $\vx'$. Since $\vx'$-coordinates implicitly define a comoving lookback time $\tau(\vx') = |\vx'|/\clight$ (where $\clight$ denotes the speed of light), we omit spelling out the time explicitly and write $A(\vx')=A(\vx',\tau)$ for any quantity $A$ that is defined on the light cone. The operation that links quantities on the light cone to their corresponding quantities in 4D spacetime can be encoded in a projection operator with kernel
\beq
\label{eq:lc}
\mathcal{C}_{\vx'(\vx,\tau)}=\DiracDelta\(\vx'-\vx\) \DiracDelta\(\tau-|\vx|/c\).
\eeq

Third, we employ a set of coordinate systems on the light cone, in which each system is orientated such that one axis points into the direction of a source galaxy.
These line of sight coordinate systems are centered on the observer and spanned by the vectors $(\hat{\vr}_0^i,\hat{\vr}_1^i,\hat{\vr}_2^i)$, where $\hat{\vr}_0^i(\vx')$ points into the direction of the $i$th source galaxy and the normal vectors $(\hat{\vr}_1^i,\hat{\vr}_2^i)$ span the two-dimensional plane perpendicular to $\hat{\vr}_0^i$. The radial comoving distance of each galaxy from the observer is denoted $r^i=|\vr^{i}|\equiv|\vr_0^{i}|$.
The transformation from the light cone system into the line of sight (LOS) system of a source galaxy $i$ is achieved by a rotation of the $x'_0$-axis into the $i$th line of sight, while $x'_1$ and $x'_2$ get aligned with the image coordinates of the galaxy observation. We label the corresponding transformation operators ${\mathcal{R}^i}_{\vr \vx'}$.

\section{Prior model}
\label{sec:PriorModel}

The comoving density $\rho(\vx,\tau)$ can be split into its time-independent spatial mean $\bar{\rho}$,  and a perturbation $\delta\rho(\vx,\tau)=\bar{\rho}\,\delta(\vx,\tau)$. The fractional overdensity $\delta(\vx,\tau)$ is commonly modeled as a homogeneous, isotropic Gaussian field with zero mean and power spectrum $P_\delta(k,\tau)$. This is an excellent description at early times where fluctuations are very small, as e.g., shown by observations of the cosmic microwave background radiation (CMB). At linear level, valid for $\delta \ll 1$, the time evolution of the density field can be described by
\beq
\label{eq:rho_linear}
\rho(\vx,\tau)=\bar{\rho}\[1+\delta(\vx,\tau)\] = \bar{\rho}\[ 1+ \int_{\vy}D(|\vx-\vy|,\tau,\tau_{0})\varphi(\vy,\tau_{0})\]
=\bar{\rho}\[ 1+ {{D}_{\vx\vy}(\tau,\tau_0)\varphi_{\vy}(\tau_0)} \]
.
\eeq
In the last expression we have introduced a short-hand notation that will be used in the rest of the paper: repeated indices are integrated over if they do not appear on both sides of the equation. 
In Eq.~\eqref{eq:rho_linear}, $D(|\vx-\vy|,\tau,\tau_{0})$ is the integration kernel of a linear homogeneous and isotropic, but possibly scale-dependent, growth operator. 
The field $\varphi$ is an isotropic and homogeneous random field whose values are drawn from a multivariate normal distribution with mean $\bar{\varphi}$ and covariance $\Phi$,
\beq
\label{eq:varphi_model}
\varphi \hookleftarrow \mathcal{N}(\varphi|\,\bar{\varphi},\Phi).
\eeq
Here, it describes the three dimensional density fluctuations at time $\tau_{0}$ and is translated to other times $\tau$ by the growth operation $D_{\vx\vy}(\tau,\tau_{0})$. This implies $\bar{\varphi} = 0$ for this linear Gaussian model.

The description Eq.~\eqref{eq:rho_linear} breaks down when $\delta \not \ll 1$ such that non-linearities become important. A possible way to account for non-linearities is to include higher-order terms from a perturbation series expansion of the full non-linear evolution equations. However, a further shortcoming of the model~\eqref{eq:rho_linear} is that it allows arbitrarily negative density contrasts, which physically can not be smaller than -1. To obtain a strictly positive density field, we instead modify Eq.~\eqref{eq:rho_linear} by an exponential:
\beq
\label{eq:rho}
\rho(\vx,\tau)=\bar{\rho}\[1+\delta(\vx,\tau)\]= \bar{\rho} \exp\[{{D}_{\vx\vy}(\tau,\tau_0)\varphi_{\vy}(\tau_0)}\].
\eeq
Since the expectation value of the density $\rho(\vx,\tau)$ must equal $\bar{\rho}$, the mean of $\varphi$ must be set to 
\beq
\bar{\varphi}_{\vx}(\tau)=-\frac{1}{2}D^{-1}_{\vx\vy}(\tau,\tau_0)\[D(\tau,\tau_0) \Phi(\tau_0) D(\tau,\tau_0)\]_{\vy\vy},
\eeq
for every time $\tau$. Note that we integrate over the index $\vy$ in this expression, i.e. the diagonal of the composite operator in square brackets is treated as a field.

For a local growth operator, $D_{\vx\vy}(\tau,\tau_0)=D(\tau,\tau_0)\DiracDelta(\vx-\vy)$, this mean correction simplifies to 
\beq
\bar{\varphi}_\vx(\tau)=-\frac{1}{2}D(\tau,\tau_0) \hat{\Phi}_\vx(\tau_0),
\eeq
where we defined $\hat{\Phi}_\vx(\tau_0)\equiv \Phi_{\vx\vx}(\tau_0)$.

The Gaussian field $\varphi$ and the growth operator $D$ can be related to known quantities. To see this, consider the expansion of the Fourier modes of $\delta$ in Eulerian perturbation theory (see e.g.~\cite{Bernardeau2002}),
\beq
\delta(\vk,\tau)=\sum_{n=1} D^{(n)}(\tau)\delta^{(n)}(\vk),
\eeq
where $\delta^{(n)}$ are convolutions of $n$ initial fields $\delta(\vk,\tau_0)$ with an integration kernel that changes from order to order. The first term in this series is $D^{(1)}(\tau)/D^{(1)}(\tau_0)\delta_0(\vk,\tau_0)$, where $D^{(1)}(\tau)$ is the growing solution to the linearized growth equation~\citep{2003MNRASLinder}.

We use this analogy to motivate the simplest possible form of the growth operation in the lognormal model and write
\begin{align}
\label{eq:D_restrict}
D_{\vx\vy}(\tau)&=\DiracDelta(\vx-\vy)D^{(1)}(\tau)
\end{align}
where we have set $D^{(1)}(\tau_0)=1$.
This approximation erases any a-priori assumption of scale-dependent growth and mode-coupling of the log field $\varphi$. Such a simplification is viable since the model in Eq.~\eqref{eq:D_restrict} describes only our prior assumptions about the density field $\rho$. The algorithm will find the most probable realization of $\varphi$ for a fixed growth operator $D$ given the data. If a scale-dependence is favored by the data, it will be recovered, at least partially, in the estimate of $\varphi$. 

Our algorithm also allows to incorporate a more general growth operation at the expense of computation time and memory usage. The application of the most general $D_{\vx\vy}(\tau)$ generates a four-dimensional field: three-dimensional spatial comoving volumes for every time-slice $\tau$. This very large volume is then restricted to a three-dimensional cut by application of the light cone operator [Eq.~\eqref{eq:lc}]. The prior model for the matter overdensity on the light cone then becomes
\beq
\delta_{\vx'}=\mathcal{C}_{\vx'(\vx,\tau)}\mathrm{exp}\[D_{\vx\vy}(\tau)\varphi_\vy(\tau)\]-1.
\eeq

For the purpose of comparing the lognormal to the Gaussian density model we choose the simplest form of $D$, Eq.~\eqref{eq:D_restrict}, since it can be applied to the three-dimensional field on the light cone
\beq
\delta_{\vx'}=\mathrm{exp}\[D_{\vx'\vy'}\varphi_{\vy'}\]-1.
\eeq
More complicated models, e.g. motivated by perturbation theories for large-scale structure, could be envisioned but lie beyond the scope of this work. A growth model that is better informed about the spatial correlations in the density field could potentially improve the reconstruction in regions with low signal to noise (see also the discussion in Sec.~\ref{sec:conclusions}.

\section{Data model}
\label{sec:datamodel}
In this section we establish the analytic relation between the signal field $\varphi$, the field of overdensities $\delta$ that we aim to reconstruct, and the data that is obtained from a weak lensing measurement.

Weak galaxy lensing surveys produce galaxy image ellipticities that can be quantified, e.g., by a complex number $\epsilon=\epsilon_1+i\epsilon_2$ such that $\epsilon=(a-b)/(a+b)\exp(2\ii\eta)$ for an ellipse with major axis $a$, minor axis $b$, and position angle $\eta$~\citep[e.g.][]{Blandford1991}.
We use the common approximation that the components of the intrinsic source galaxy ellipticity $\ve^{\mathrm{s}}=(\epsilon^{\mathrm{s}}_1,\epsilon^{\mathrm{s}}_2)$, which define the shape that would be observed in the absence of lensing\footnote{We use vector notation, e.g. $\ve$, to denote the tuple of real and imaginary part $(\epsilon_1,\epsilon_2)$ of a complex number $\epsilon{=}{\epsilon_1}+i{\epsilon}_2$.}, follow a global bivariate Gaussian distribution with zero mean and variance $\sigma_{\epsilon}^2$ per component:
\beq
\label{eq:shape_noise}
\ve^{\mathrm{s}}\hookleftarrow \mathcal{N}(\ve^{\mathrm{s}}|\,0,N^{\mathrm{s}}),\qquad N^{\mathrm{s}}_{ij}=\delta_{ij}\sigma^2_{\epsilon}.
\eeq
This approximation has shortcomings (see e.g.~\citep{2013MNRASMiller3}), but serves for the proof of our concept, since we create the mock data on which we test the algorithm with exactly this shape noise model. In the future, more elaborated (Bayesian hierarchical) shear estimators, that e.g. take into account galaxy properties, can be incorporated into the algorithm~\citep{2007MNRASMiller1,Schneider2015,2014MNRASBernstein}.

Lensing distorts the galaxy images in shape and size~\citep{1967Gunn,1990Kochanek,Blandford1991,1994Seitz,1995ApJKaiser}. If the distortion is small, i.e. in the limit of weak lensing, the relation between intrinsic source ellipticity and observed ellipticity can be linearized and simplifies to~\citep{SeitzSchneider1997_3,KrauseHirata2010}
\beq
\epsilon=g + \epsilon^{\mathrm{s}},
\eeq
where $g$ is the reduced shear. The reduced shear combines the effect of anisotropic lensing distortions, encoded in the shear $\gamma=\gamma_1+i\gamma_2$,  and the isotropic distortion, encoded in the convergence $\kappa$ 
\beq
g=\frac{\gamma}{1-\kappa}\approx \gamma.
\eeq
If $\kappa\ll 1$, which is often the case for galaxy lensing, the reduced shear can be approximated by the shear itself $g\approx\gamma$.

The shear and convergence at angular position $\vtheta$ are related to the lensing potential $\psi$ by 
\beq
\label{eq:def_gam_kap}
{\gamma}_1(\vtheta)=\frac{1}{2}\(\partial_{1}^2-\partial_{2}^2\)\psi(\vtheta);
\qquad 
{\gamma}_2(\vtheta)=\partial_{1}\partial_{2}\psi(\vtheta);
\qquad 
{\kappa}(\vtheta)=\frac{1}{2}(\partial_{1}^2+\partial_{2}^2)\psi(\vtheta).
\eeq
The lensing potential $\psi$ is an integrated measure of scalar perturbations to the background metric along the perturbed photon geodesic. Integrating the perturbations along the unperturbed geodesic turns out to be an excellent approximation~\cite{ShapiroCooray2006,KrauseHirata2010}. Working in this so-called Born approximation, choosing the Newtonian gauge~\citep{1992Mukhanov} and assuming no anisotropic stress, the lensing potential can be written as a weighted projection of the peculiar Newtonian gravitational potential $\phi$ along the line of sight. For a source at LOS distance $r^i$, this integration reads
\beq
\label{eq:Phi-Psi}
\psi(\vtheta)=
\frac{2}{\clight^2}\int\limits_{0}^{r^i} \dd r \,\frac{r^i-r}{r r^i} \phi(r, r \theta_1, r \theta_2),
\eeq
where we have assumed a spatially flat Universe.
Applying the angular derivatives in Eq. \eqref{eq:def_gam_kap} to the expression for the lensing potential in Eq. \eqref{eq:Phi-Psi}, we get
\beq
\label{eq:tidal_int}
\partial_k\partial_l\psi(\vtheta) 
=
\frac{2}{\clight^2}\int\limits_{0}^{r^{i}} \dd r \, W(r;r^{i}) \partial_{r_k}\partial_{r_l}\phi\bigl(r, r \theta_1, r \theta_2 \bigr),
\eeq
where $k,l \in (1,2)$, and the lensing efficiency
\beq
\label{eq:Kernel_redef}
W(r;r^i) = \frac{r(r^i-r)}{r^i}.
\eeq
In practice the distance to the source $r^i$ cannot be determined directly but follows from the photometrically measured redshift $z^i$. Photometrically measured redshifts are associated with a relatively high error, $\sigma_z/(1+z)\approx 0.04-0.06$~\citep{2012MNRASHilde}. In its most simple form the algorithm ignores this uncertainty. We will use this simplified model to validate the functionality of the algorithm in terms of reconstructing non-linear structures in the lognormal approximation. Redshift uncertainties will be included later in Section \ref{sec:z-marg-Likelihood}.

The lensing shear is completely determined by the second derivatives of the lensing potential perpendicular to the LOS.
The tidal tensor $\partial_{r_k} \partial_{r_l} \phi(\vr)$ along the LOS of the $i$th source galaxy is obtained by rotating the tidal tensor on the global coordinate grid $\vx'$
\beq
\label{eq:tidal_global}
{T}_{ij}(\vx')=\partial_{x'_i} \partial_{x'_j} \phi(\vx'),\qquad i,j \in (0,1,2),
\eeq
into the specific coordinate system (with coordinates $\vr^i$) that points into the direction of this $i$th galaxy and projecting it onto the $(r^i_1-r^i_2)$-plane perpendicular to the LOS.

The last relation required to connect the data to the density fluctuations is Poisson's equation. It relates the potential $\phi(\vx')$ to the density fluctuations $\delta(\vx')$,
\beq
\label{eq:Poisson}
\nabla^2 \phi(\vx')=\frac{3}{2} \OmegaMatter H_0^2 \frac{\delta(\vx')}{a(|\vx'|/\clight)},
\eeq
where $H_0$ denotes the Hubble constant (which will be parametrized by $h = H_0 / (100 \, \kms\Mpc^{-1})$ in our test simulations), and $a(|\vx'|/\clight)=a(\tau)$ denotes the scale factor at the time $\tau$ corresponding to LOS distance $r^i(x')$.

\section{Implementation}
\label{sec:impldetails}
The implementation, not only of the data model, but of the entire algorithm, is based on NIFTy~\citep{2013A&ASelig}, a versatile software package for the development of inference algorithms. We further compute cosmology-dependent quantities, like power spectra and distance-redshift relations, with the publicly available CLASS code~\citep{CLASS}. 
To summarize the data model we introduce short-hand notations for each operation in terms of operators. 

In its most general form, the prior and data model, that connect the Gaussian field $\varphi$ with a data vector of $N_s$ measured source ellipticities, are as follows:
The growth operator $D_{\vx\vy}(\tau)$ imprints a growth structure on the Gaussian field. The resulting four-dimensional field is plugged into the exponential of $E(\cdot)=\exp(\cdot)-1$ [see Eq.~\eqref{eq:rho}] to obtain the fractional overdensity $\delta_\vx(\tau)$. The overdensity induces the potential $\phi_\vy(\tau)$ by the Poisson equation, encoded in the operator $\mathcal{P}_{\vy\vx}(\tau)=\Delta^{-1}\frac{3}{2} \OmegaMatter H_0^2 \delta(\vx')/a(|\vx'|/\clight)$ [Eq.~\eqref{eq:Poisson}]. The potential can be computed efficiently in Fourier space. The gravitational potential is then restricted to the light cone of the observer by the light cone operator $\mathcal{C}_{\vy'(\vy,\tau)}$ [Eq.~\eqref{eq:lc}]. We compute the tidal tensor of the resulting 3D field [(Eq.~\eqref{eq:tidal_global}] by application of a global differential operator, which we denote $\mathcal{T}_{\vz'\vy'}$. The resulting tidal tensor is then rotated into each galaxy's LOS coordinate system by a rotation operator, $\mathcal{R}^{i}_{\vr\vz'}$. An integration operator $\mathcal{I}_{j\vr}$, which applies the integration in Eq.~\eqref{eq:tidal_int}, integrates the components of each of the resulting $N_s$ tidal tensors along the unperturbed photon geodesic. For this operator, we use an adapted version of the implementation that was already successfully used in a similar reconstruction method~\citep{2016MaksimTomo}. The application of $\mathcal{I}_{j\vr}$ yields derivatives of the lensing potential for each galaxy location. From this we can compute the shear components by a linear operator $\mathcal{L}_{ij}$ that comprises the equations in Eq.~\eqref{eq:def_gam_kap}. Rotation and integration map the three dimensional continuous signal space into the discrete space (one point for every galaxy) of the data. The shear components are thus automatically computed at the locations of the galaxies. 

In the simplified implementation that we use for this work, we avoid the 4D coordinate grid $(\vx,\tau)$ and work on the three-dimensional light cone from the beginning. In the prior, we model the Gaussian log-density $\varphi$ with the power spectrum of matter fluctuations today $P_{\mathrm{lin}}(k,a{=}1)$, where $a$ denotes the scale factor. The growth operator is diagonal in configuration space and only a function of comoving distance to the observer $D_{\vx'\vy'}=D^{(1)}(\tau)\DiracDelta(\vx'-\vy')$, where $D^{(1)}(\tau)$ is the growing solution of the linearized growth equation. The Poisson operator is split into two parts. First, a multiplication with $3/2\, \OmegaMatter H_0^2/a(|\vx'|/\clight)$, i.e. an operation that is diagonal in configuration space. Second, the inverse Laplace operation $\Delta^{-1}$, which is diagonal in Fourier space. The inverse Laplacian in the Poisson equation is a non-local operation that should strictly be applied to 3D spatial volumes at fixed time. Here, we apply it on the light cone noting that the induced error in radial direction should be small (roughly of order $a^2$) if we apply the first part of the Poisson operator first and if $D^{(1)}(\tau)$ is approximately proportional to the scale factor. In this case, the first order term of the exponential expression
\beq
\exp \(D_{\vx'\vy'}\varphi_{\vy'}\)-1= D_{\vx'\vy'}\varphi_{\vy'}+\mathcal{O}(\varphi^2)\approx a(|\vx'|)\DiracDelta(\vx'-\vy') \varphi_{\vy'}+\mathcal{O}(\varphi^2),
\eeq
and therefore the first order time dependence of the overdensity field will partly be canceled by the $1/a$ in the Poisson equation before the inverse Laplacian rescales the field. This cancellation corresponds to the commonly known fact that the comoving gravitational potential is constant in a matter dominated Universe.

After computing the six independent entries of the tidal shear tensor, we integrate each of its components along the LOS.
Only after the integration we rotate the resulting tensors into their LOS coordinate systems and project out all entries which are not in the plane perpendicular to $\vr$. This change of order allows to efficiently combine three operations, that is the rotation into the LOS, projecting out non-perpendicular components of the tidal tensor, and the computation of shear components and convergence, in a single linear operation. We denote the corresponding operator $\mathcal{G}$ (Gamma-Projection-Rotation).

For one data point the implemented data model in operator notation reads
\beq
d_i=\ve_i=(R(\varphi))_i+\ve^{\mathrm{s}}_i+\mathbf{n}_i=\mathcal{G}_{ij} \mathcal{I}_{j\vr'} \mathcal{T}_{\vr'\vz'} \mathcal{P}_{\vz'\vy'} \[\mathrm{exp}\(D_{\vy'\vx'}\varphi_{\vx'}\)-1_{\vy'}\]+\ve^{\mathrm{s}}_i+\mathbf{n}_i.
\eeq
The total data vector ${d}$ has dimensions $2\times N_s$, i.e. two ellipticity components for each of the $N_s$ source galaxies. 
We use the letter ${R}$ to encode the total response of the data to a signal $\varphi$, i.e. the composite action of all operators. We have also added an experimental noise $\mathbf{n}$ here for completeness. In general this will be subdominant to the shape noise $\ve^{\mathrm{s}}$ and we ignore it in the following. Note, however, that the formalism allows for an incorporation of several independent noise sources.
\section{MAP estimator}
\label{Sec:MAPesti}
Our aim is to obtain a maximum a posteriori (MAP) estimate of the signal field $\varphi$. The posterior distribution is related to the likelihood and the prior by Bayes' theorem
\beq
\label{eq:Bayes}
P(\varphi|d)=\frac{P(d|\varphi)P(\varphi)}{P(d)}.
\eeq
The prior probability $P(\varphi)$ is modeled as a Gaussian distribution with covariance $\Phi$. In a first simple approximation $\Phi$ can be taken to be diagonal in Fourier space with the usual power spectrum of matter overdensities $P_\mathrm{lin}(k,a)$. 

To obtain the likelihood, we marginalize over all realizations of the shape noise
\beq
\label{eq:marg_likeli}
P(d|\varphi)=\int \mathcal{D}\ve^{\mathrm{s}}\, P\(d|\varphi,\ve^{\mathrm{s}}\)P\(\ve^{\mathrm{s}}\)=\mathcal{N}\[d|R(\varphi),N\],
\eeq
where the covariance $N$ of the shape noise was defined in Eq.~\eqref{eq:shape_noise} and we neglect any other sources of measurement noise.
With this, the negative log-posterior becomes
\beq
\label{eq:posterior}
-\ln P(\varphi|d)\,\widehat{=}\,\frac{1}{2}\[d-{R}(\varphi)\]^\dagger N^{-1}\[d-{R}(\varphi)\]+\frac{1}{2}\(\varphi-\bar{\varphi}\)^\dagger \Phi^{-1}\(\varphi-\bar{\varphi}\),
\eeq
where we have dropped most terms that are independent of the field of interest, $\varphi$. 

The maximum of the posterior distribution is found by minimizing the expression in Eq.~\eqref{eq:posterior}. Note that the posterior distribution is not Gaussian. Due to the exponential in the response ${R}$ it is not quadratic in $\varphi$. To find the minimum of the negative log-posterior we apply a Newton-like minimization scheme~\citep{BFGS}. This requires the derivative of the negative log-posterior with respect to $\varphi$
\beq
-\frac{\delta \ln {P}(\varphi|d)}{\delta \varphi_{\mathbf{u}}}=\Phi^{-1}_{\mathbf{u}\vq}\(\varphi-\bar{\varphi}\)_{\vq}+\[d-{R}(\varphi)\]_{i}N^{-1}_{ij}\mathcal{G}_{jk} \mathcal{I}_{k\vr'} \mathcal{T}_{\vr'\vz'} \mathcal{P}_{\vz'\vy'} \[\mathrm{exp}(D_{\vy'\vx'}\varphi_{\vx'})*D_{\vy'\mathbf{u}}\],
\eeq
where the star denotes a pointwise product in position space, i.e. $(\alpha * \beta)_{\vx}=\alpha_{\vx}\beta_{\vx}$.
\section{Redshift-marginalized likelihood and posterior}
\label{sec:z-marg-Likelihood}
We can take into account the source redshift uncertainty by generalizing the marginalized likelihood in Eq.~\eqref{eq:marg_likeli} to include the probability of the source redshifts $z_s$ given their measured photometric redshifts $z_{\mathrm{photo}}$. This probability is given by the posterior redshift distribution function $P(z^s|z_{\mathrm{photo}})$, 
where $z^s$ and $z_{\mathrm{photo}}$ are vectors of the redshifts of all sources. 
Including this redshift posterior in the likelihood yields,
\beq
\label{eq:z_likeli}
P(d|\varphi)=\int \mathcal{D}\ve^{\mathrm{s}} \int \mathcal{D} z^s\, P(d|\varphi,\ve^{\mathrm{s}},z^s)P(\ve^{\mathrm{s}})P(z^s|z_{\mathrm{photo}})=\int \mathcal{D} z^s\, \mathcal{N}[d| R^{z^s}(\varphi),N]P(z^s|z_{\mathrm{photo}}).
\eeq
In most cases, this integration cannot be done analytically and the resulting distribution will in general not be Gaussian. In the absence of further information about the general shape of this distribution, a Gaussian characterized by the first and second central moment of the full distribution is the most conservative approximation one can use (it is the maximum entropy approximation and therefore adds the least additional information). This motivates the use of a Gaussian to model the redshift-marginalized likelihood.

To obtain the first and second moment, we need to calculate $\langle d\rangle_{P(d|\varphi)}$ and $\langle d d^\dagger\rangle_{P(d|\varphi)}$.
For the expectation value of the data, we obtain
\begin{align}
\langle d \rangle_{P(d|\varphi)}&=\int \mathcal{D}z^s\, \int \mathcal{D}d\, d\, \mathcal{N}[d| R^{z^s}(\varphi),N]\, P(z^s|z_{\mathrm{photo}})
=\int  \mathcal{D} z^s\, R^{z^s}(\varphi)\,P(z^s|z_{\mathrm{photo}})\\
\label{eq:exp_data}
\langle d_i \rangle_{P(d|\varphi)}&=\mathcal{G}_{ij} \[\int \dd z^s\, P(z^s|z_{\mathrm{photo}})\,\mathcal{I}_{j\vr}^{z^s}\]\, \bar{R}(\varphi)_{\vr}=\mathcal{G}_{ij}\[\int\limits_{0}^{\infty}\dd r\, \int\limits_{r}^{\infty}\dd r^j\, W(r,r^j) P(z^s_j|z_{\mathrm{photo}_{j}})\frac{dz^s_j}{dr^j}\]\bar{R}(\varphi)_{\vr}\\
\nonumber
&\equiv\[\mathcal{G}\bar{\mathcal{I}}\bar{R}\](\varphi)_{i}\equiv\tilde{R}(\varphi)_{i},
\end{align}
where $\bar{I}$ denotes the redshift averaged integration operator defined in the square brackets in the first line of Eq.~\eqref{eq:exp_data} and we have introduced $\bar{R}(\cdot)\equiv\mathcal{T}\mathcal{P}\(\mathrm{exp}\[D(\cdot)\]-1\)$ to summarize the action of all source-redshift independent operators.

The second moment is 
\beq
\label{eq:2ndmom}
\langle d d^\dagger\rangle_{P(d|\varphi)}=N+\langle(R(\varphi))(R(\varphi))^\dagger\rangle_{P(z|z_{\mathrm{ph}})}.
\eeq
Non-diagonal elements of the second term in Eq.~\eqref{eq:2ndmom} read
\beq
\langle\((R(\varphi))_i(R(\varphi))_j\)\rangle_{P(z|z_{\mathrm{ph}})}=\langle {(R(\varphi))}_i \rangle_{P(z|z_{\mathrm{ph}})}\langle{(R(\varphi))}_j\rangle_{P(z|z_{\mathrm{ph}})}=(\tilde{R}\varphi)_i(\tilde{R}\varphi)_j
\eeq
and diagonal elements are
\begin{align}
\label{eq:RR}
\langle (R(\varphi))_i(R(\varphi))_i \rangle_{P(z|z_{\mathrm{ph}})}
&=\mathcal{G}_{ij}\mathcal{G}_{ij}\[ \int\limits_{0}^{\infty}\dd r \int\limits_{0}^{\infty}\dd r'\, \int\limits_{\mathrm{max}(r,r')}^{\infty}\dd r^j\, W(r,r^j)  W(r',r^j) P(z^s_j|z_{\mathrm{photo}_{j}})\frac{dz^j}{dr^j} \]\bar{R}(\varphi)_{\vr}\,\bar{R}(\varphi)_{\vr'}\\ \nonumber
 &\equiv\mathcal{G}_{ij}\mathcal{G}_{ij}\tilde{\mathcal{I}}_{j \vr,\vr'}\bar{R}(\varphi)_{\vr}\,\bar{R}(\varphi)_{\vr'},
\end{align}
where the new operator $\tilde{\mathcal{I}}$ denotes the squared average of the integration operator, i.e. the square brackets in the first line of Eq.~\eqref{eq:RR}.

The Gaussian approximation to the likelihood is then $\mathcal{N}[d|\tilde{R}(\varphi),\tilde{N}]$, where $\tilde{N}=N+Q$ and $Q$ is 
\begin{align}
Q=\langle (R(\varphi))_i(R(\varphi))_j \rangle_{P(z|z_{\mathrm{ph}})}-\langle (\tilde{R}\varphi)_i \rangle_{P(z|z_{\mathrm{ph}})} \langle (\tilde{R}\varphi)_j\rangle_{P(z|z_{\mathrm{ph}})}=
\langle (R(\varphi))_i(R(\varphi))_i \rangle_{P(z|z_{\mathrm{ph}})}-(\tilde{R}\varphi)_i (\tilde{R}\varphi)_i.
\end{align}
This expression is still signal-dependent and we approximate it further by replacing $\varphi$ by its posterior mean $\langle\varphi\rangle_{P(\varphi|d)}$. Since this mean depends on $\tilde{N}$, the resulting set of equations must be solved iteratively.

To estimate the effect of the additional redshift-marginalization, we compare the effective lensing efficiency that appears in the redshift-marginalized data model (Eq.~\ref{eq:exp_data}), 
\beq
W_\mathrm{eff}(r,r^s)= \int\limits_{r}^{\infty}\dd r^s\, W(r,r^s) P(z^s|z_{\mathrm{photo}})\frac{dz^s}{dr^s},
\eeq
to the exact lensing efficiency $ W(r,r^s)$. 

\begin{figure}
\begin{center}
\includegraphics[scale=0.5,valign=t]{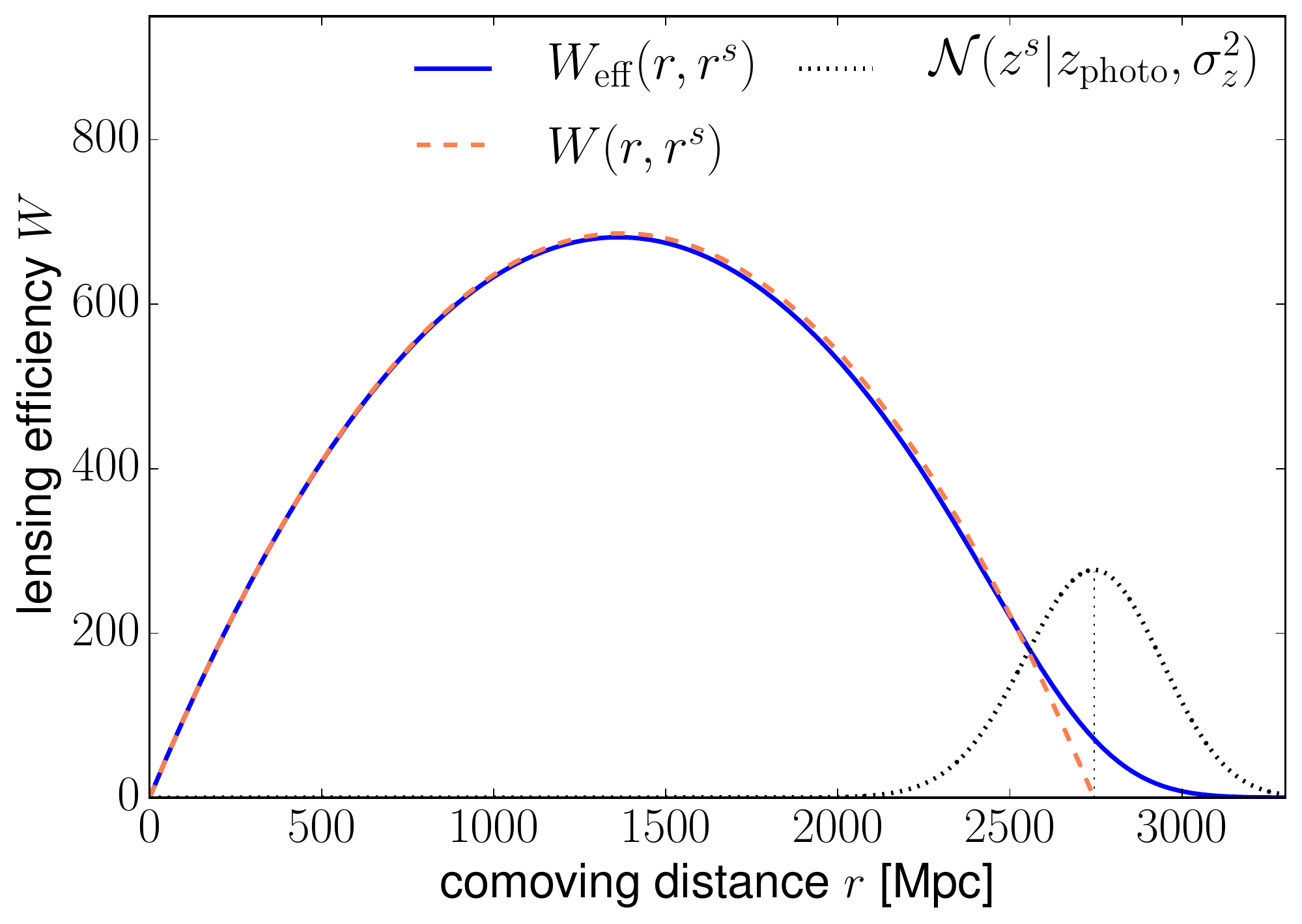}
\includegraphics[scale=0.5,valign=t]{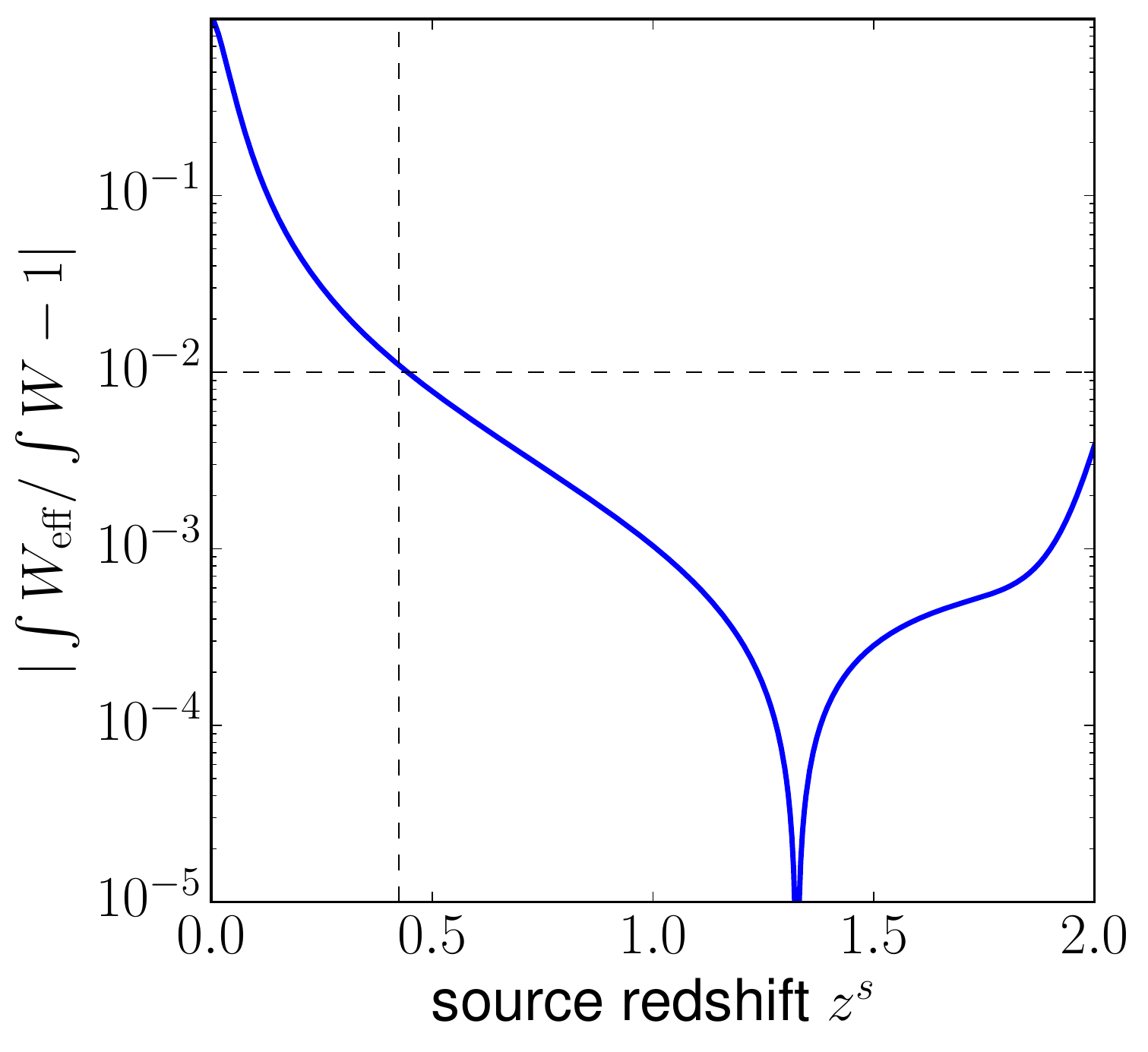}
\end{center}
\caption{\label{fig:EffzKernel}Left Panel: In the presence of redshift uncertainties the expected measured galaxy shape (i.e. the mean of its likelihood) is obtained by replacing the lensing efficiency $W\left[r,r^s(z^s)\right]$ with known source redshift $z^s$ (blue) by the redshift-marginalized effective lensing efficiency $W_{\mathrm{eff}}$ (orange-dashed) in the data model (Eq.~\eqref{eq:exp_data}). In the example shown, we have placed a source at redshift $z^s=0.8$ and assume a Gaussian photometric redshift uncertainty characterized by $\sigma_z/(1+z)\approx 0.04$. The redshift distribution function $P(z^s|z_\mathrm{photo})$ is shown for comparison (not normalized). In this example, the pointwise difference between the kernels is below $4\%$ for all $r<0.9r^s$. Right panel: Relative ratio of the integration volumes for different source redshifts assuming the same redshift error. In this idealized case (we assume a homogeneous density contrast along the line of sight), the effect of replacing the lensing kernel is sub-percent for source redshifts $z^s>0.43$ as indicated by the dashed lines.}
\end{figure}

In the left panel Fig.~\ref{fig:EffzKernel} we plot both kernels for a source at redshift $z^s=0.8$ and a Gaussian redshift posterior with variance $\sigma_z=0.04(1+z^s)$. For this particular configuration we find that both integration kernels are very similar over most of the distance to the source, i.e. their relative difference never exceeds 4\% for $r<0.9 r^s$. Only very close to the galaxy the kernels start to differ significantly. At this point the amplitude of the kernels has already dropped by more than $80\%$. The contribution to the lensing signal from structures with $r>0.9r^s$ is therefore strongly suppressed.

In the right panel of Fig.~\ref{fig:EffzKernel} we plot the relative difference of the integration volumes of the kernels assuming a uniform density contrast along the line of sight and the same Gaussian redshift posterior with $\sigma_z=0.04(1+z^s)$. We find that the difference between the kernels is sub-percent for sources with $z^s>0.43$.

These tests suggest that the redshift uncertainty should change the lensing signal of individual galaxies on average at the percent level.

We caution, however, that these tests are very simplified and only serve to get a rough estimate of the effect of redshift uncertainties, especially since we have ignored catastrophic outliers or the fact that redshift posteriors can be bimodial. For the proof of the concept of our reconstruction algorithm, we ignore redshift uncertainties. Following a test of their exact impact (e.g. by comparing a reconstruction with exact spectroscopic redshifts to one with redshifts obtained from the corresponding photometric redshift posteriors), they could be included in the way that has been outlined in this section.

\section{Validation and Tests}
\label{sec:validation}
To validate the implementation and assess the goodness of the tomographic reconstruction, we perform a number of increasingly realistic tests, which we denote tests A, B and C. In tests of type A we employ an even source distribution over the entire box. These tests are the least realistic ones and serve to validate the correctness of the implementation. Test B uses a survey-like source-redshift distribution and test C adds realistic shape noise. 

In all of these tests, we place the observer in the center of the bottom of the computational box\footnote{This position is not fixed by the algorithm. For a reconstruction from a survey that covers a significant fraction of the sky, the observer can be placed in the center of the box, for example.} and resolve the underlying and reconstructed overdensity fields with $128^3$ pixels. Depending on the test, we allow the physical sizes of the box to differ. The current pixel resolution is limited due to computation time and memory usage. A higher resolution will be accessible after parallelization and adaption of the code for the usage on a high-performance cluster.

The minimization is achieved through 300 steps of an LBFGS algorithm~\cite{BFGS} followed by a steepest descent algorithm. The latter is set to have reached convergence if the maximal pixel-wise relative difference between two subsequent field estimates is smaller than $10^{-4}$ for three iteration in a row.  If the steepest descent does not converge within 200 iterations, the same combination of LBFGS and steepest descent minimizers is repeated. We find that, at latest in this second run, all of our example reconstructions meet the convergence criterium. In the current serial implementation, the entire minimization takes between 1 day (for idealized tests at fixed redshifts) and 6 days (for a realistic source distribution and shape noise) on a 2.3 Ghz core of an Intel Xeon E5 processor.

Most of the tests are based on mock data that we create by applying the data model described in Sec.~\ref{sec:datamodel} to non-linear density fields obtained from N-body simulations.
For tests at fixed redshifts we use snapshots of the Millennium-XXL simulation~\cite{2012MNRASAngulo} from which we take the entire volume of size $[3 h^{-1}\,\Gpc]^3$. For more realistic test cases we construct a light cone of size $[500 h^{-1}\,\Mpc, 500 h^{-1}\,\Mpc, 4000 h^{-1}\,\Mpc]$ by joining snapshots of the Millennium Run \cite{2005NatureSpringel}. The Millennium box measures $500 h^{-1}\,\Mpc$ along each side and we shift it every time $500 h^{-1}\,\Mpc$ in the $z$-direction have been constructed. This procedure ensures that a LOS is unlikely to hit the same structure repeatedly.
In the resulting light cone we achieve a resolution of $3.9 h^{-1}\,\Mpc$ in the $x$- and $y$-directions corresponding to mildly non-linear scales and reach a redshift of $z=2.2$ in the $z$-direction. Note that the physical size of the pixels is eight times longer along the $z$-axis, meaning that we obtain a poorer resolution in radial direction.

We map the simulations onto our pixel grid by a nearest grid point interpolation. This procedure suppresses power on small scales and modifies the power spectrum by a kernel consisting of cardinal sine functions (see e.g.\cite{Jing2005}). To be consistent with this modification, we apply the same kernel to the theory power spectra which we use in the prior distributions. 

Both the Millennium-XXL and the Millennium simulation use a flat $\Lambda$CDM cosmology with $\OmegaMatter=0.25$ and $h=0.73$.

\subsection{Fidelity of the Reconstruction}
\label{sec:fidel}

The BFGS algorithm is a quasi-Newton method. While Newton methods rely on an exact evaluation of the inverse Hessian at each step, quasi-Newton methods estimate the curvature from previous evaluations of the gradient. The 'L' in L-BFGS refers to the limited memory version of this algorithm in which the inverse Hessian is approximated by a sparse matrix instead of a dense one. This additional approximation becomes necessary for problems with many degrees of freedom such as this. In principle the final estimate of the inverse Hessian could be used as an approximation for the pixel-pixel covariance, but it depends on the correlated previous steps of the algorithm and only probes the curvature along the search directions. 
The covariance, $C$, of a Gaussian posterior distribution, $\mathcal{G}(\varphi|d)$, with maximum $m$ can be estimated by sampling. A straightforward algorithm to obtain such an estimate is outlined in the following.
\begin{enumerate}
\item Draw random realizations of signal and noise from the Gaussian prior and noise distributions.
\beq
\varphi'\hookleftarrow \mathcal{G}(\varphi,\Phi) \ \mathrm{and}\ n'\hookleftarrow \mathcal{G}(n,N).
\eeq
\item Use the resulting realizations to model a data vector $d'$
\beq
d'=R\varphi'+n'.
\eeq
\item From this random data vector reconstruct the Wiener filter estimate $m'$, i.e. the maximum of the Gaussian posterior distribution, 
\beq
m'=C R^\dagger N^{-1} d', \ \mathrm{ with }\ C=[R^\dagger N^{-1} R + \Phi^{-1}]^{-1}.
\eeq
\item Repeat steps 1.-3. $N$ times and estimate $C$ from
\beq
\label{eq:cov_esti}
\hat{C}=\frac{1}{N} \sum_i^N (\varphi'_i-m'_i)(\varphi'_i-m'_i)^\dagger.
\eeq
\end{enumerate}
Computing the full covariance is computationally infeasible since it would require $\sim9$ Tb to store a full covariance matrix of size $1/2 n(n+1)$ for a map with $n=128^3$ pixels. Numerically feasible are, e.g., estimates of its diagonal (the pixel-wise covariance), or of the full covariance of 1D slices of the reconstruction box.

The above algorithm yields an unbiased estimator of the posterior covariance only if the data model is linear in the density field. The lognormal model is inherently non-linear and therefore the posterior non-Gaussian. However, for sufficiently small perturbation $\Delta m$ around the MAP estimate $m$ the Laplace approximation should be applicable, meaning that we can approximate the negative log posterior as a Gaussian,
\begin{align}
\label{eq:LapAp}
- \ln P(\varphi=m+\Delta m|d) & \approx \frac{1}{2}(\tilde{d}-R'\tilde{\varphi})^\dagger N^{-1} (\tilde{d}-R'\tilde{\varphi})+\frac{1}{2} {\tilde{\varphi}}^\dagger \Phi^{-1} \tilde{\varphi} \\
& \hat{=} -\ln \mathcal{G}(\tilde{\varphi}|d).
\end{align}
To obtain Eq.~\eqref{eq:LapAp}, we have linearized the response operator
\beq
R' = \frac{\delta R}{\delta \varphi}|_{\varphi=m},
\eeq
redefined the data vector $\tilde d=d-R(m)+R'(m-\bar{\varphi})$ and shifted the signal $\tilde{\varphi}=m+\Delta m-\bar{\varphi}$.  The covariance $C'$ of this Gaussian approximation to the posterior is
\beq
\label{eq:cov}
C'=\[R'^\dagger N^{-1}R' + S^{-1}\]^{-1}.
\eeq
This approximate covariance can be evaluated as outlined above. As an example, we estimate $\mathrm{diag}({C'})$, which can be interpreted as an uncertainty map for the reconstruction, for one of our test cases in the next section.

\subsection{Tests A: Simple geometry}
In a first series of tests we distribute 500 000 sources evenly in the box and slightly beyond ($\pm 100 h^{-1}\,\Mpc$). We place sources outside of the reconstructed volume in order to increase the area in which we can recover the underlying density field with high resolution. The quality of the reconstruction decreases with increasing distance to the observer because of increasing distances between lines of sight and a decreasing number of background sources that help to break the LOS degeneracy of the lensing kernel. We further only add unphysically low, negligible shape noise in this first test set.

We perform three different tests of type A for which we create data from increasingly realistic input fields:

1) A self-consistency test, where we create an overdensity field from a random realization of the Gaussian field $\varphi$ [Eq.~\eqref{eq:varphi_model}] in a cubic box of side length $1000 h^{-1}\,\Mpc$. We then apply the implemented data model and use the algorithm to recover this input field. We do not show any results from this test, since they would not provide any additional insights compared to tests A2 and A3, but simply state that we can recover the input field with high fidelity which means that the implementation is in itself consistent.

\begin{figure} 
\begin{center}
\includegraphics[scale=0.36,valign=t]{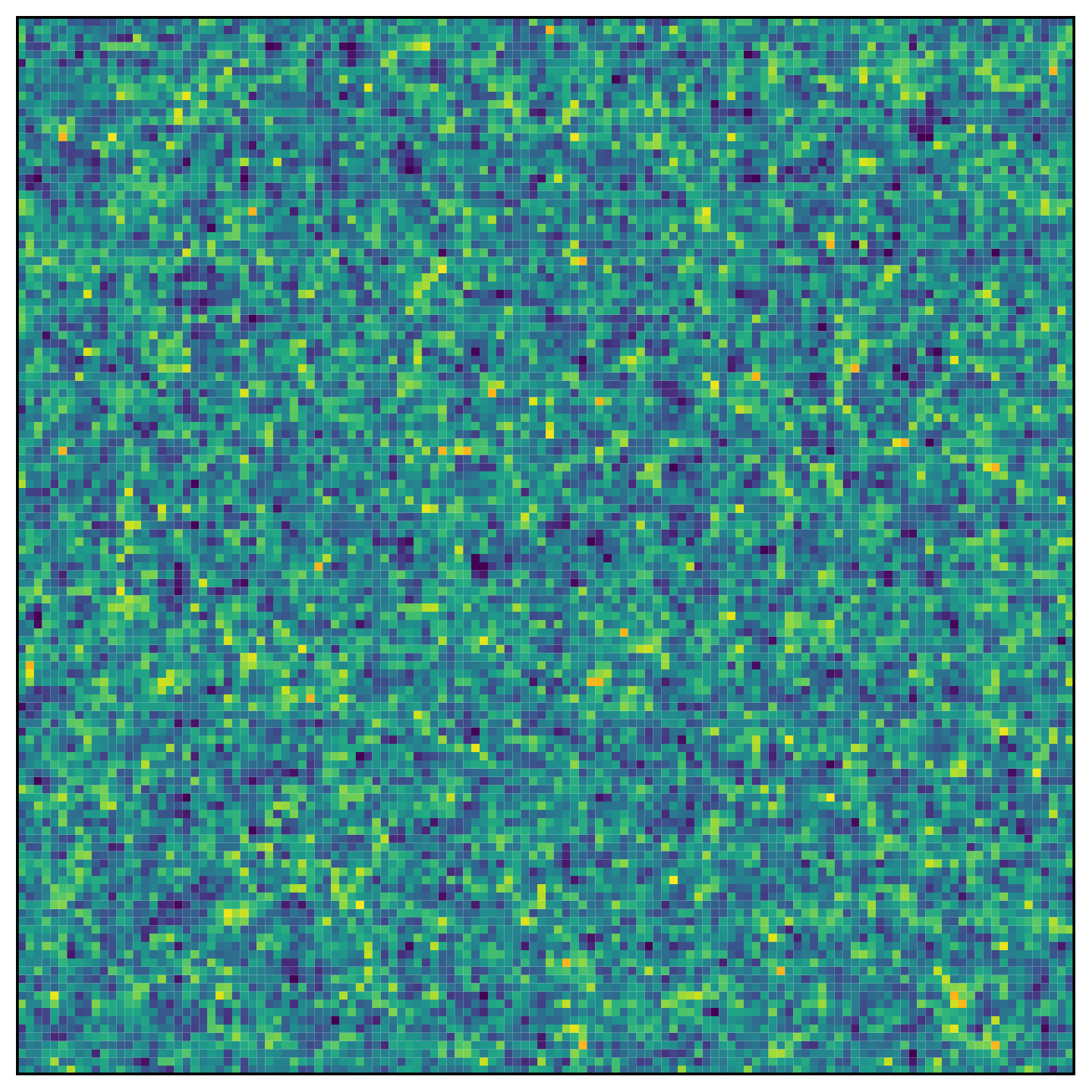}
\includegraphics[scale=0.36,valign=t]{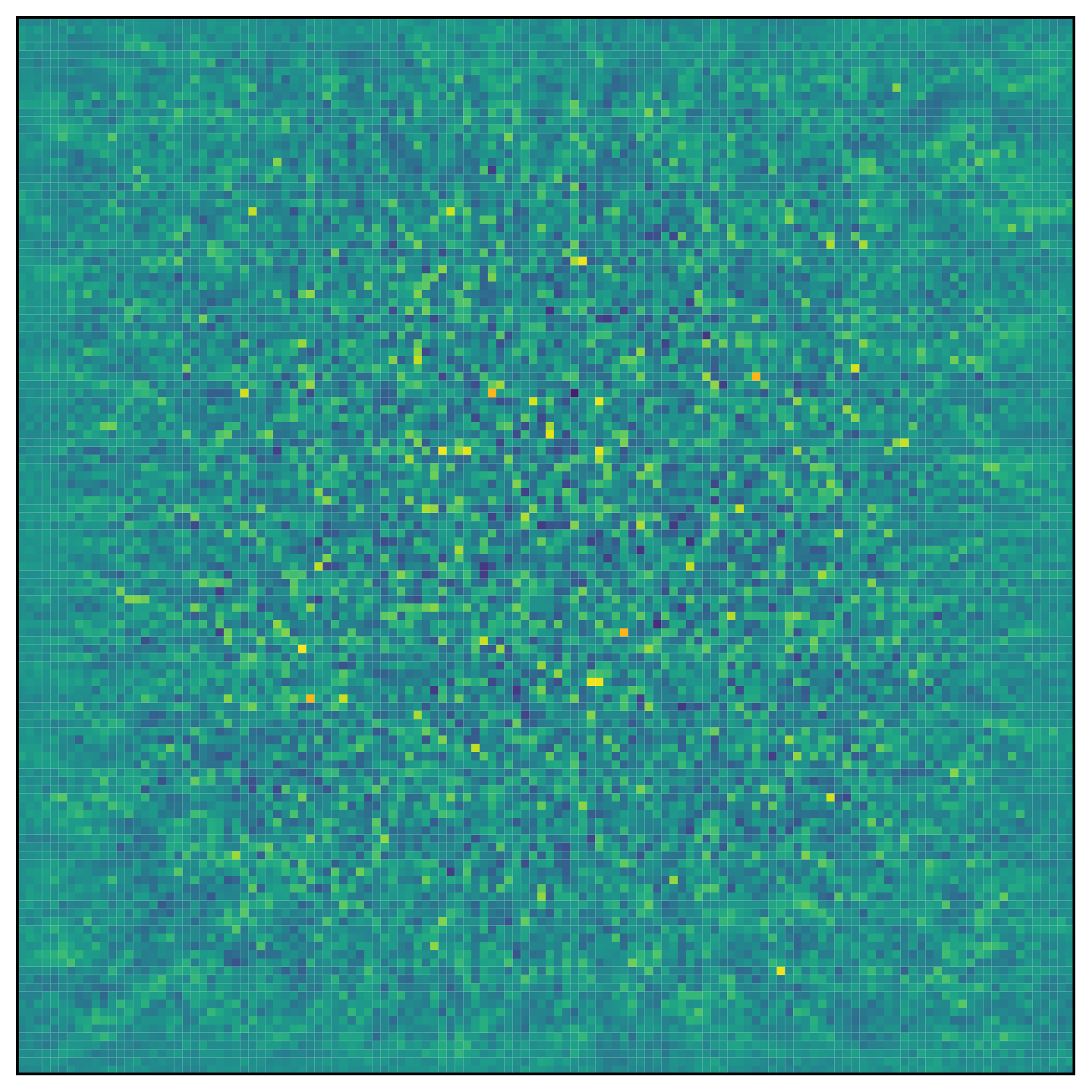}
\includegraphics[scale=0.36,valign=t]{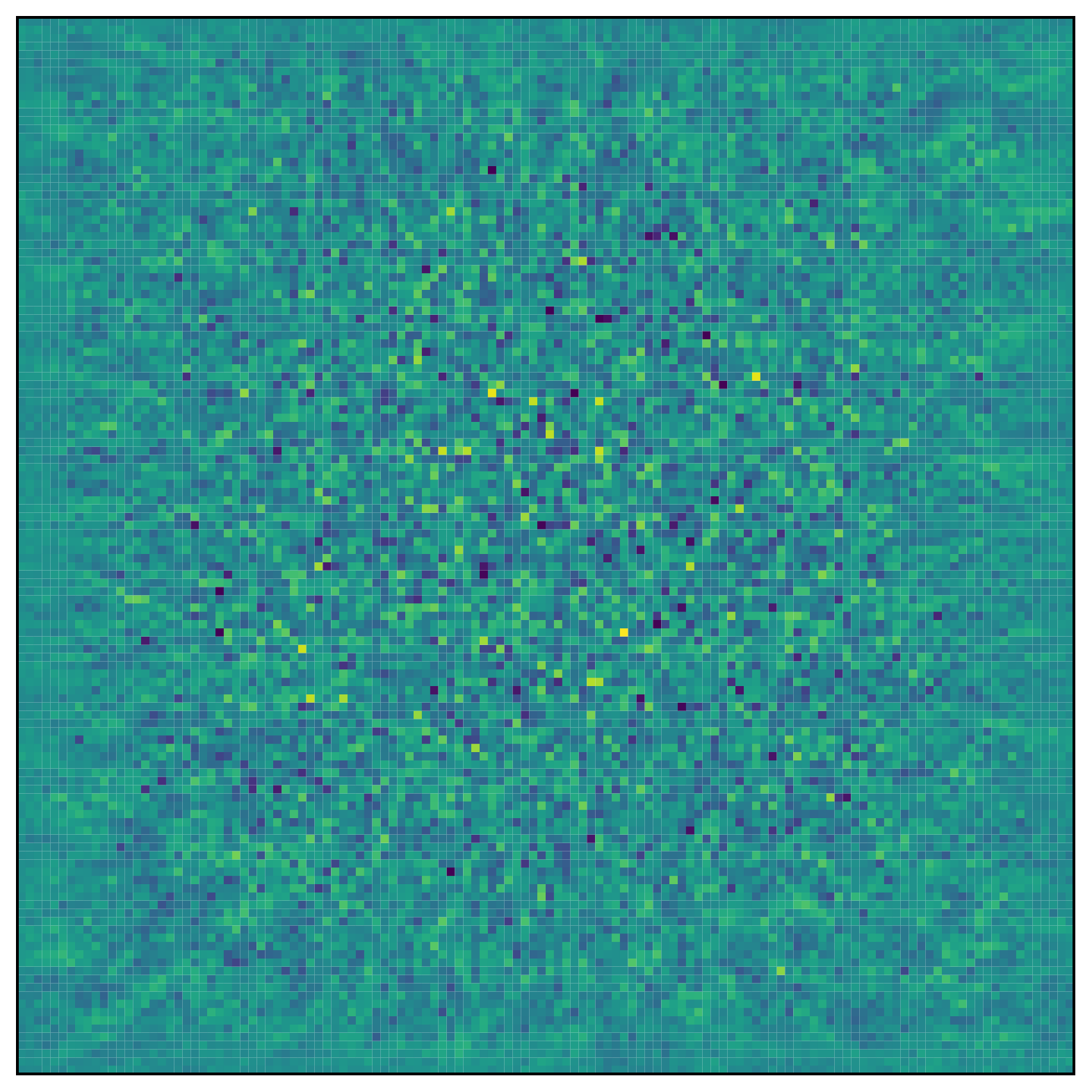}

\includegraphics[scale=0.36,valign=t]{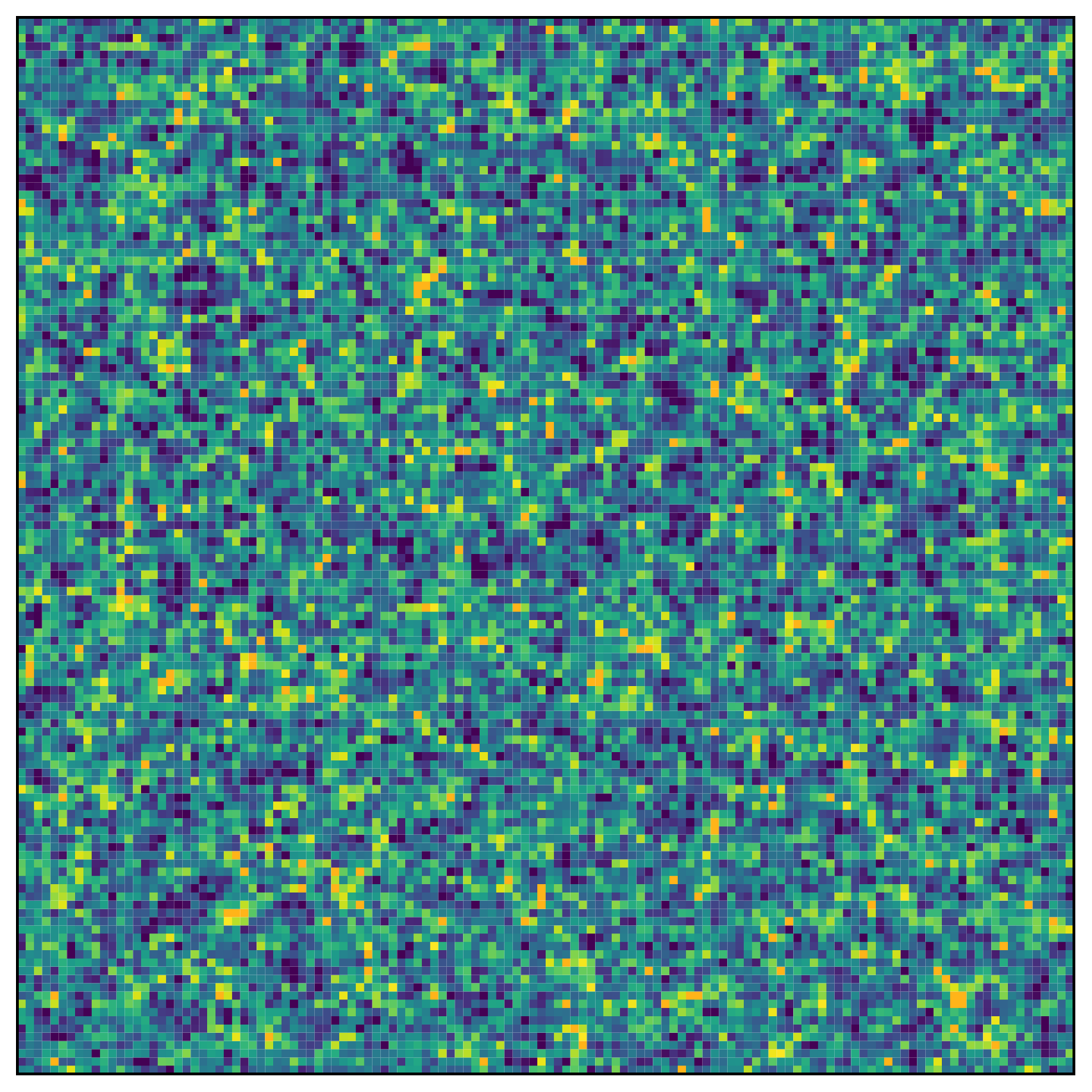}
\includegraphics[scale=0.36,valign=t]{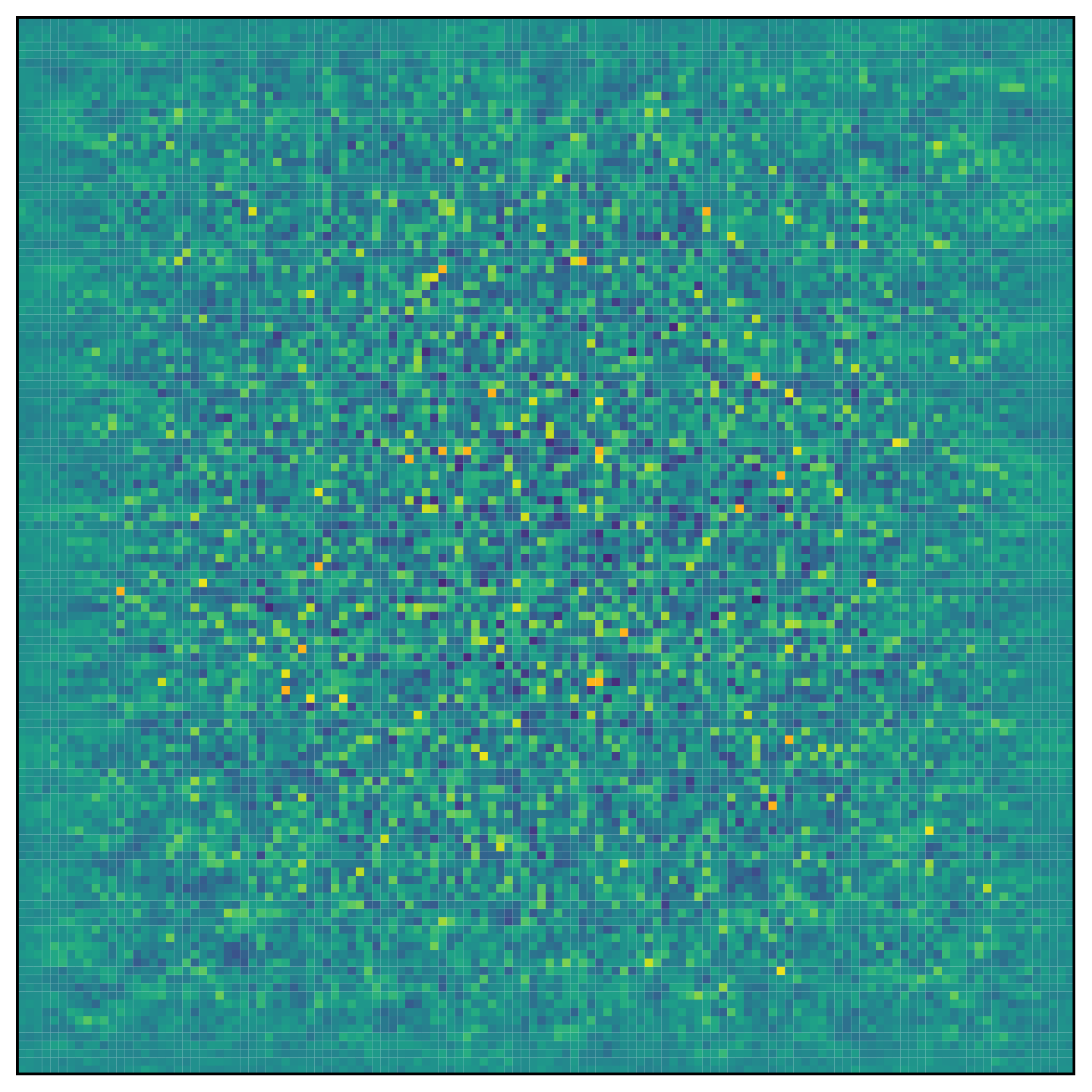}
\includegraphics[scale=0.36,valign=t]{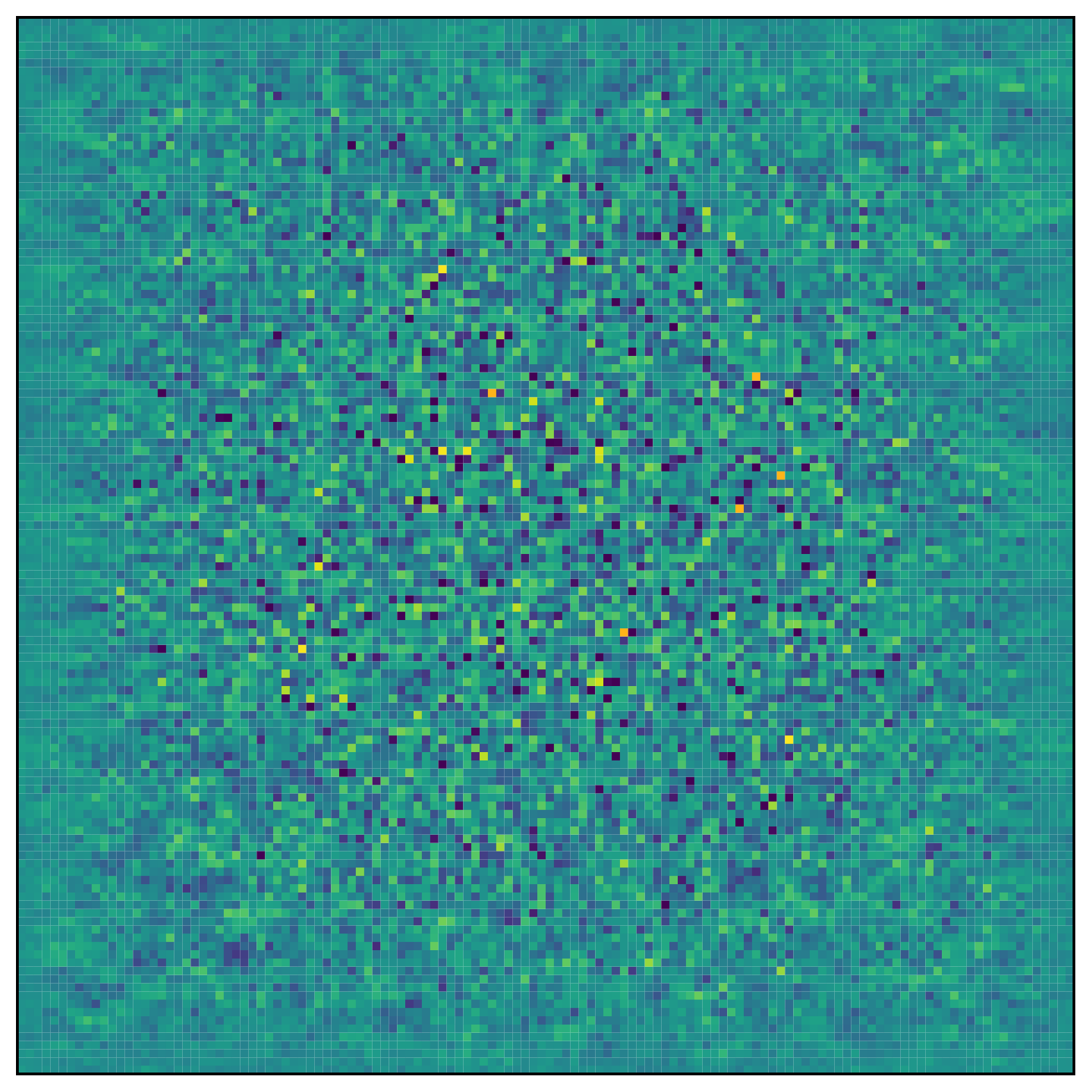}

\includegraphics[scale=0.36,valign=t]{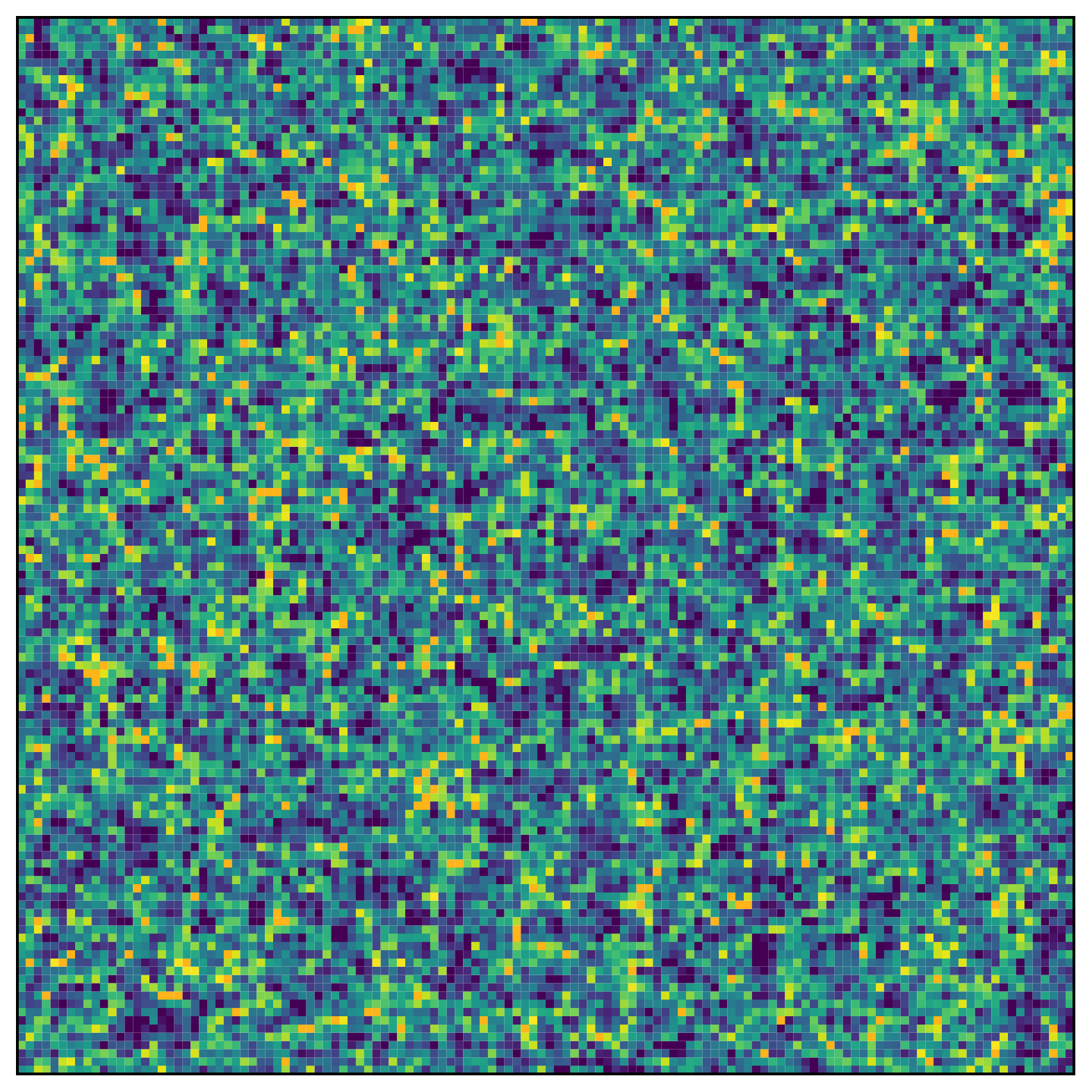}
\includegraphics[scale=0.36,valign=t]{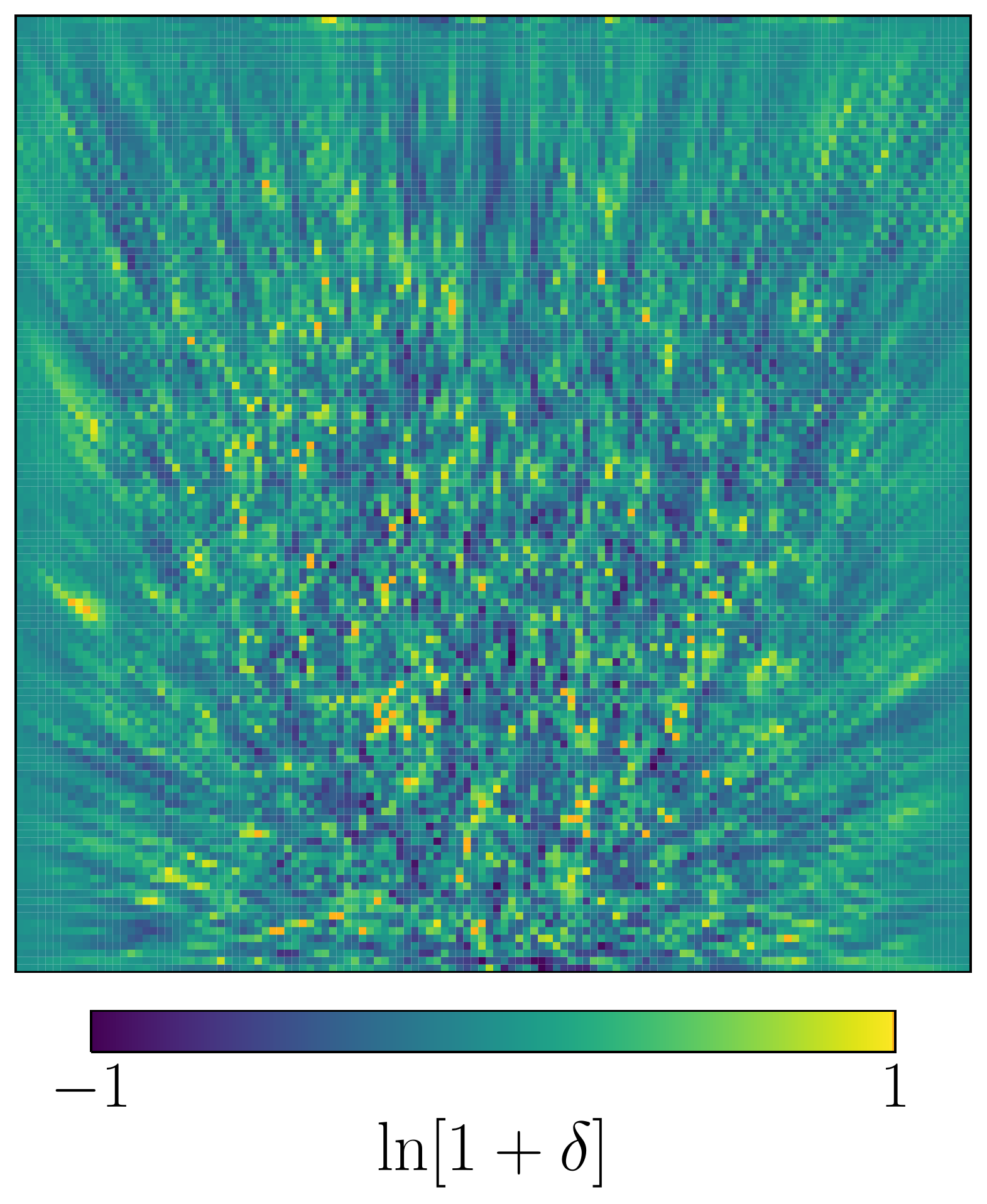}
\includegraphics[scale=0.36,valign=t]{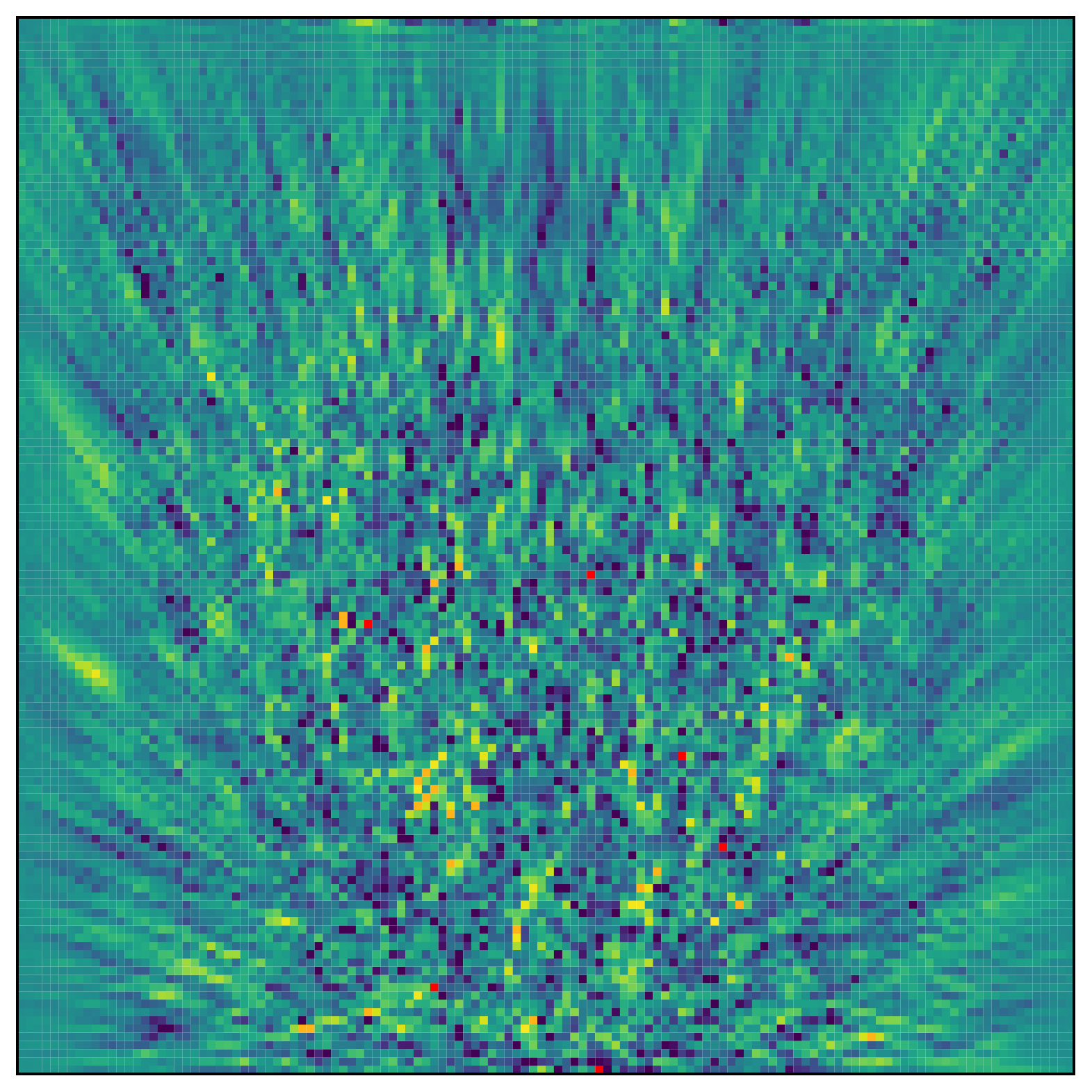}
\end{center}
\caption{\label{fig:TestA2Pics} Qualitative comparison of reconstruction methods in the test series A2, for which we create mock data from individual simulation snapshots, apply an even source distribution and add negligible shape noise. We show central slices through the 3-dimensional fields. The observer is located in the center of the $(x-y)$-plane at the origin of the $z$-axis. The redshift of the snapshot decreases from top to bottom, $z=[1,\,0.25\,,0.]$. In the two upper rows we show the $(x-y)$ plane at $z=\mathrm{max}(z)/2$, in the last row we look at the $(x-z)$-plane at $y=\mathrm{max}(y)/2$. The fields are from left to right: the underlying density field from the simulation, its reconstruction with a lognormal prior and its reconstruction with a Gaussian prior (Wiener filter). We plot $\ln[1+\delta]$ and mark unphysical negative densities in the Wiener Filter reconstruction in red. The resolution of the reconstruction decreases with distance to the observer. This is because 1) the density of lines of sight decreases 2) there are less sources behind the point we want to reconstruct 3) the information from these sources is suppressed by the shape of the integration kernel. In all cases, the lognormal reconstruction is superior in capturing the highest values of the density field and avoids unphysically low density contrasts below -1. The Gaussian prior seems better in identifying low density regions.}
\end{figure}

\begin{figure}
\begin{center}
\subfloat[$z=1$]{
\scalebox{0.4}{\includegraphics{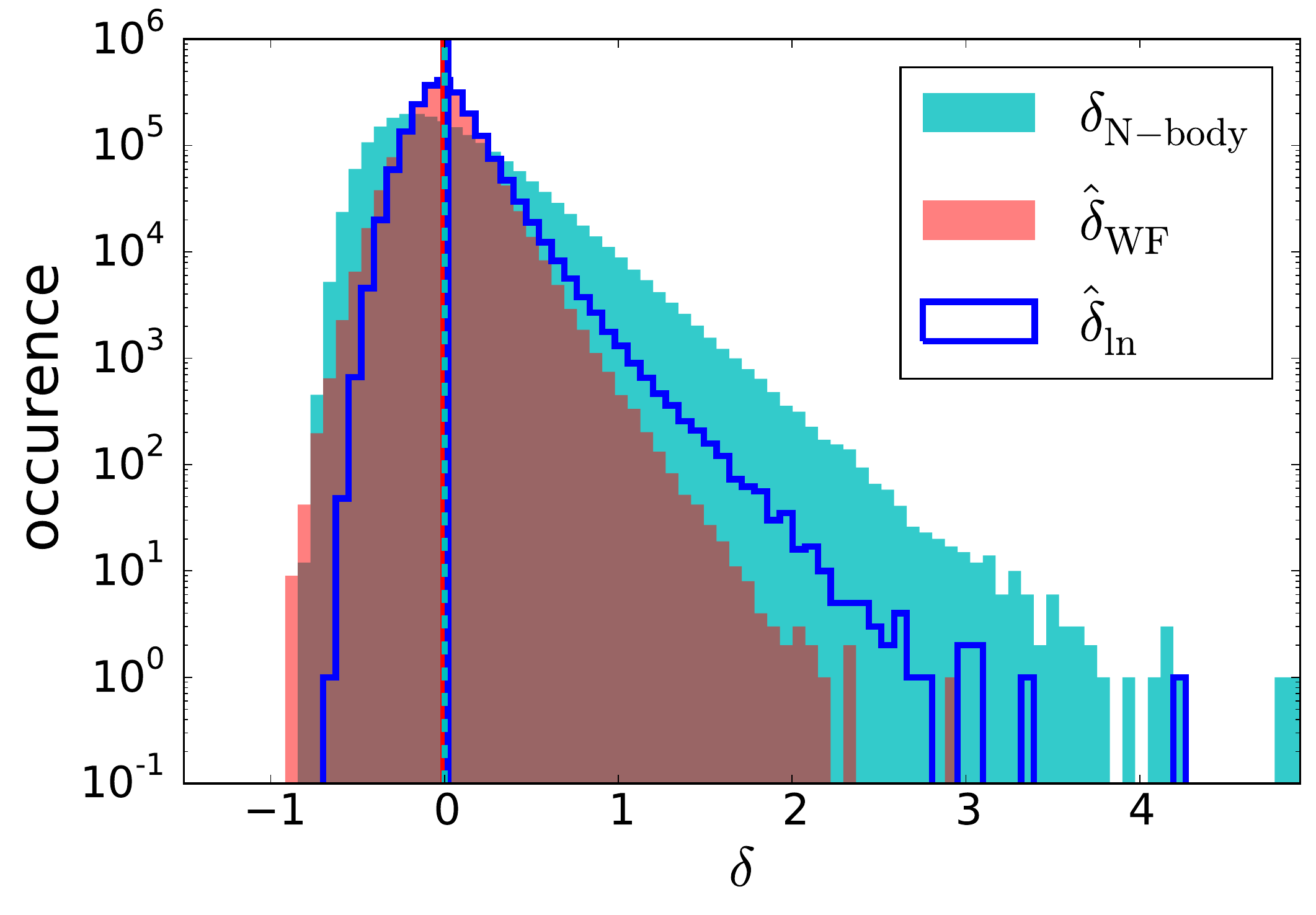}}}
\subfloat[$z=0.25$]{
\scalebox{0.4}{\includegraphics{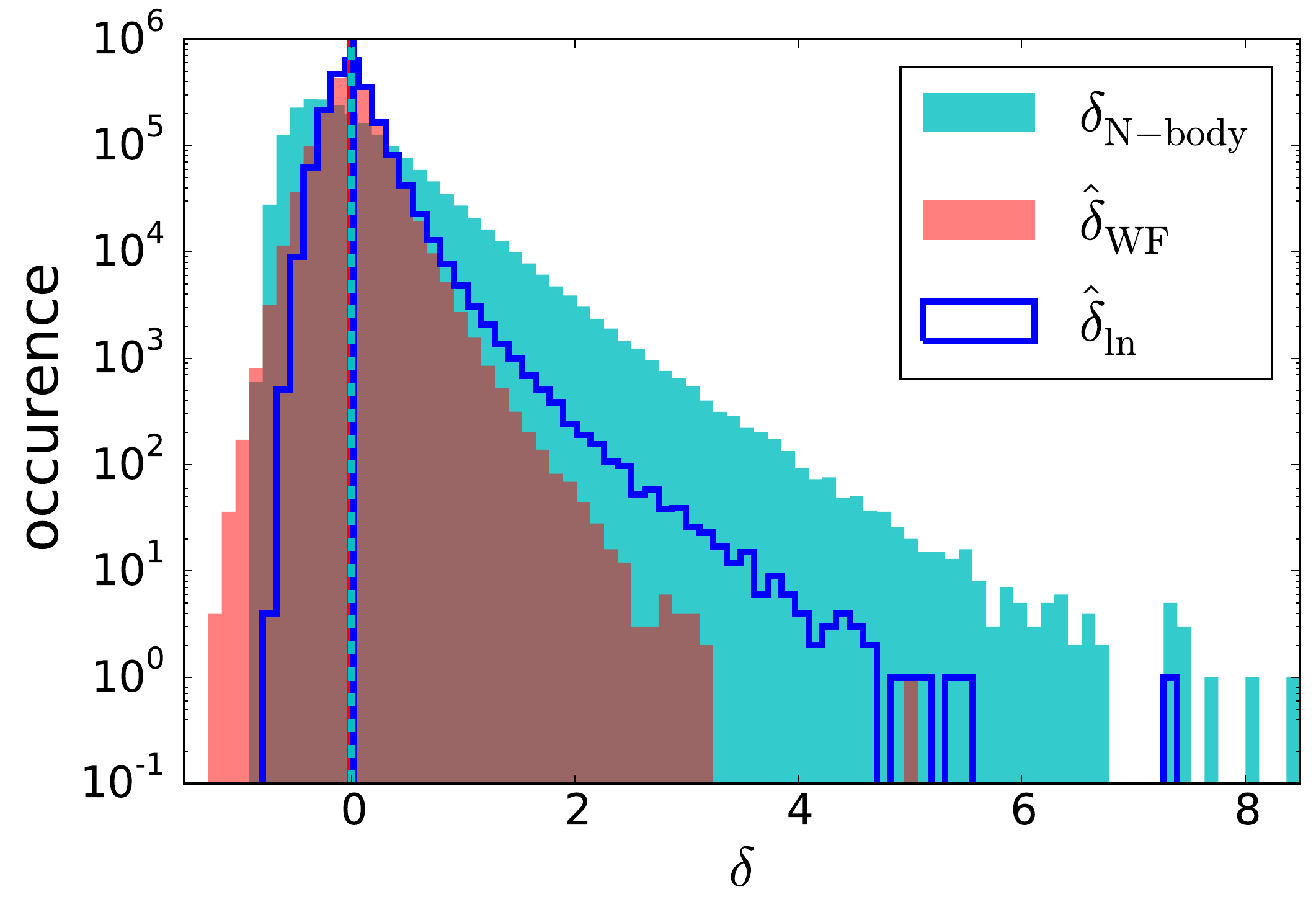}}}

\subfloat[$z=0$]{
\scalebox{0.4}{\includegraphics{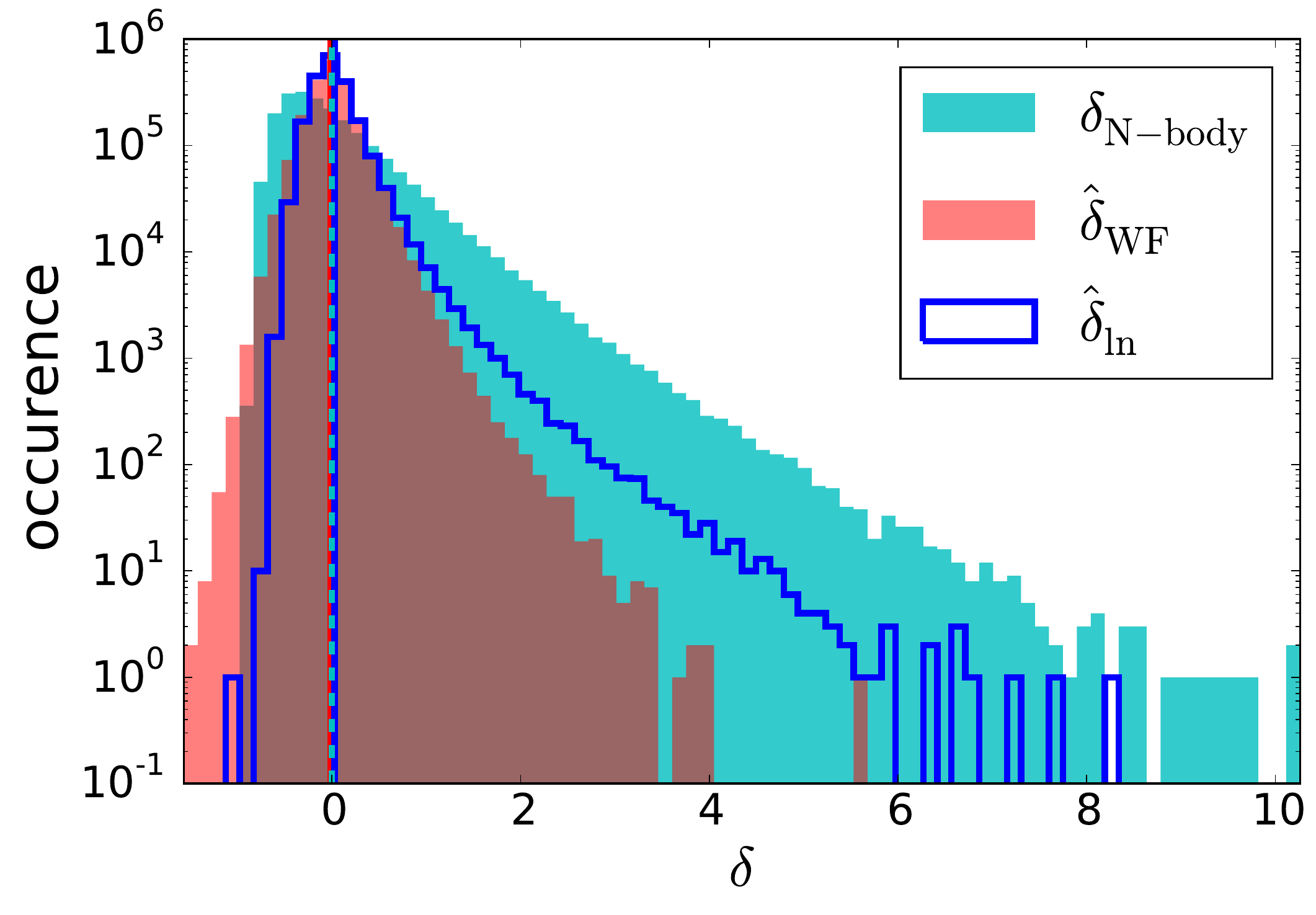}}}
\end{center}
\caption{\label{fig:TestA2PDF} 1-point PDFs of the simulated density contrasts (in cyan) and their Wiener filter and lognormal reconstructions (red and dark blue) in tests on mock data created from different snapshots of an N-body simulation with an even source distribution and negligible shape noise (A2). The PDF of the lognormal reconstruction is slightly more skewed than the Wiener filter PDF, which is closer to symmetric in all cases. Both the Wiener filter and the lognormal reconstruction are equally biased in the position of the peak, however, the mean values of both reconstructions agree well with the mean value of the underlying density. The mean values of the density contrast are indicated by vertical lines in the same color as their corresponding distribution. We also find that the Wiener filter produces unphysical density contrasts below -1 in snapshots at $z=0.25$ and $z=0$.}
\end{figure}

\begin{figure}
\begin{center}
\includegraphics[scale=0.37,valign=t]{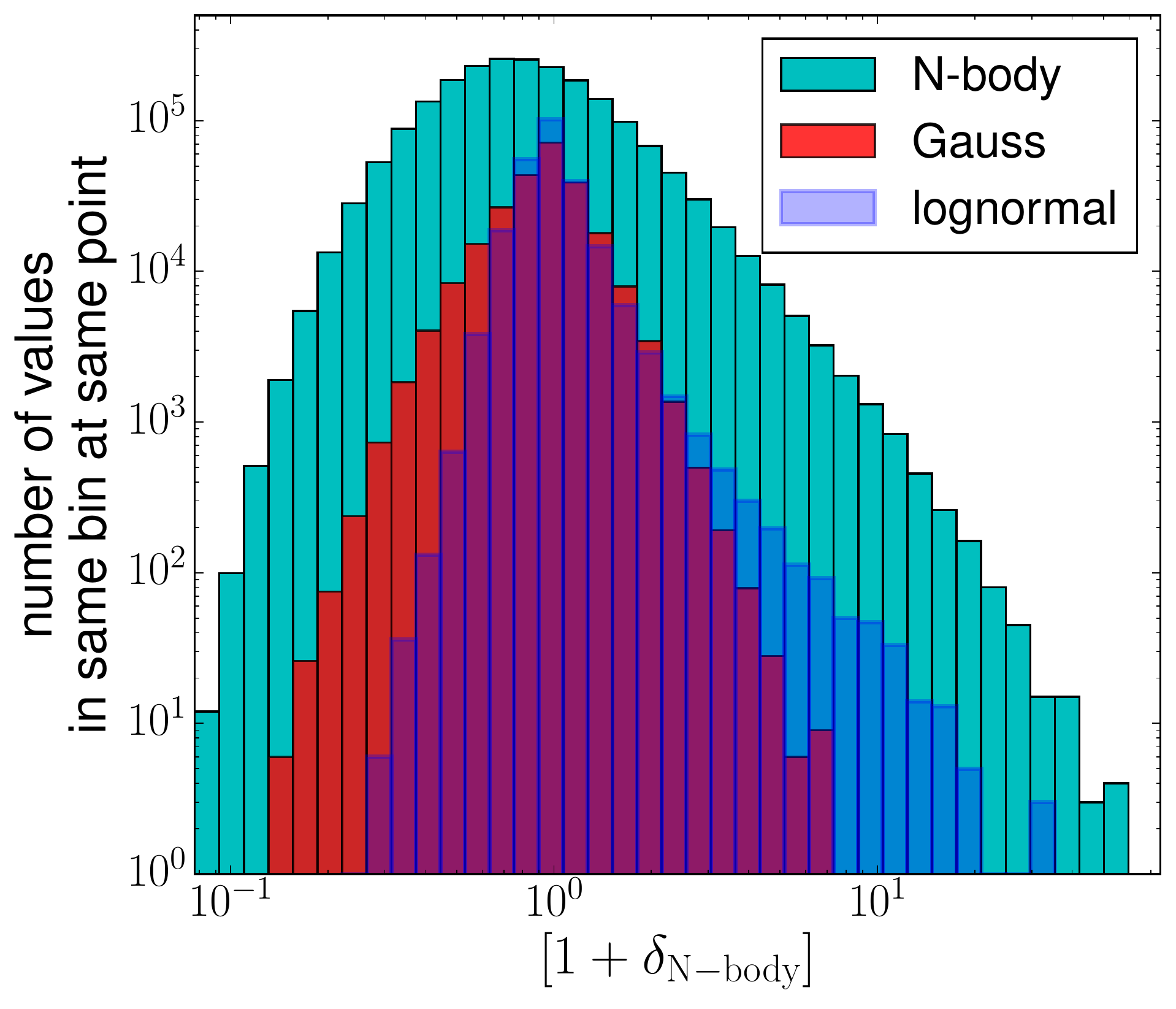}
\includegraphics[scale=0.37,valign=t]{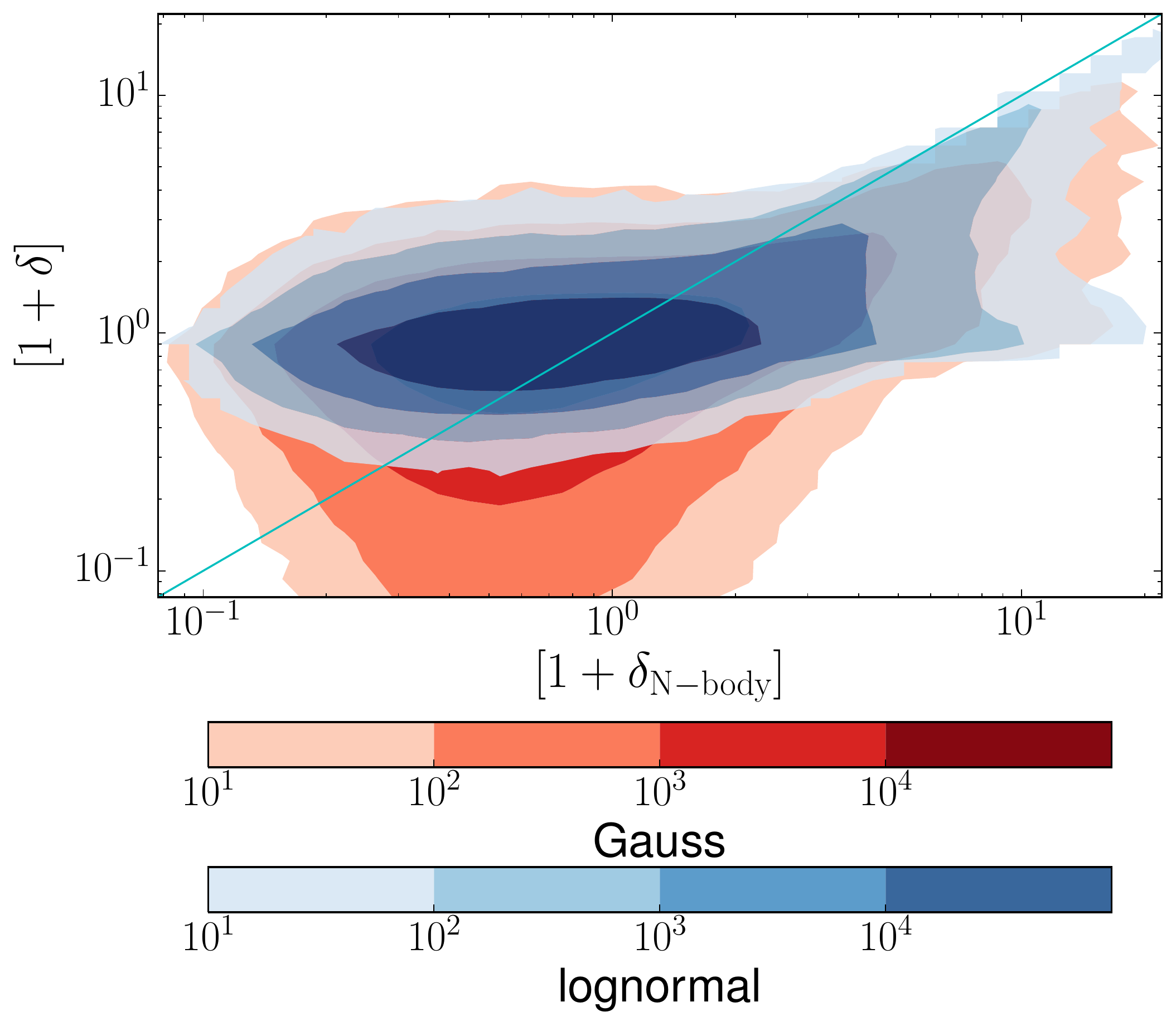}
\end{center}
\caption{\label{fig:TestA3PicsQuant} Comparison of reconstruction methods in test A3: In the left panel, we have binned the underlying density into 40 bins of equal width (0.17) in ln-space and count the points in the reconstructions whose density lies in the same bin at the same location. The total number of points in a density bin in the original density field is shown for comparison in the background (cyan). This plot shows that the lognormal model is better in capturing overdensities, while it is worse than the WF reconstruction in capturing underdensities (note the log scaling). In the right panel, we plot the full distributions of the reconstructions against the underlying density. A perfect reconstruction would follow the diagonal. The lognormal reconstruction achieves this better in the high density tail and prefers values around the mean in poorly constrained regions. The Gaussian prior tends to identify more low density regions.}
\end{figure}

2) A test in which we create shears from overdensity fields taken from an N-body simulation at fixed redshifts. We use three different snapshots at individual redshifts $z=1$, $z=0.25$ and $z=0$. This allows to assess the algorithm's ability to recover increasingly non-linear fields. 
In each case, we compare the lognormal reconstruction to a Wiener filter (WF) reconstruction, which uses the same data model and source distribution but a Gaussian prior on the overdensity field $\delta$. Comparisons between input and reconstructed fields in each case are shown in Fig.~\ref{fig:TestA2Pics}. The resolution of the reconstruction decreases with distance to the observer. This is because 1) the density of lines of sight decreases 2) there are less sources behind the point we want to reconstruct 3) the information from these sources is suppressed by the shape of the integration kernel. In Table~\ref{tab:TestAb} we compare the minimal and maximal values in the reconstructions to the extremal values in the true density field. The lognormal reconstruction is superior in capturing the highest values of the density field and avoids unphysically low density contrasts below -1. Table~\ref{tab:TestAa} summarizes pixel-wise quadratic differences and the pixel-wise Pearson correlation between underlying and reconstructed fields. For all redshifts tested the lognormal prior yields higher correlations with the original field than the Gaussian prior. The difference between lognormal and WF reconstruction increases as the input field becomes more non-linear. This is also reflected in the 1-point probability distribution functions (PDFs), which we show in Fig.~\ref{fig:TestA2PDF}. While the Wiener Filter PDF is closer to symmetric in all cases, we can capture part of the skewness of the input field by applying the lognormal prior. An notable feature of the 1-point PDFs is that the maximum value is slightly biased in both reconstructions. The distributions of the reconstructions peak at zero, i.e. at their mean, while the underlying field peaks below. This is a feature of both priors since they prefer the mean density (or a zero density contrast) if the data does not contain enough information on the density.\footnote{To be precise one should note that the maximum of the lognormal posterior lies slightly below the mean density. However, the difference to the mean turns out to be small in all test cases. For the light cone box we have calculated that the prior distribution peaks at $\delta=-0.1$. This value is consistent with the lognormal reconstructions that we obtain in our test cases. Closely inspecting the right panels of Figs.~\ref{fig:TestA3PicsQuant} and ~\ref{fig:TestB2Picsb}, one can see that the bulge of the reconstructed densities lies slightly below $1$, at $\delta+1\approx0.9$).}

\begin{table}[htb]
\centering
\begin{tabular}{l| l l l | l l l }
redshifts & \multicolumn{3}{c|}{$\mathrm{max}(\hat{\delta})$, $\mathrm{max}({\delta})$} & \multicolumn{3}{c|}{$\mathrm{min}(\hat{\delta})$, $\mathrm{min}({\delta})$}\\
\hline
 & lognormal & WF & N-body & lognormal & WF & N-body  \\
\hline
1. 	& 4.22 & 2.95 & 4.92 & -0.63 & -0.91 & -0.79\\
0.25& 7.38 & 5.00 & 8.49 & -0.77 & -1.28 & -0.87\\
0.0 & 8.32 & 5.55 &10.26 & -0.99 & -1.57 & -0.89\\
light cone & 32.90 & 20.81 & 70.20 & -0.98 & -4.28 & -0.92
\end{tabular}
\caption[Comparison of Lognormal to Normal Reconstruction - Extremal Values]{\label{tab:TestAb} Quantitative comparison of reconstruction methods for tests A2 and A3. We compare the minimal and maximal values of the underlying and reconstructed fields ($\delta$ and $\hat{\delta}$, respectively).}
\end{table}

\begin{table}[htb]
\centering
\begin{tabular}{l| l  l | l l}
redshifts & \multicolumn{2}{c|}{$\langle(\hat{\delta}-\delta)^2\rangle/\sigma^2_\delta$} &\multicolumn{2}{c|}{$\sqrt{\langle \hat{\delta} \delta \rangle /(\sigma_\delta \sigma_{\hat{\delta}})}$}\\
\hline
 & lognormal & WF & lognormal & WF \\
\hline
1. 		& 0.66 & 0.69 & 0.58 & 0.56\\
0.25 	& 0.73 & 0.77 & 0.52 & 0.49\\
0.0 	& 0.73 & 0.77 & 0.53 & 0.48\\
light cone & 0.65 & 0.73 & 0.61 & 0.51
\end{tabular}
\caption[Comparison of Lognormal to Normal Reconstruction - Correlations]{\label{tab:TestAa} Quantitative comparison of reconstruction methods for tests A2 and A3. We compare the lognormal to the Gaussian prior in terms of the mean square pixel-wise difference (first column) and the Pearson-correlation coefficient (second column) between the reconstruction and the underlying field (which we denote $\hat{\delta}$ and $\delta$, respectively).}
\end{table}

\begin{figure}
\begin{center}
\includegraphics[scale=0.36,valign=t]{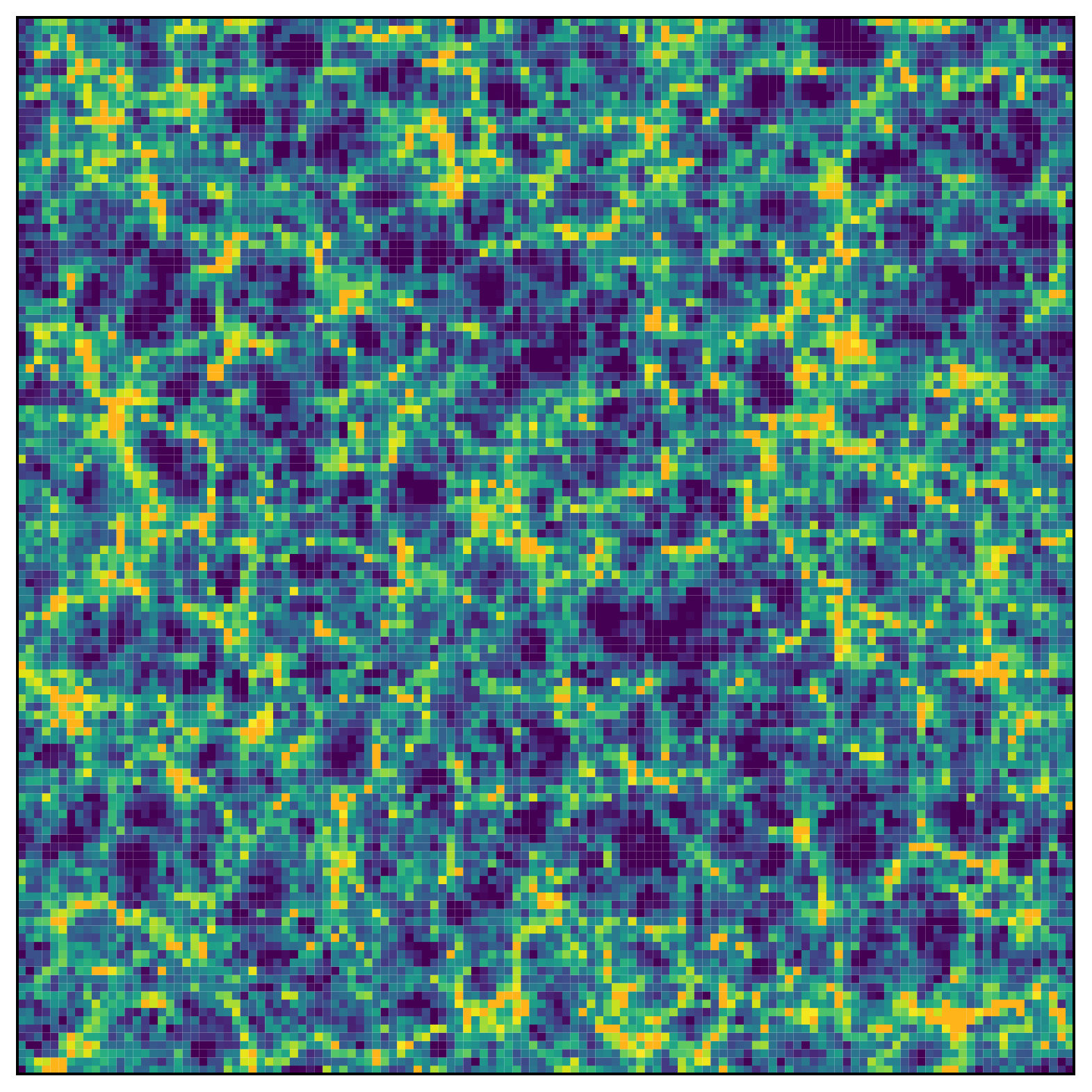}
\includegraphics[scale=0.36,valign=t]{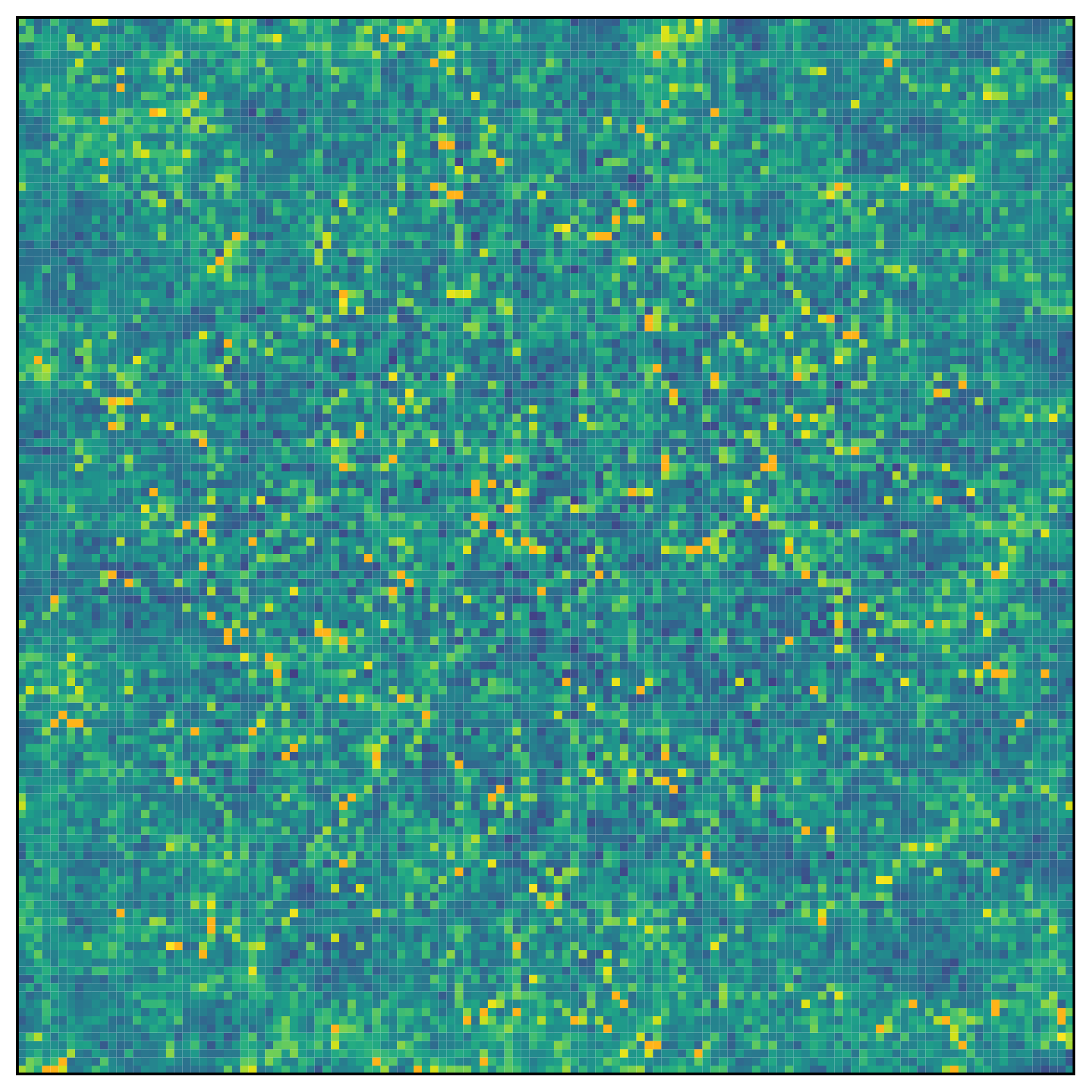}
\includegraphics[scale=0.36,valign=t]{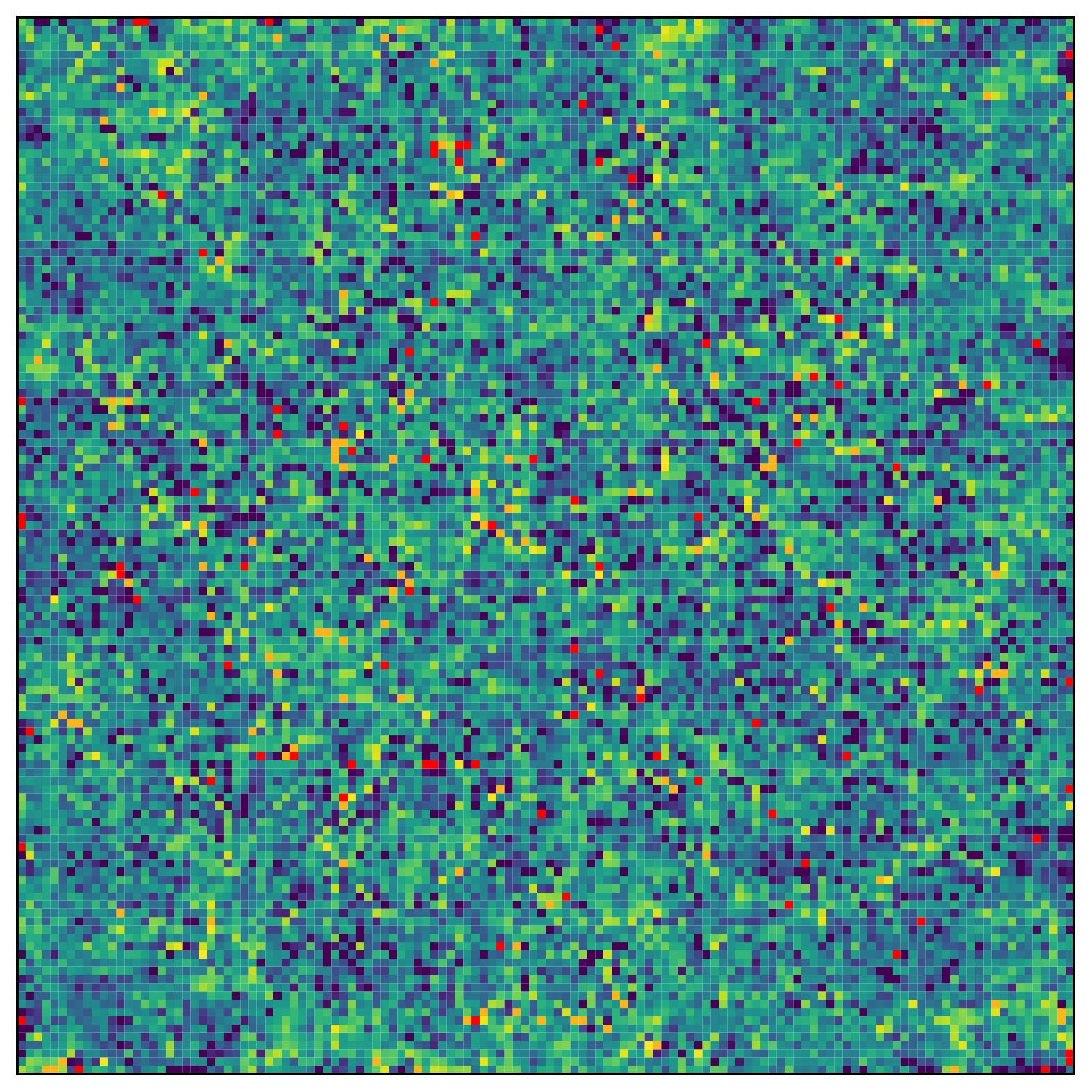}

\includegraphics[scale=0.36,valign=t]{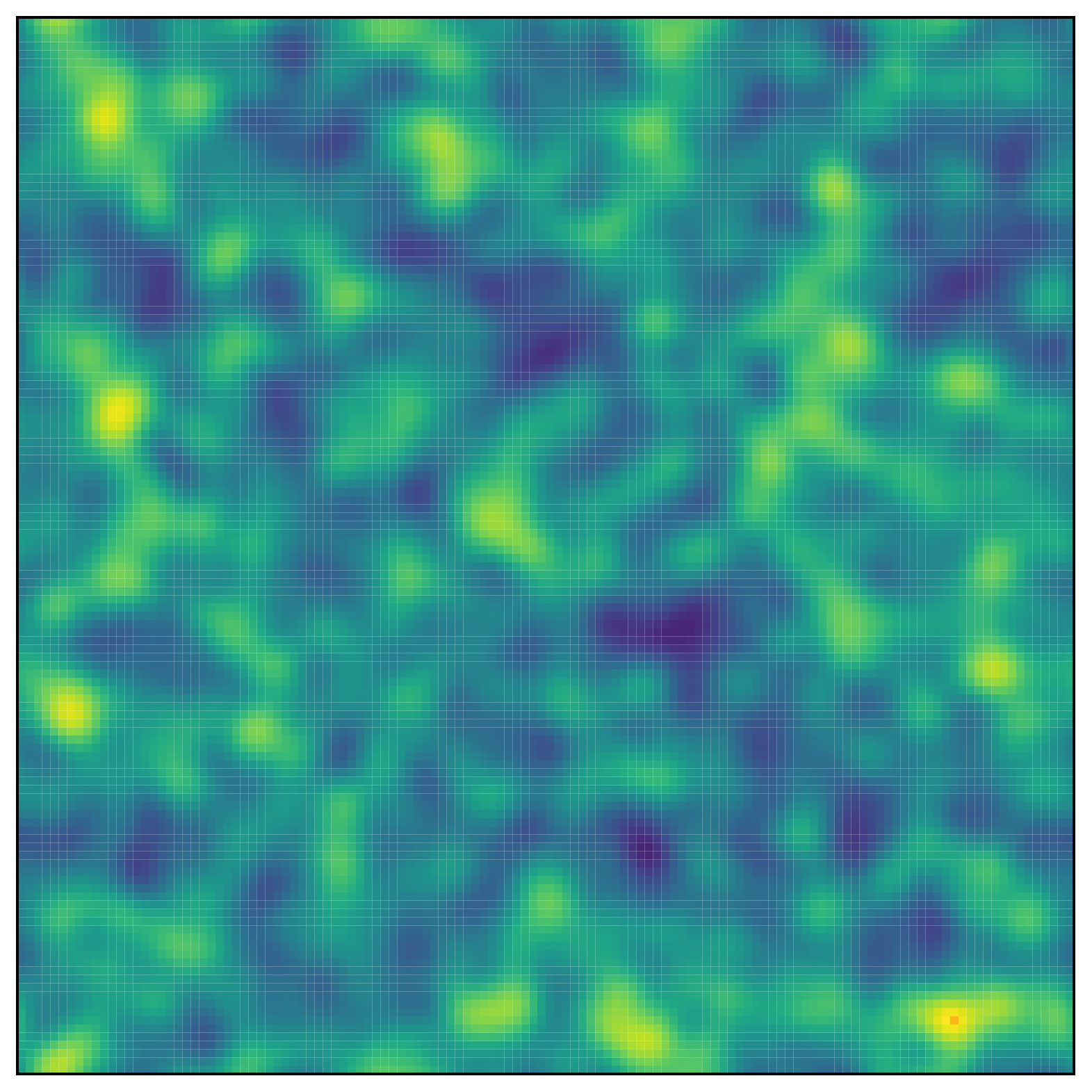}
\includegraphics[scale=0.36,valign=t]{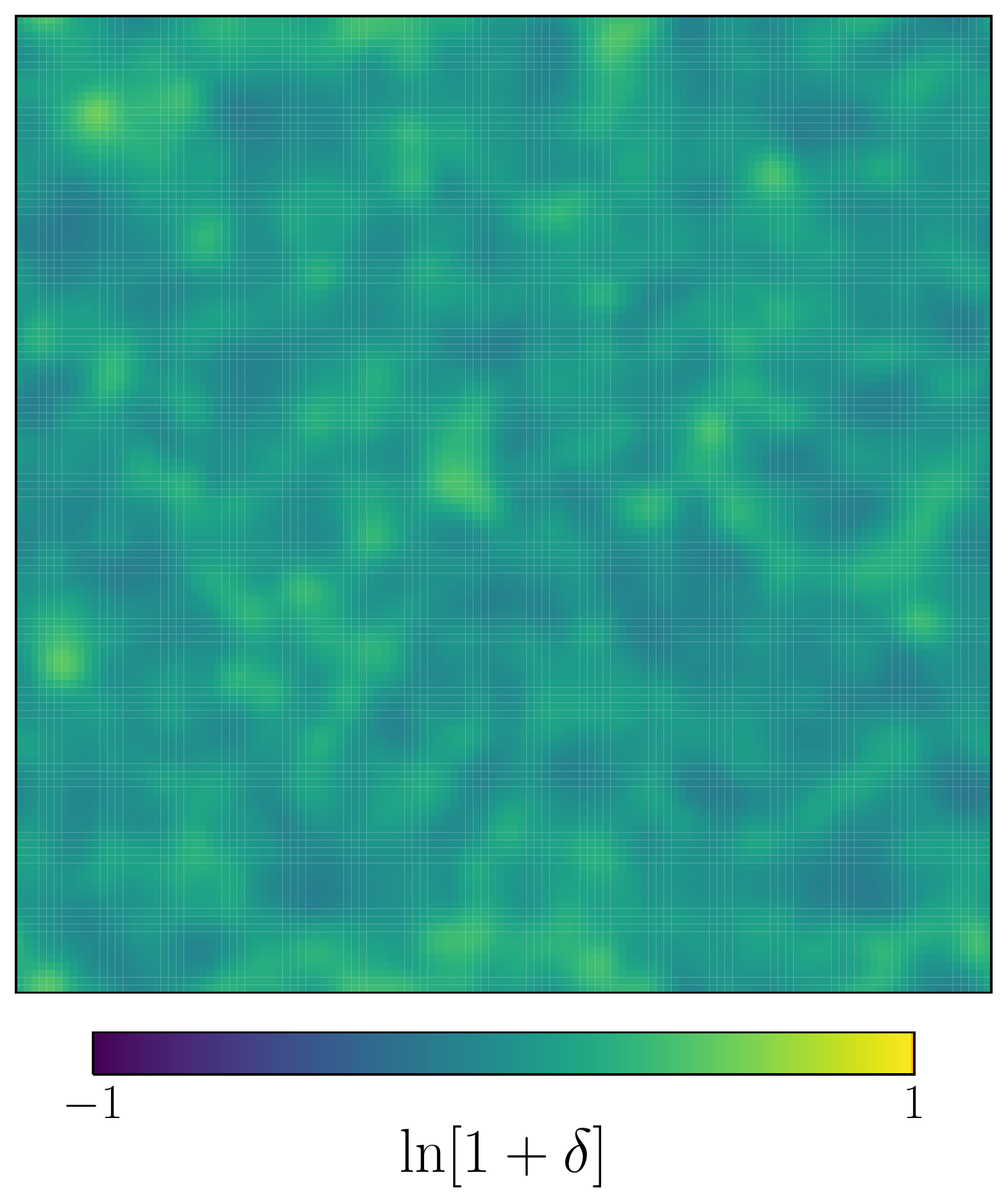}
\includegraphics[scale=0.36,valign=t]{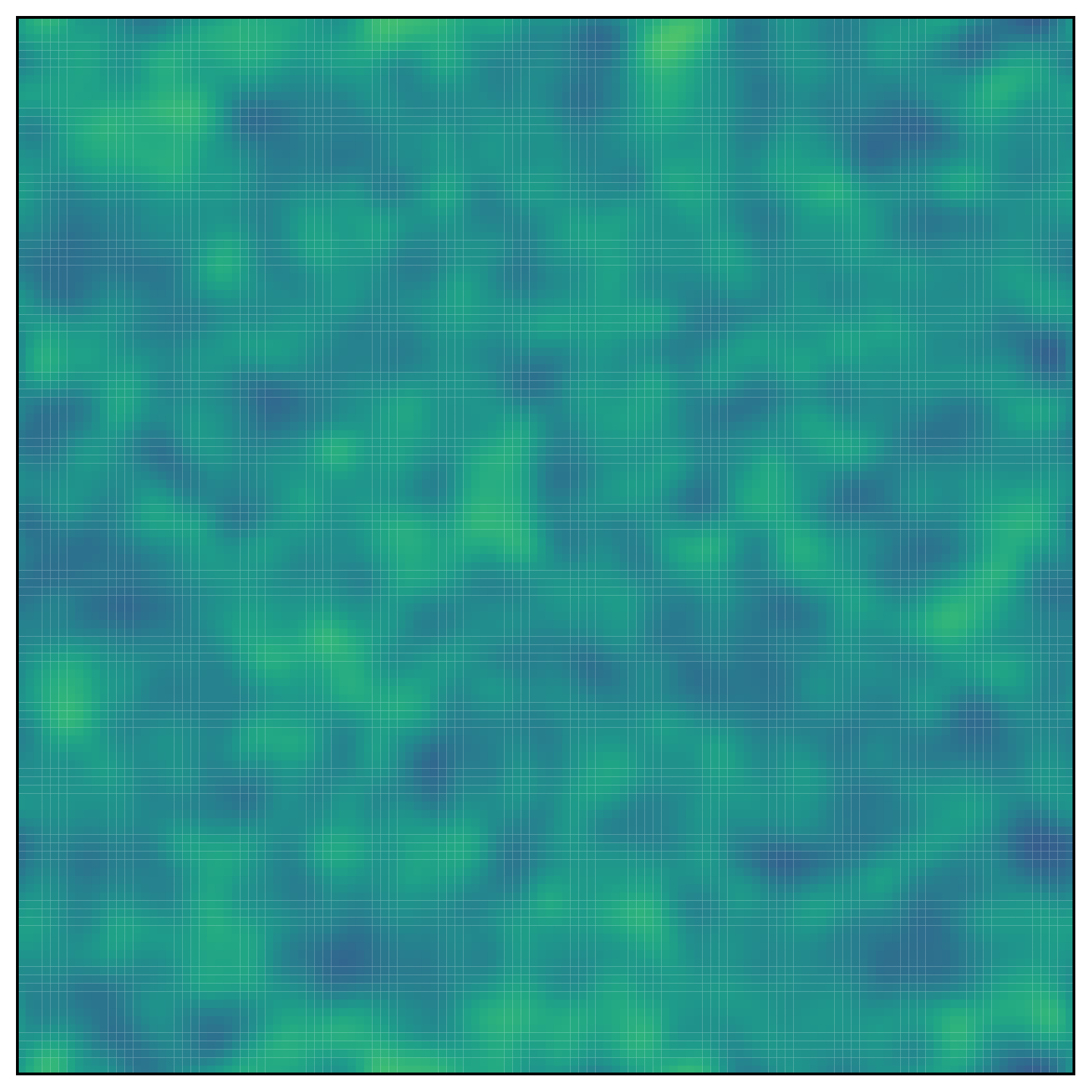}
\end{center}
\caption{\label{fig:TestA3PicsQual} Qualitative comparison of reconstruction methods in test A 3. For this test, we create mock data from a simulated density field on a light cone, use an even source distribution and add negligible shape noise. The fields are from left to right: the underlying overdensity field from the simulation, the reconstructed overdensity field using a lognormal prior and the reconstructed overdensity using a Gaussian prior (Wiener filter). We show the central $(x-y)$-plane of the reconstruction box in each panel. The observer is located in the center of the $(x-y)$-plane at the origin of the $z$-axis. Note that we plot $\ln[1+\delta]$ and mark negative densities in red. In the second row, we show a smoothed version of the fields. The smoothing is performed with a Gaussian kernel with $\sigma=8h^{-1}\,\Mpc$.}
\end{figure}

3) A test similar to A2 but with a redshift dependent density field constructed from different snapshots. The physical size of the box spans $[500h^{-1}\,\Mpc]^2$ in the $(x-y)$-plane and $4000h^{-1}\,\Mpc$ into the $z$-direction corresponding to a maximum redshift of $z=2.2$.\footnote{The light cone is constructed by merging planes from different snapshots along the $z$-direction of the box, but the algorithm assumes that the distance and redshift increase in radial direction from the observer. This leads to a mismatch between the distance to a (source) position assumed by the algorithm and the distance or rather redshift that this position corresponds to in the simulation. The distance in the simulation is $\cos\theta$ times the distance assumed by the algorithm, where $\theta$ is the angle between the LOS and the $z$-axis. This mismatch should only become relevant for relatively large angles $\theta>\pi/2$. Lines of sight with such angles only cover a minor fraction of the box and we therefore expect the test to not be severely affected by this approximation.} Results of test A3 are shown in Fig.~\ref{fig:TestA3PicsQual} and Fig.~\ref{fig:TestA3PicsQuant}. The quantitative comparison between the lognormal and the corresponding WF run is listed in the last rows of Tables~\ref{tab:TestAb} and \ref{tab:TestAa}, labeled \enquote{light cone}. We find a similar match between the underlying and reconstructed fields as in tests A2 and again a superiority of the lognormal prior over the Wiener Filter reconstruction in terms of correlation coefficients, pixel-wise differences and extremal values. The distributions of the pixel-wise matching between input and reconstructions in the left panel of Fig.~\ref{fig:TestA3PicsQuant} imply that the lognormal prior is more likely to reconstruct overdensities correctly, while the Wiener Filter is better in tracing the underdensities. The full distributions in the right panel of the same Figure suggest that the better reconstruction of low densities with the Wiener Filter is a consequence of the symmetry of the Gaussian prior: The Gaussian prior suppresses both positive and negative offsets from the mean density by an equal amount. Because of this property it also allows negative densities. The lognormal prior strongly suppresses deviations towards low densities, while it is less sensitive to deviations towards higher densities (since it suppresses deviations equally on a logarithmic scale). As a result, the lognormal model refrains from assigning very low densities to regions with small signal-to-noise.

\begin{figure} 
\begin{center}
\includegraphics[scale=0.36,valign=t]{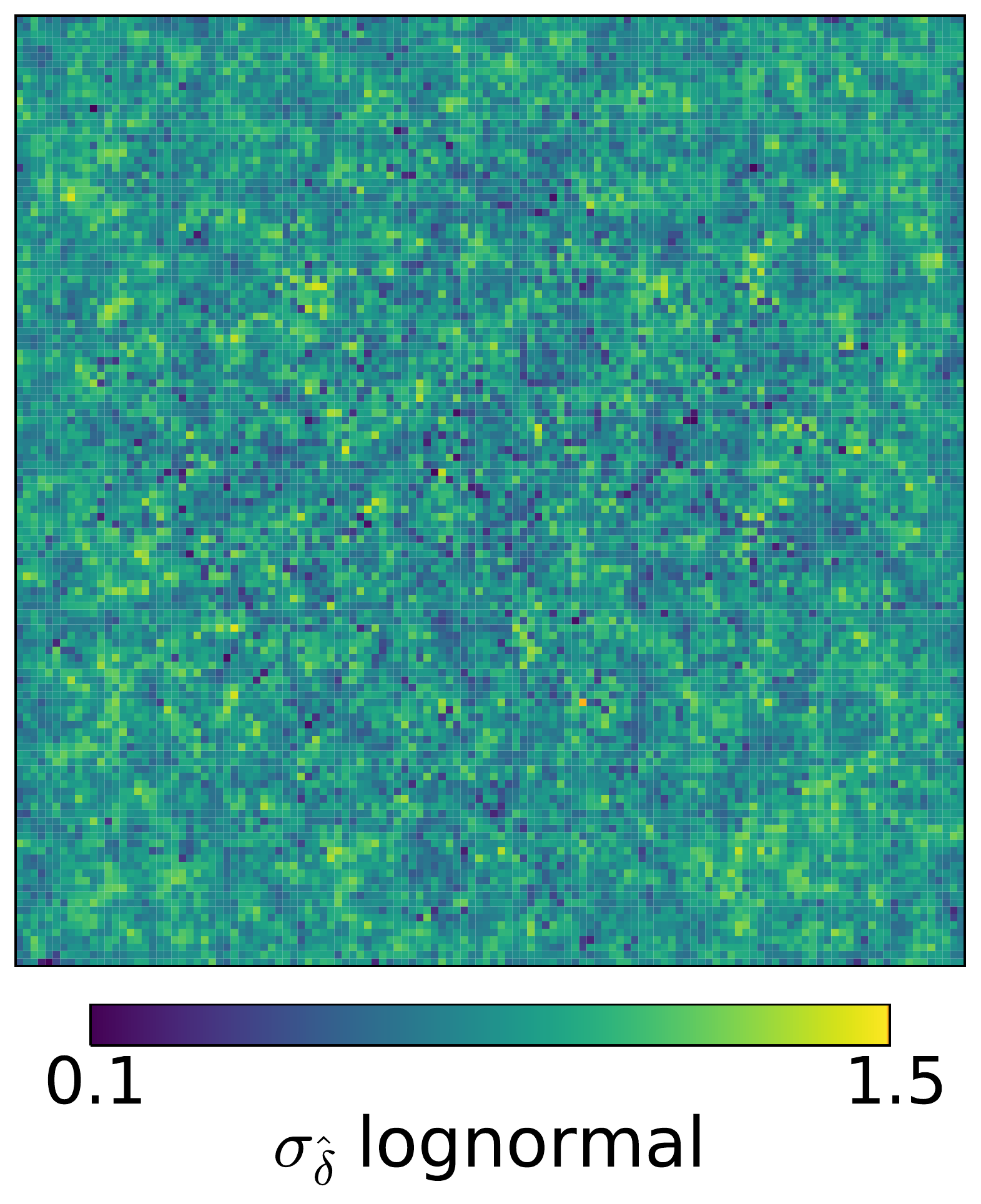}
\includegraphics[scale=0.36,valign=t]{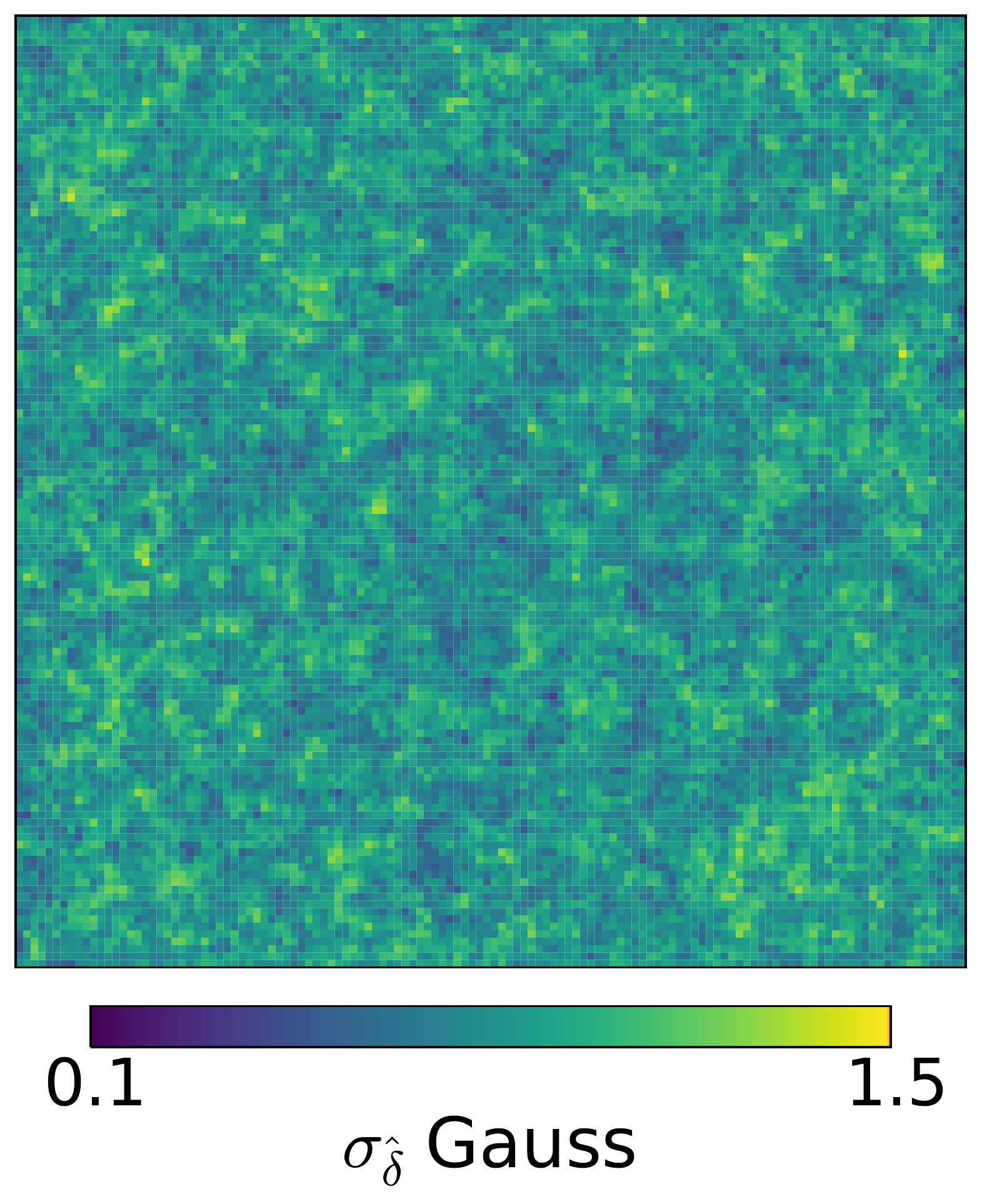}
\includegraphics[scale=0.36,valign=t]{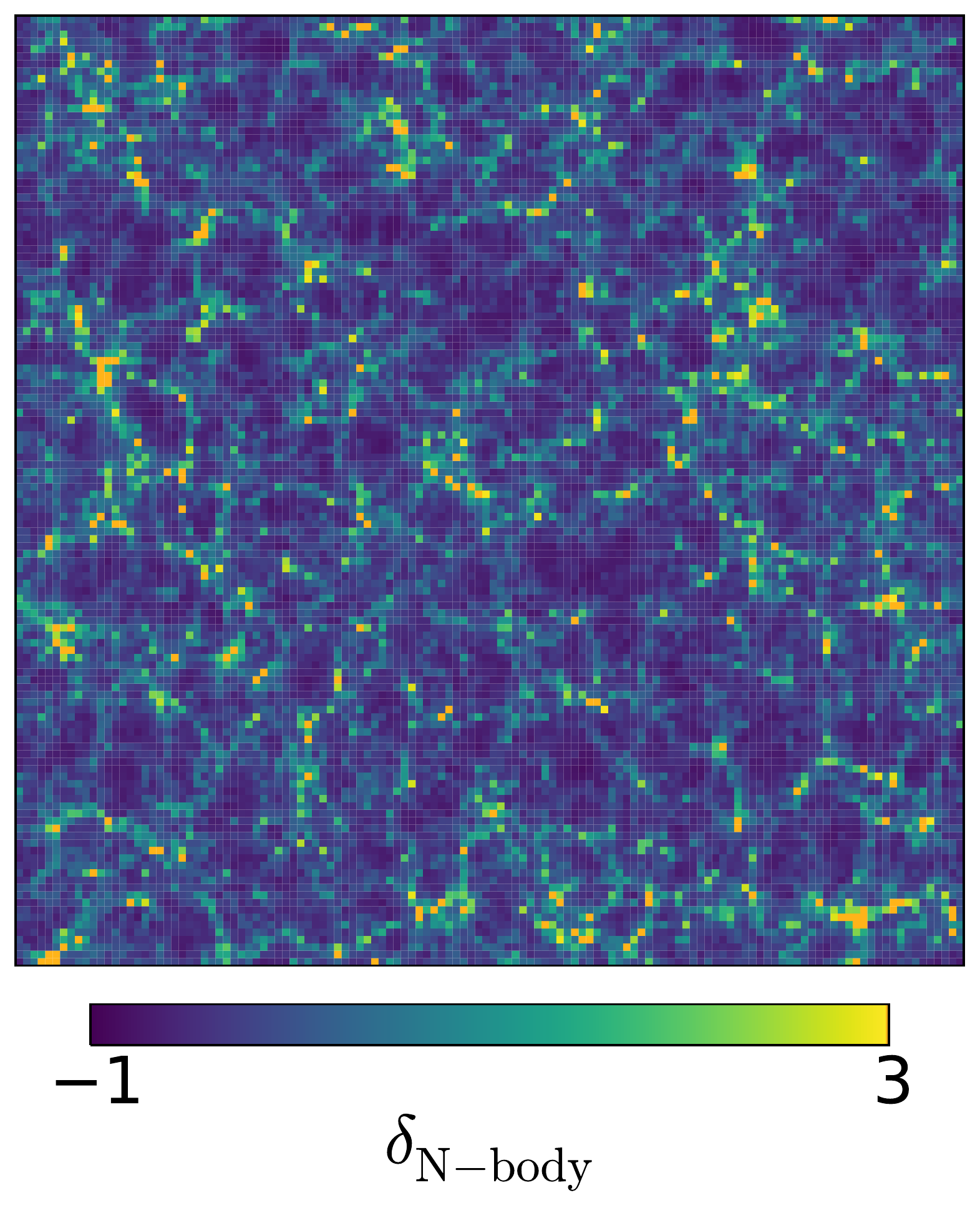}

\includegraphics[scale=0.36,valign=t]{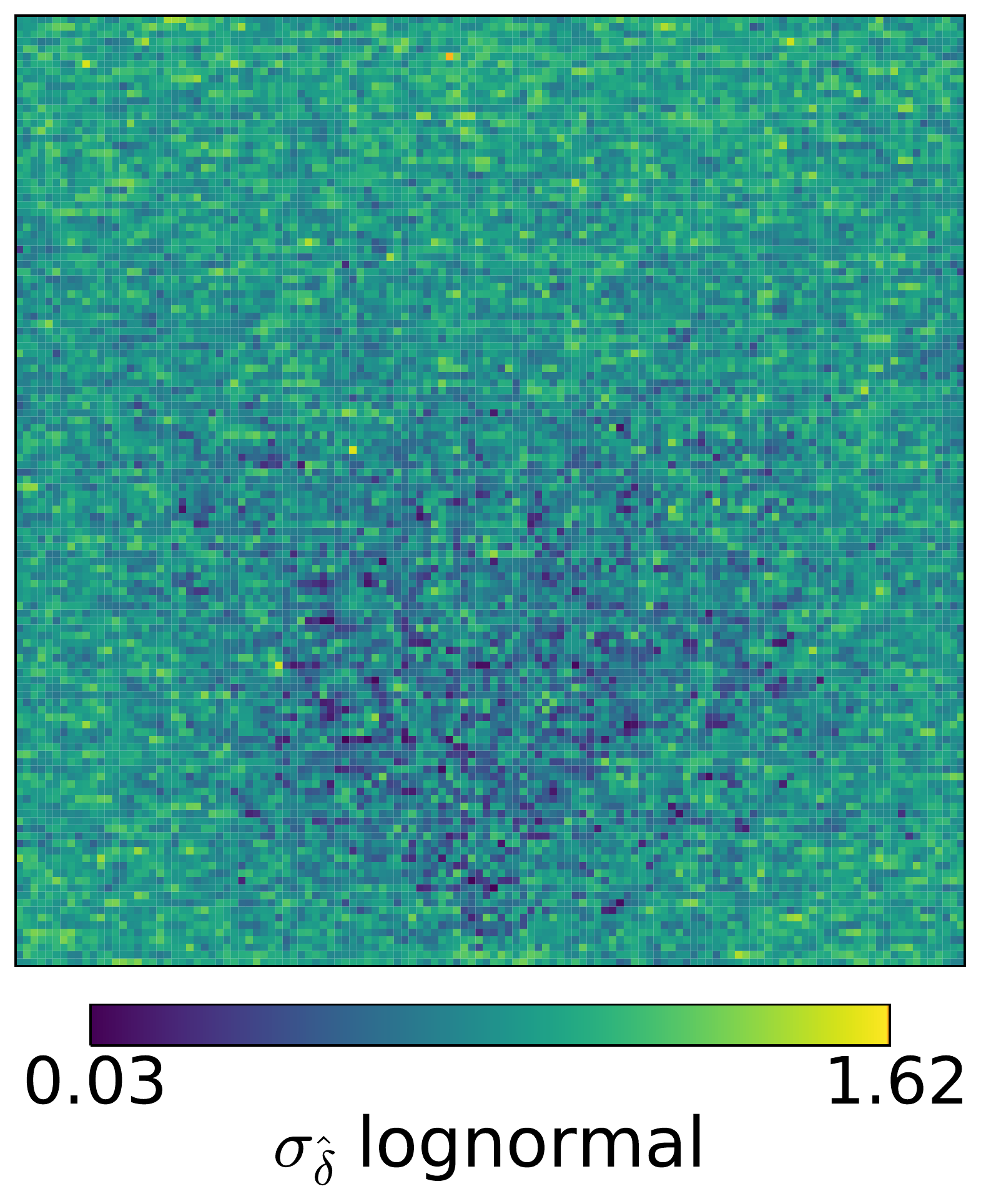}
\includegraphics[scale=0.36,valign=t]{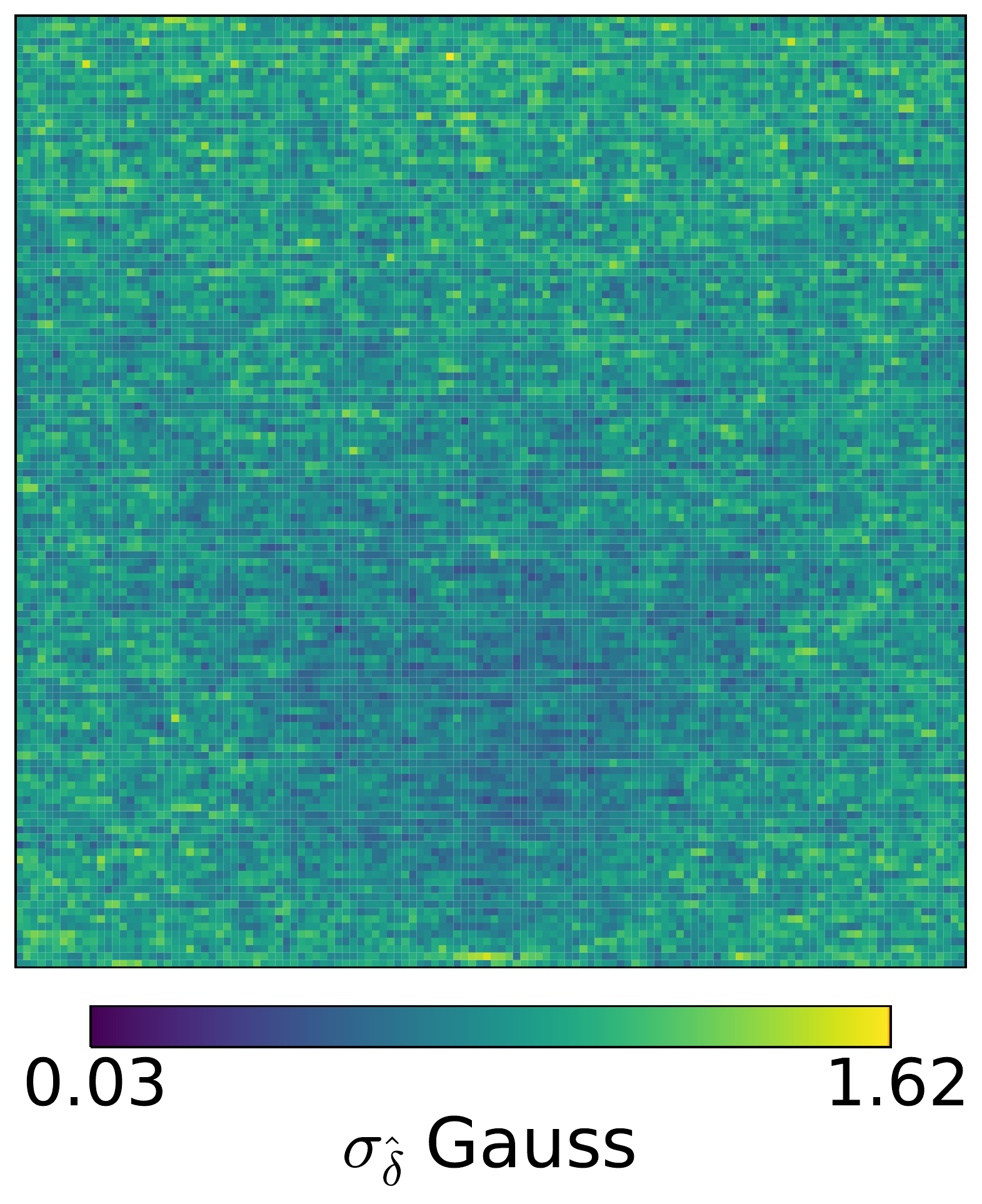}
\includegraphics[scale=0.36,valign=t]{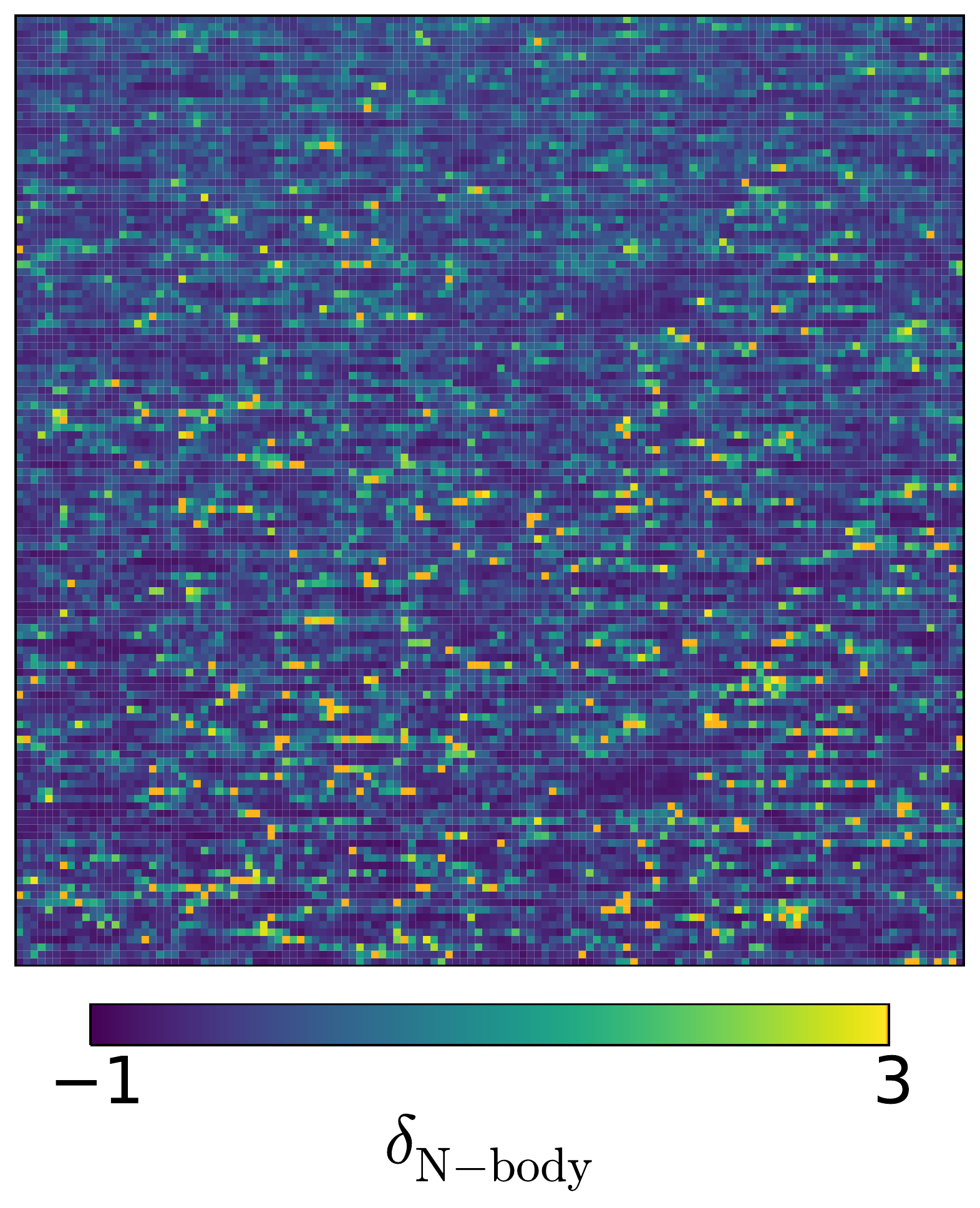}

\end{center}
\caption{\label{fig:TestA3Cov} Estimated uncertainty maps, $\sigma_{\hat{\delta}}=\sqrt{\mathrm{diag}(\hat{C})}$ (c.p. Sec.~\ref{sec:fidel}), of the reconstructions in Test A 3) obtained with 20 samples. In the upper row we show uncertainty maps of the central slice through the $(x-y)$-plane, in the lower row through the $(x-z)$-plane. The columns are from left to right: estimate of $\sigma_{\hat{\delta}}$ of the lognormal reconstruction, the Wiener Filter reconstruction, and the input field in linear scaling for comparison.}
\end{figure}

In Fig.~\ref{fig:TestA3Cov} we show estimates of the reconstruction uncertainties in this test which were obtained by following the prescription in Sec.~\ref{sec:fidel}. We find that the estimator in Eq.~\ref{eq:cov_esti} converges very quickly and use 20 samples (from the 15th sample on the average change per pixel from adding another sample is less than 0.05, adding the 20th sample changes the uncertainty estimate on average only by 0.03 per pixel). For both priors, but especially for the lognormal prior, a comparison to the input field shows that overdensities are reconstructed with higher fidelity than low density regions. The Wiener filter reconstruction is less sensitive to overdensities but slightly more sensitive to underdensities than the lognormal reconstruction.

\begin{figure}
\begin{center}
\subfloat[$z=0$]{
\scalebox{0.4}{\includegraphics{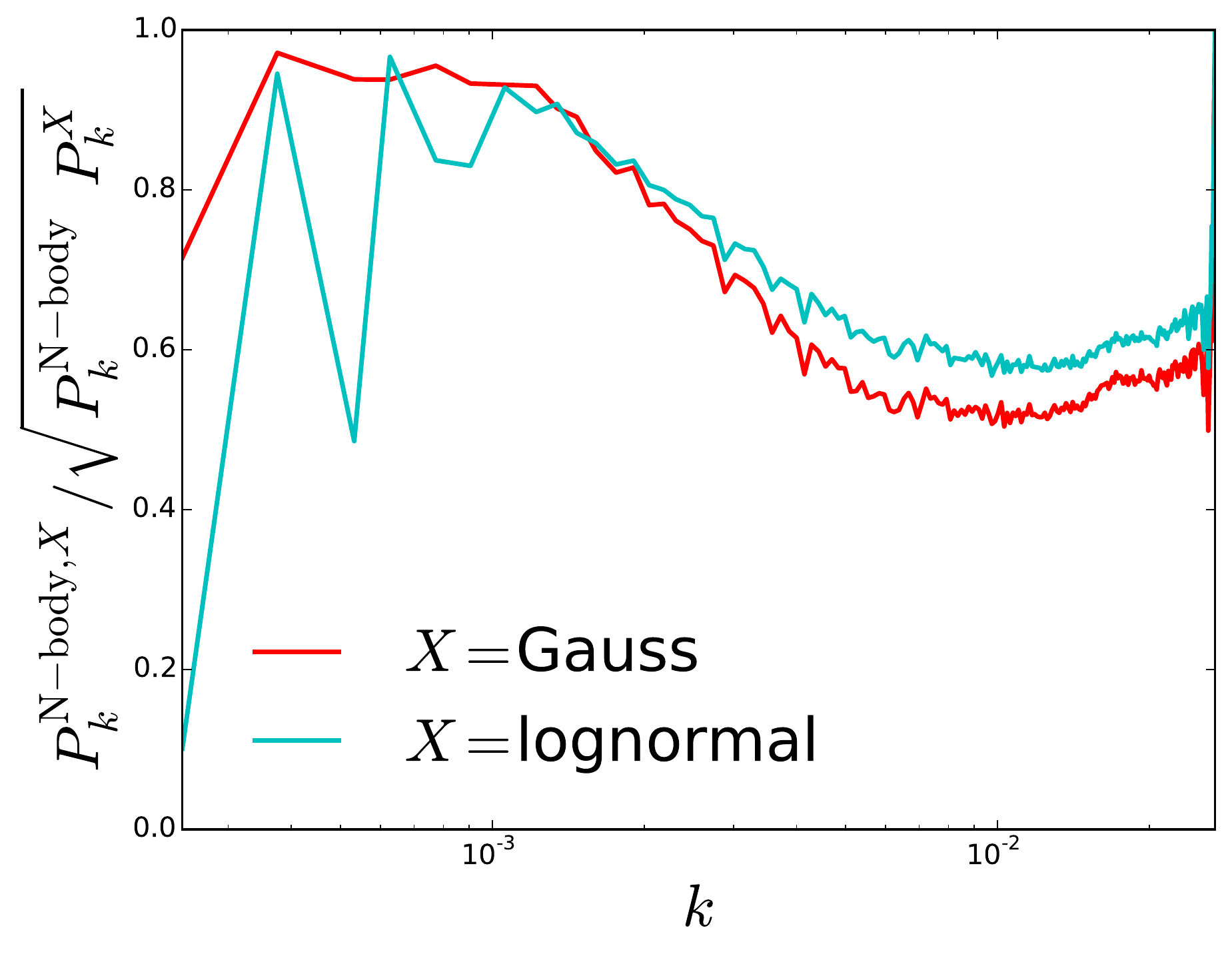}}}
\subfloat[light cone]{
\scalebox{0.4}{\includegraphics{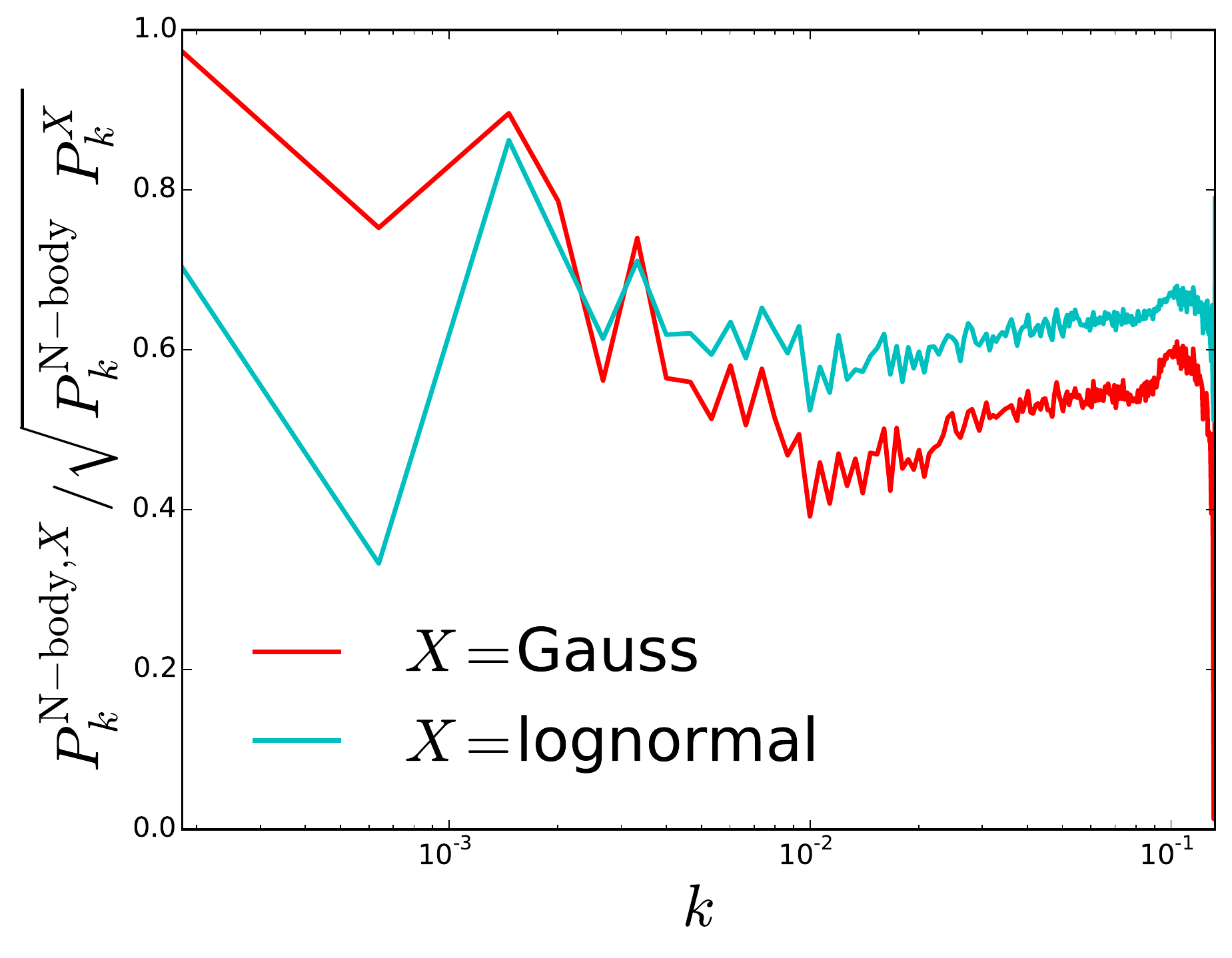}}}
\end{center}
\caption{\label{fig:TestA2Corr} Cross spectra ($2\pi\, \DiracDelta (\vk-\vk')\,  P^{Y,X}_k = \Re\langle \delta_Y(\vk) \delta_X(\vk')\rangle$) between the the underlying density field and the reconstructions divided by the square root of the auto spectra. We show results for tests of type A at $z=0$ (left panel) and on the light cone (right panel). Albeit adding only negligible noise in this test, we do not expect a perfect correlation due to the information loss caused by the sparse sampling of the underlying field by the data. We find that the Gaussian model is better in reconstructing the largest-scales, while the lognormal model shows higher correlations at low and intermediate scales.}
\end{figure}

In Fig.~\ref{fig:TestA2Corr} we compare the correlation coefficient between input field and reconstructions on different scales. The Gaussian model provides better fits on the largest scales. This could be due to the aforementioned missing power in low density regions in the lognormal reconstruction, which can be best seen by eye in the lower row of Fig.~\ref{fig:TestA3PicsQual}. On intermediate and small scales the power spectrum of the lognormal reconstruction shows higher correlations with the original field. We note that the estimator in this work was designed for the map level and is not expected to be an optimal estimator for the power spectrum.

\begin{figure}
\begin{center}
\includegraphics[scale=0.38,valign=t]{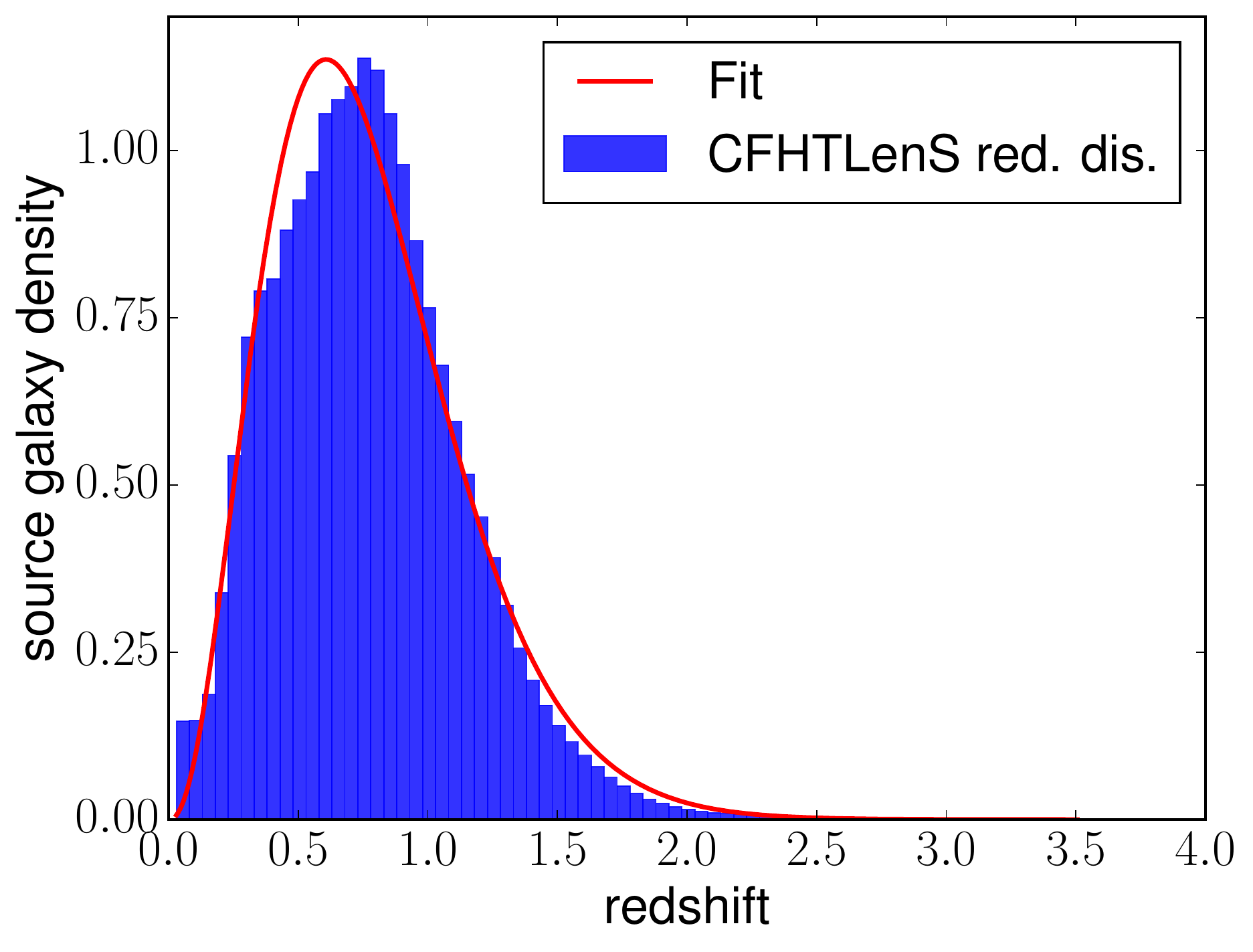}
\includegraphics[scale=0.38,valign=t]{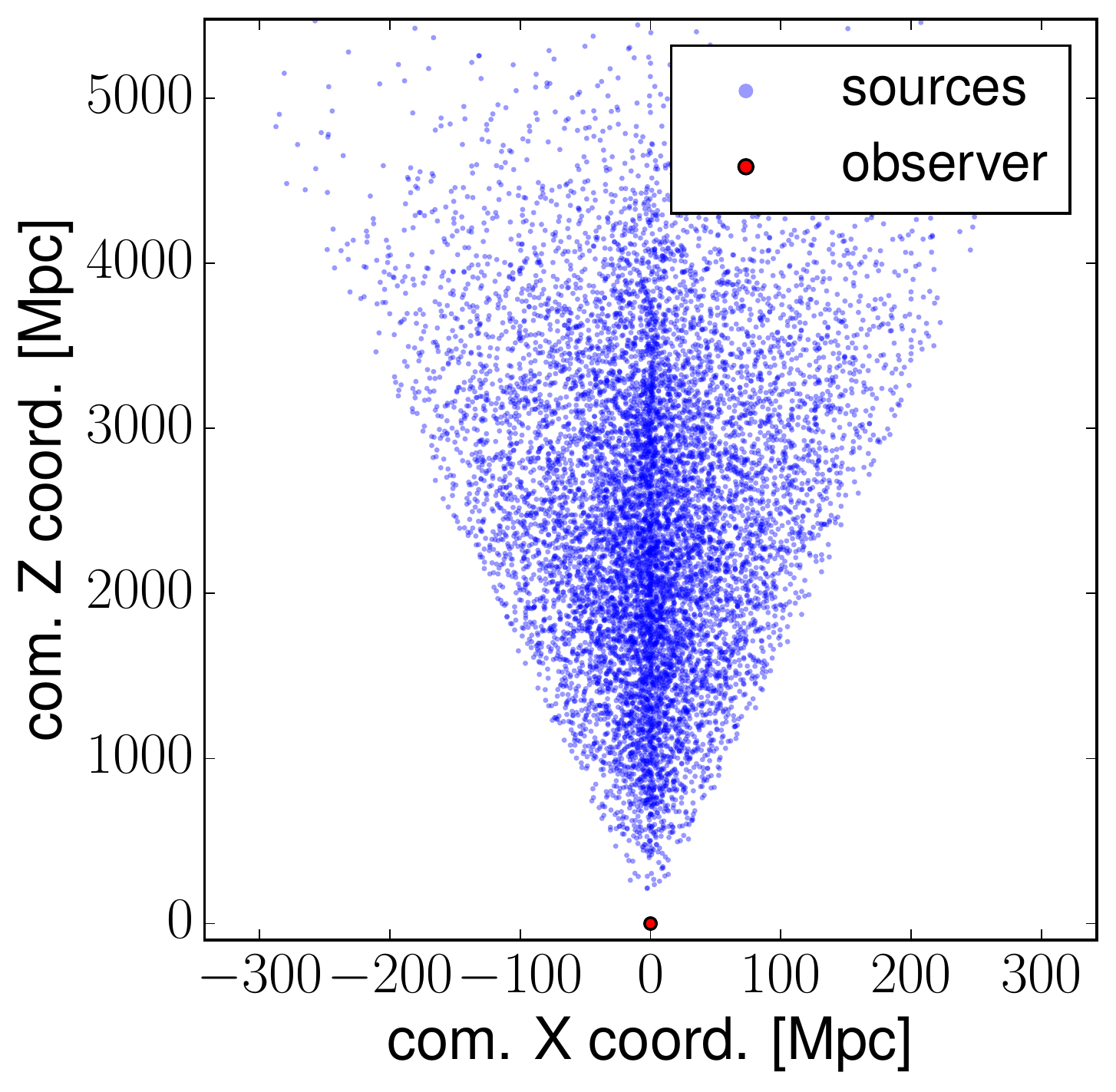}
\end{center}
\caption{\label{fig:RedDis} The source redshift distribution used for test B: In the left panel, we show the histogram of source redshifts from CFHTLenS and our fit to this distribution of the form given in Eq.~\eqref{eq:nz}. The right panel depicts the projected 3D scatter plot of the galaxy sample that is drawn from the fit and used in test B (we only show every 200th source).}
\end{figure}

\subsection{Test B: Realistic survey geometry}
\begin{figure}
\begin{center}
\includegraphics[scale=0.36,valign=t]{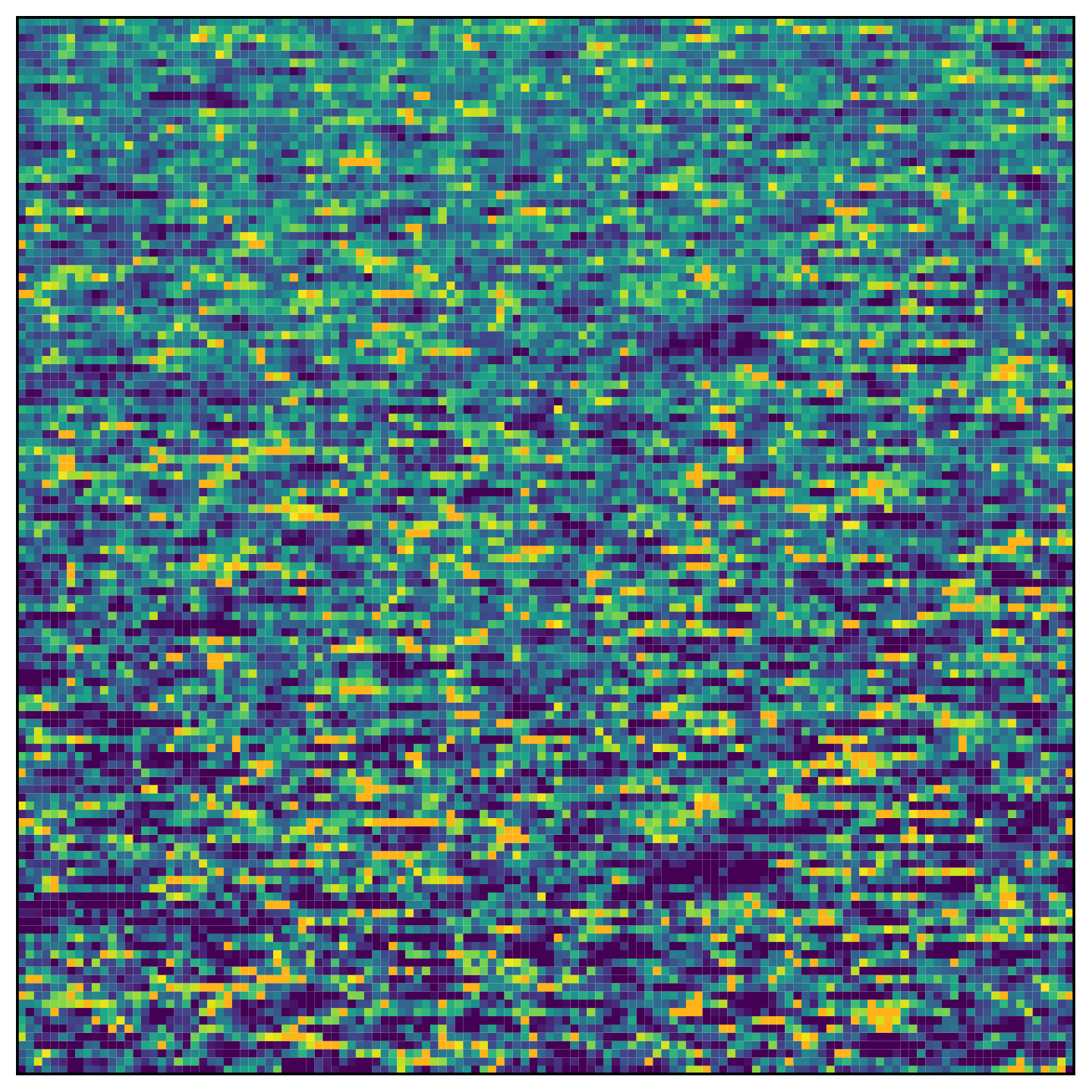}
\includegraphics[scale=0.36,valign=t]{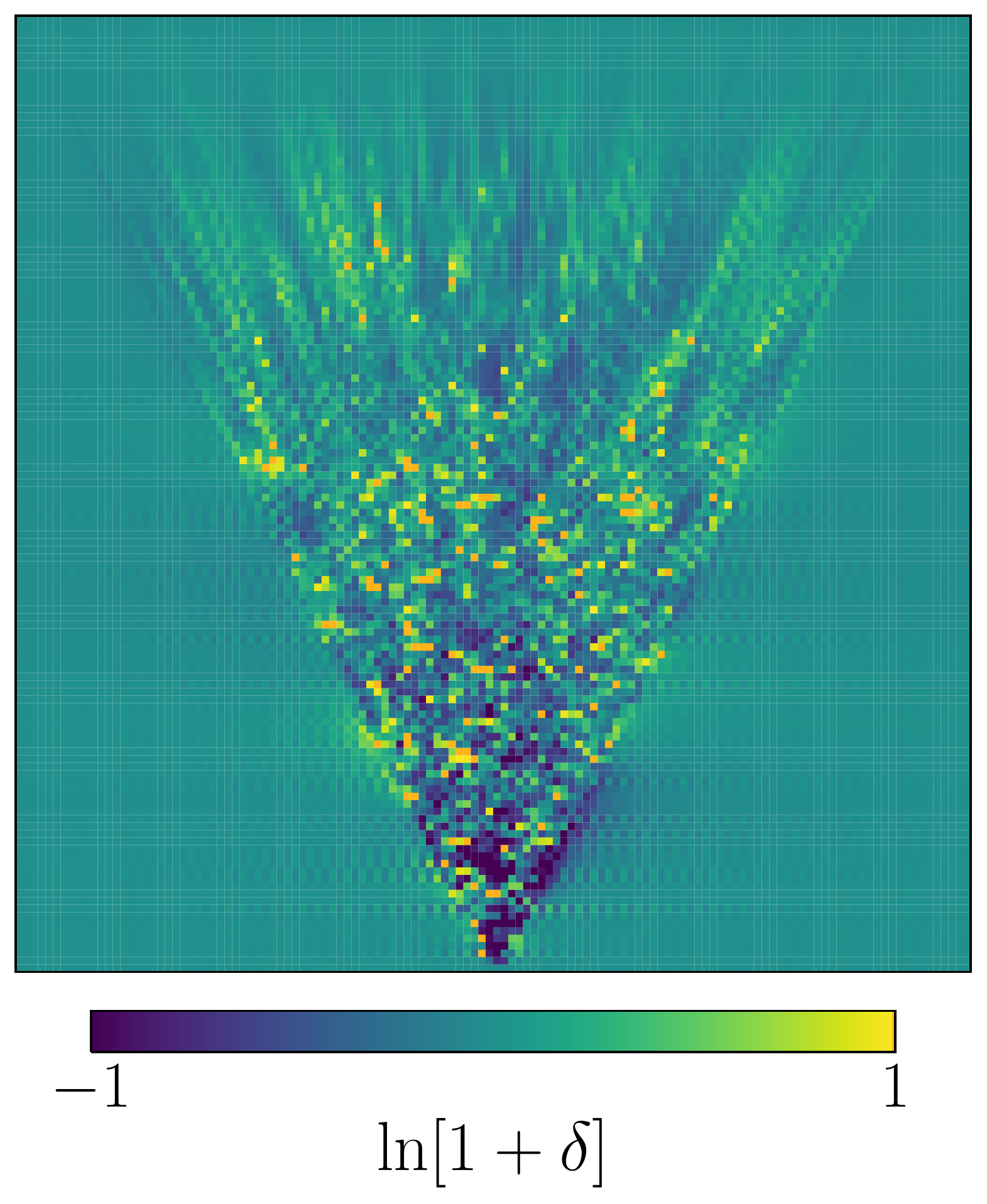}
\includegraphics[scale=0.36,valign=t]{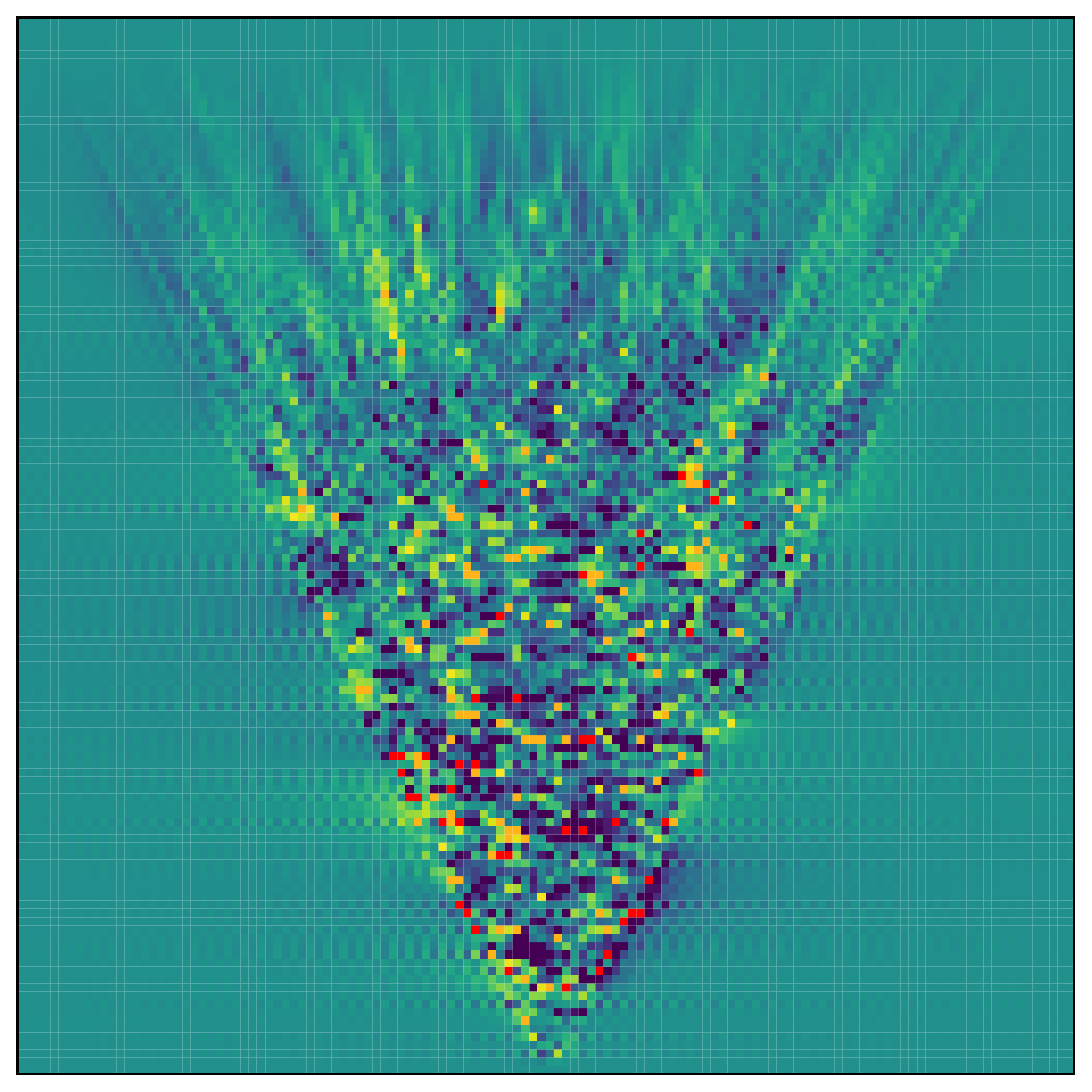}
\end{center}
\caption{\label{fig:TestB2Picsa} Results of test B, reconstructions of a redshift-dependent density field on the observer's light cone from mock data with a realistic source distribution but negligible shape noise. We show central slices through the $y$-axis of the underlying field, $\ln[\delta_{\mathrm{N-Body}}+1]$, in the left, the lognormal reconstruction, $\ln[\hat{\delta}_{\ln}+1]$, in the middle and the Wiener Filter reconstruction, $\ln[\hat{\delta}_{\mathrm{WF}}+1]$ in the right panel. Negative densities are marked in red. The apparent anisotropy of the fields stems from the fact that we have squeezed them in the $z$-direction. We use square pixels in the plot, while their physical size is eight times larger in the $z$-direction than in the $x$-direction.}
\end{figure}

In the test of type B, we employ a realistic survey geometry and source distribution. We use input overdensities from the same light cone as in test A3 to generate mock data and place sources in a cone that spans $7.15^o$. The sources are distributed according to a distribution function of form
\beq
\label{eq:nz}
n(z)=z^\alpha \mathrm{exp}[-(z/z_0)^\beta],
\eeq
where we fit $\alpha$, $\beta$, and $z_0$ to the publicly available source distribution of the CFHTLenS survey\footnote{http://www.cfhtlens.org/astronomers/cosmological-data-products}.
The resulting source distribution function is shown in the left panel of Fig.~\ref{fig:RedDis}. The fit does not exactly match the survey distribution, but the similarity is sufficient for this test. 
We then draw $\Omega[\mathrm{armcin}^2] \rho_s [\mathrm{gal}/\mathrm{arcmin}^2]$ source positions from this distribution, where $\Omega$ is the angular opening area of the cone in $\mathrm{arcmin}^2$ and $\rho_s$ the source density of the survey, which we choose to be $\rho_s=11\, \mathrm{gal}/\mathrm{arcmin}^2$ corresponding to the source density in CFHTLenS. A scatter plot of the resulting spatial source distribution is shown in the right panel of Fig.~\ref{fig:RedDis}. The reconstructions are depicted in Fig.~\ref{fig:TestB2Picsa}.

\begin{figure}
\begin{center}
\includegraphics[scale=0.37,valign=t]{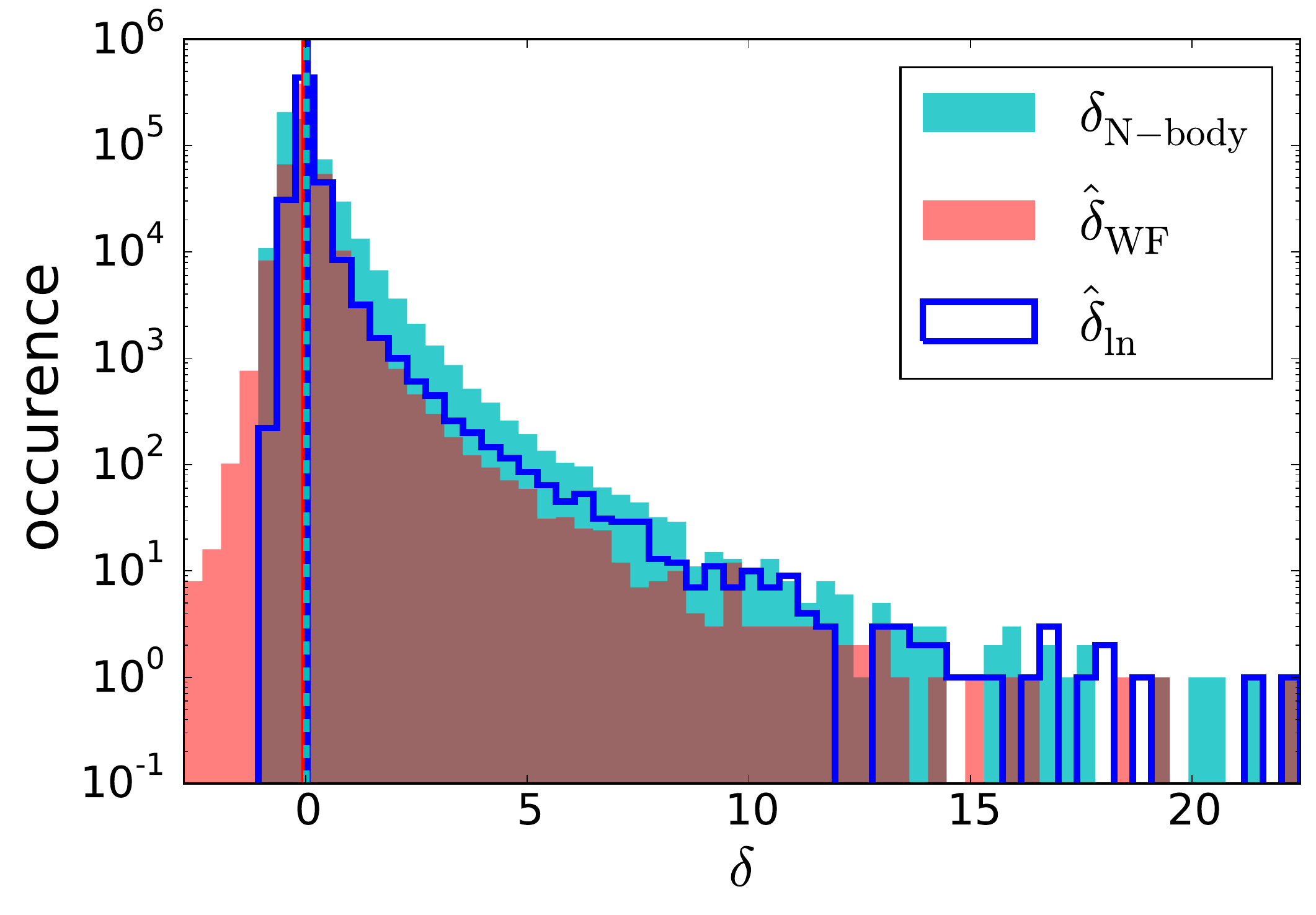}
\includegraphics[scale=0.37,valign=t]{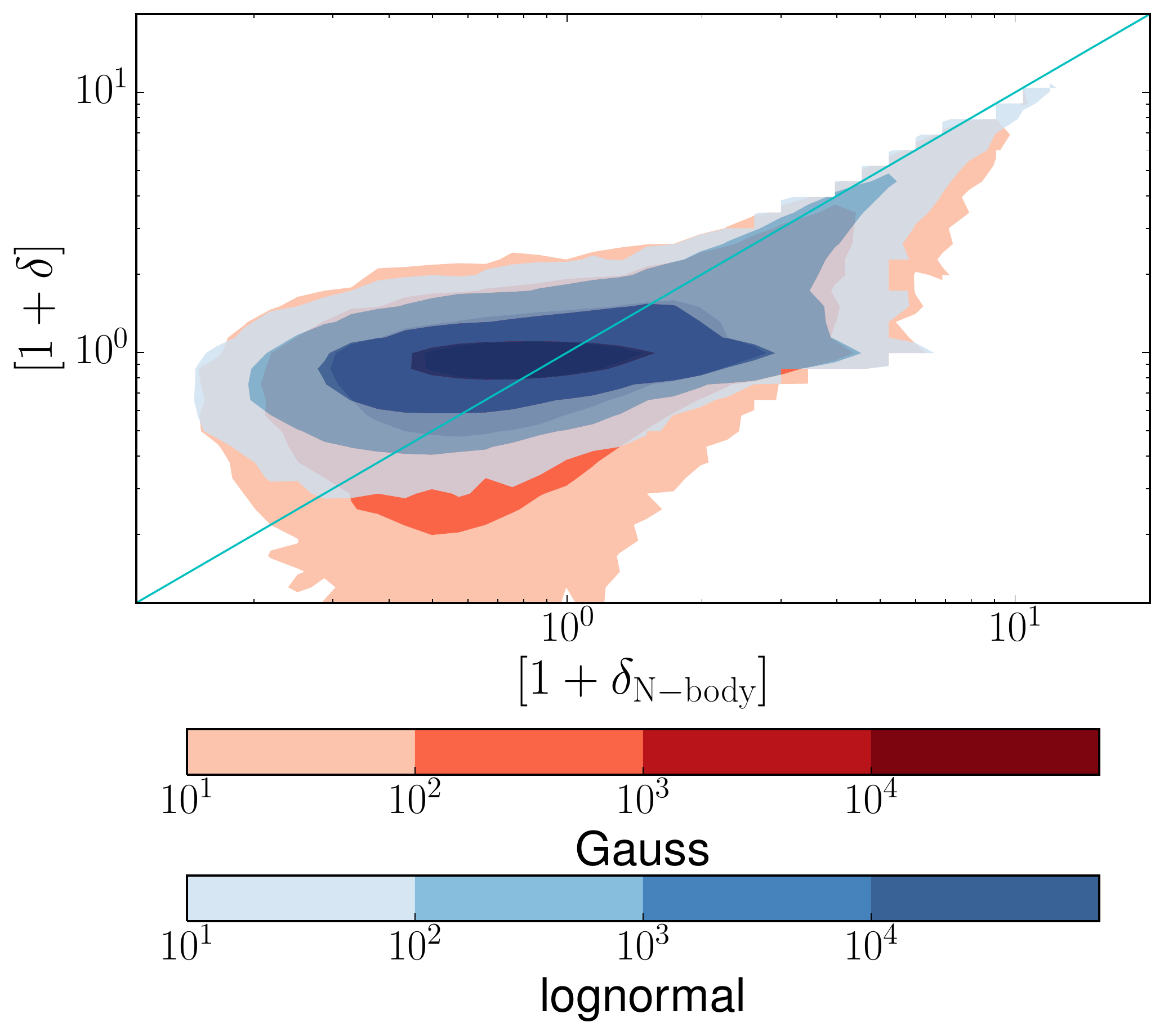}
\end{center}
\caption{\label{fig:TestB2Picsb} Results of test B, reconstructions from mock data with a CFHTLenS-like source distribution. We have restricted the analysis to the field values that lie within the observed cone. In the left panel we show the 1-point PDFs of the fields with mean values indicated by vertical lines in the same color as their corresponding distribution. The right plot shows the density distributions of the reconstructions against the underlying density. The lognormal model is clearly better in tracing the highest values of the true density distribution.}
\end{figure}

For a realistic source distribution the lognormal model seems to better reconstruct the underlying field at lower redshifts, while the Gaussian prior better traces distant low density regions. We interpret this is again a consequence of the symmetry of the Gaussian distribution, the same property which leads to the abundant reconstruction of negative densities, nicely visible in the 1-point PDF (Fig.~\ref{fig:TestB2Picsb}). 

\subsection{Test C: Realistic shape noise}
\begin{figure}
\begin{center}
\includegraphics[scale=0.4,valign=t]{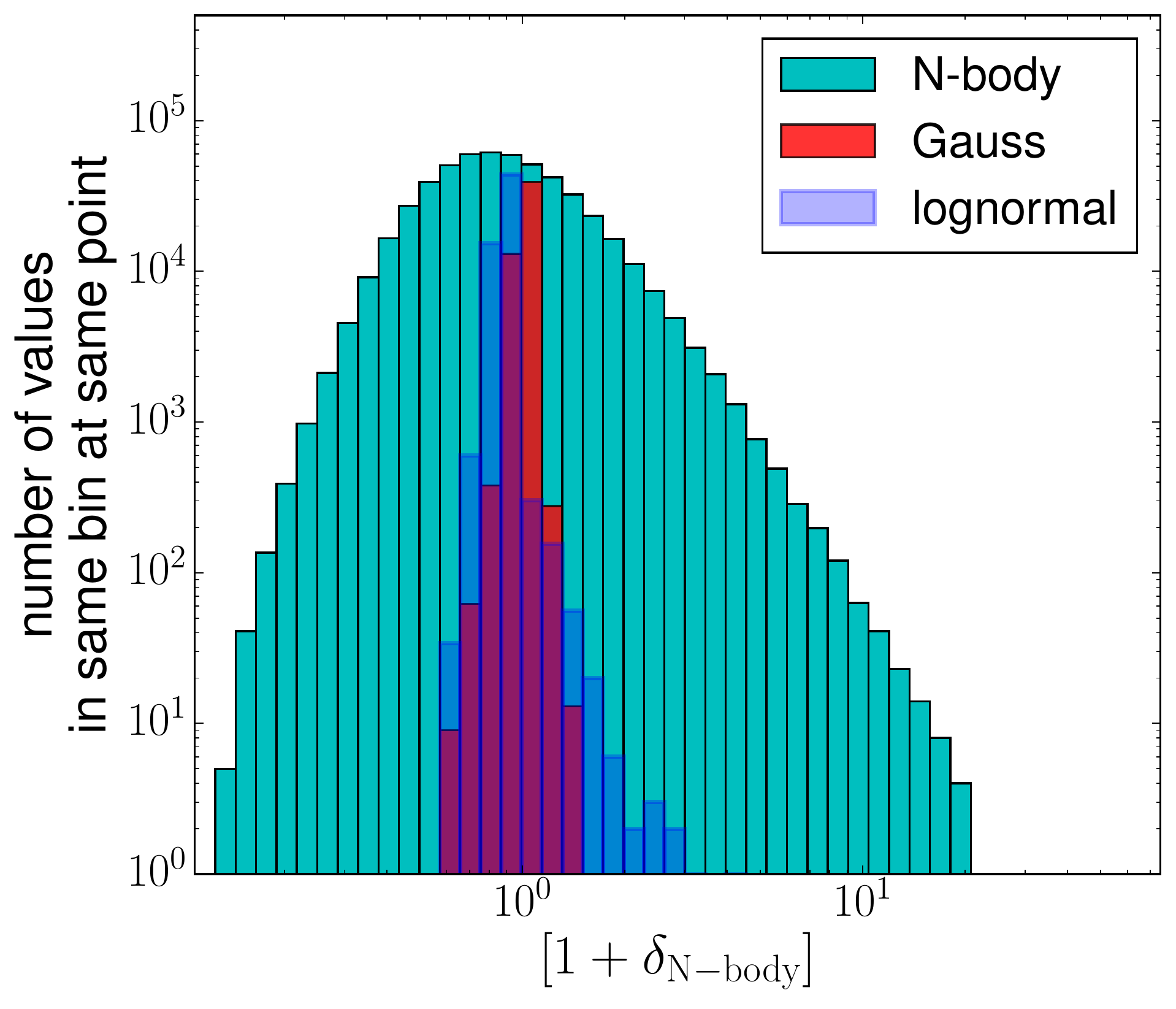}
\includegraphics[scale=0.4,valign=t]{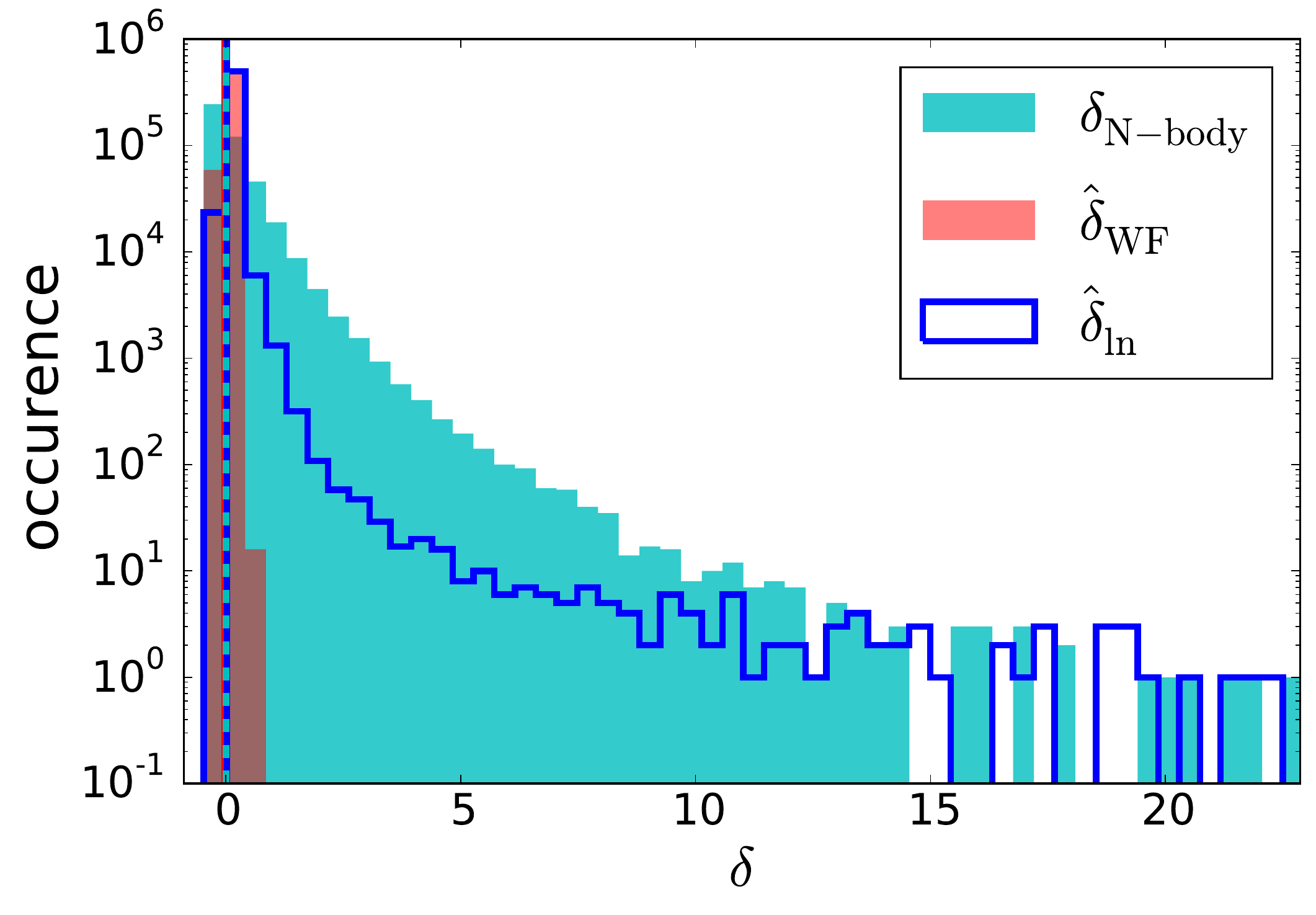}

\end{center}
\caption[PDF Comparisons - Realistic Source Distr. + typical Noise]{\label{fig:TestC3PDFs} Results of test C, reconstructions from mock data with a CFHTLenS-like source distribution and realistic shape noise. We have restricted the analysis to the field values that lie within the observed cone. In the left panel, we count the pixels in which the reconstructed densities coincide with the original density field (i.e. lie within the same density bin). Similarly to  test with low noise, we find that the lognormal prior is more likely to reconstruct overdensities correctly. In the right panel we show the 1-point PDFs of the fields with mean values indicated by vertical lines in the same color as their corresponding distribution. The lognormal reconstruction seems to trace the high density tail of the underlying distribution, while the Gaussian density estimates remain very conservative.}
\end{figure}

\begin{figure}
\begin{center}
\includegraphics[scale=0.36,valign=t]{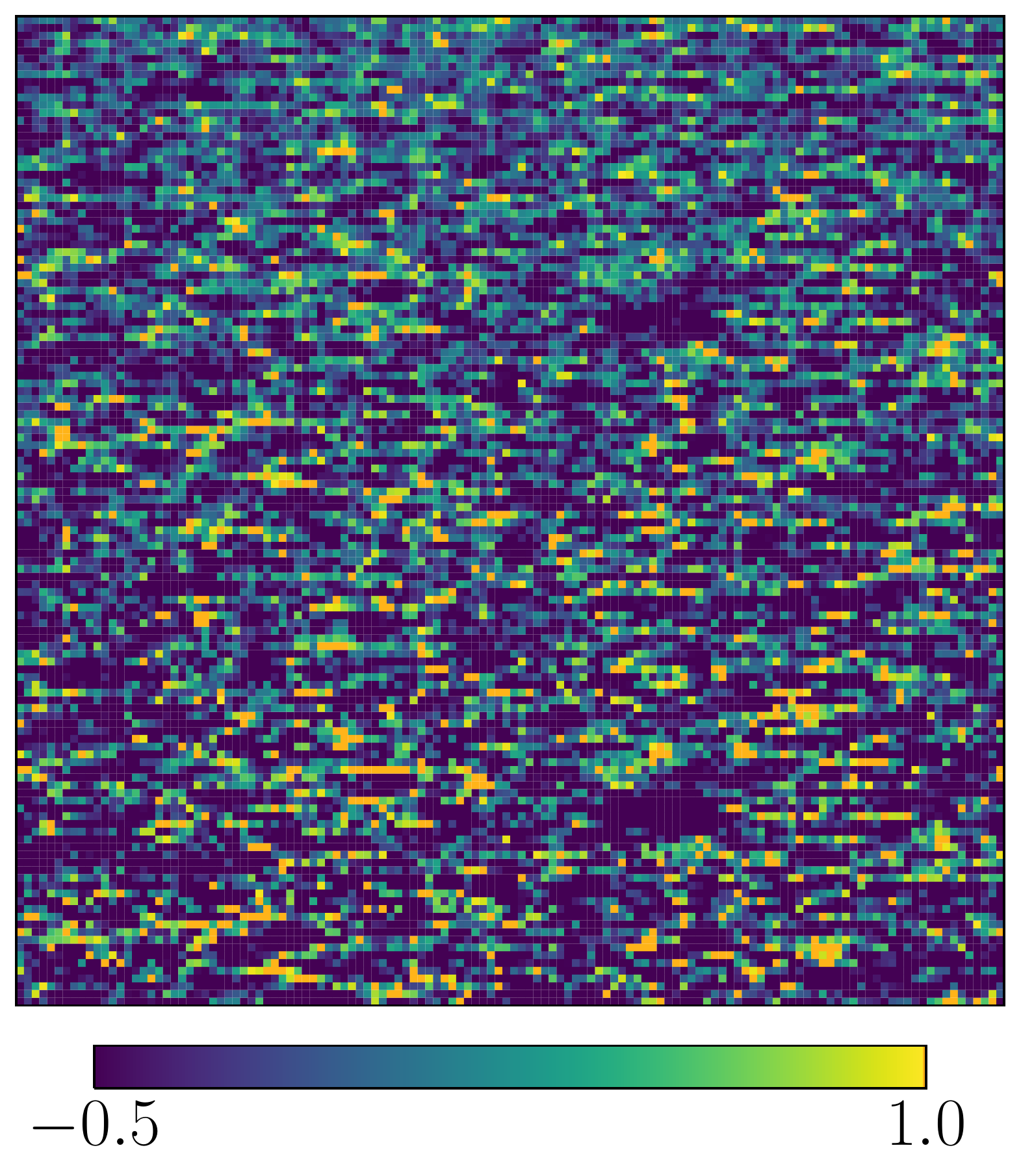}
\includegraphics[scale=0.36,valign=t]{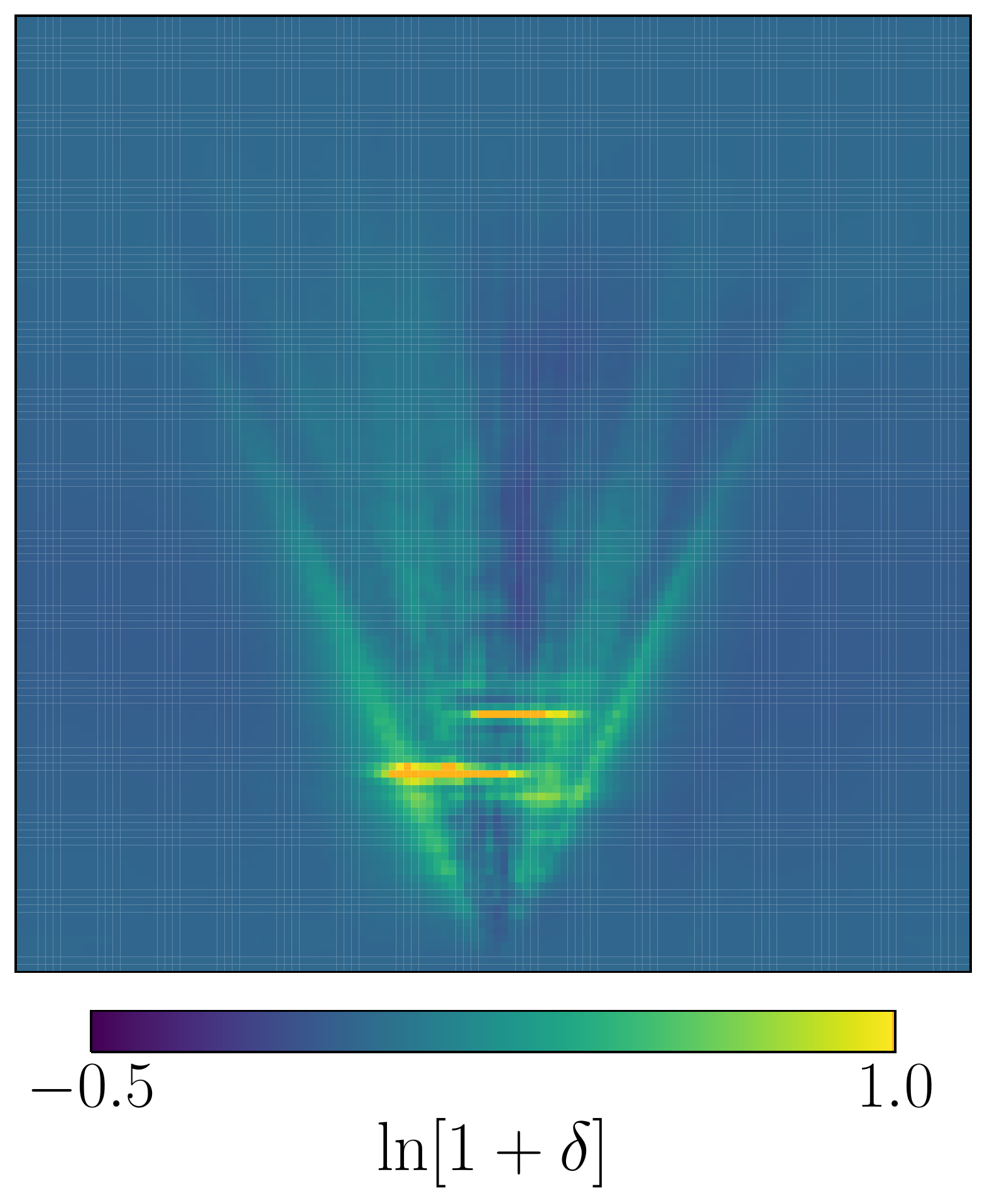}
\includegraphics[scale=0.36,valign=t]{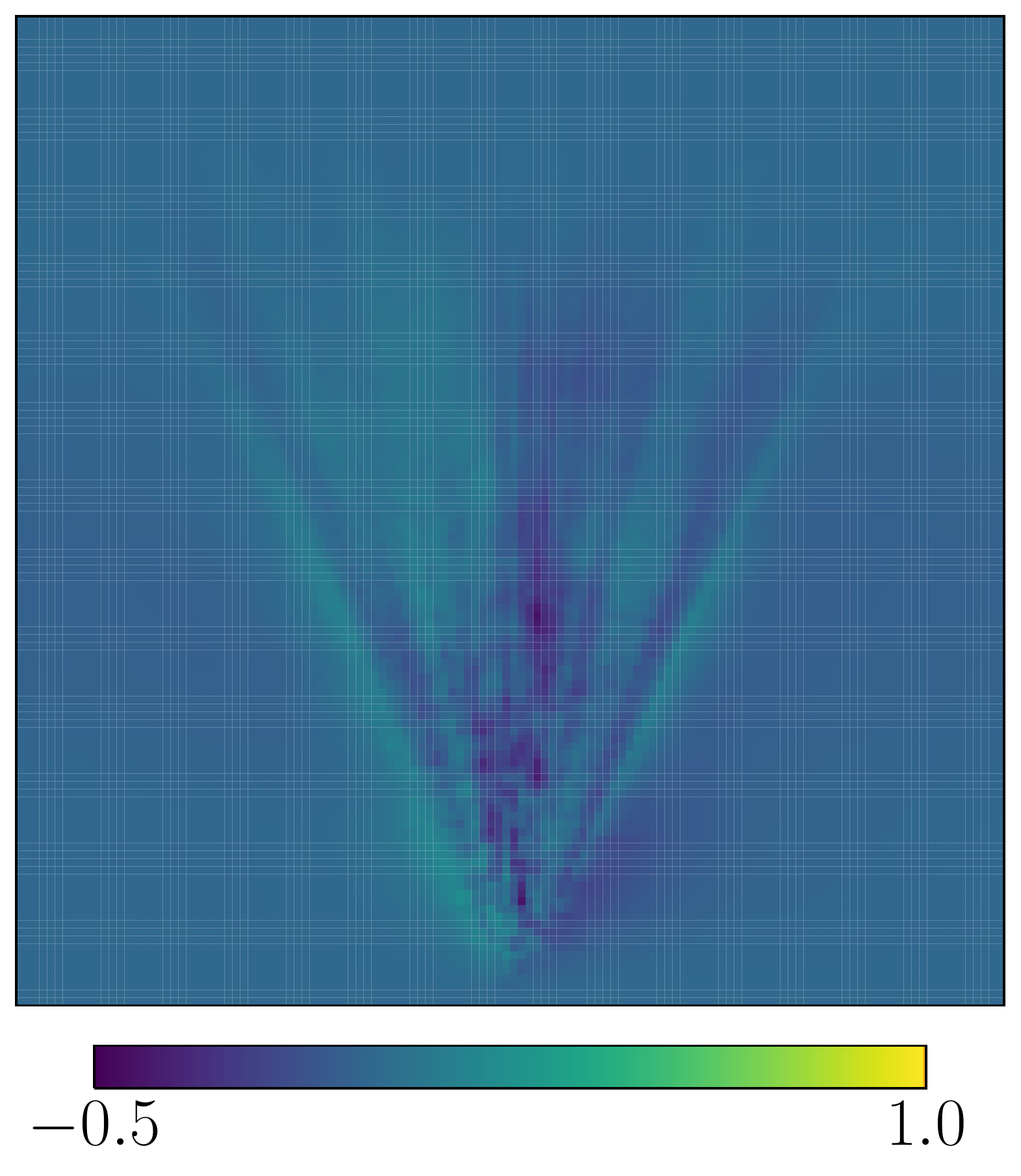}

\includegraphics[scale=0.36,valign=t]{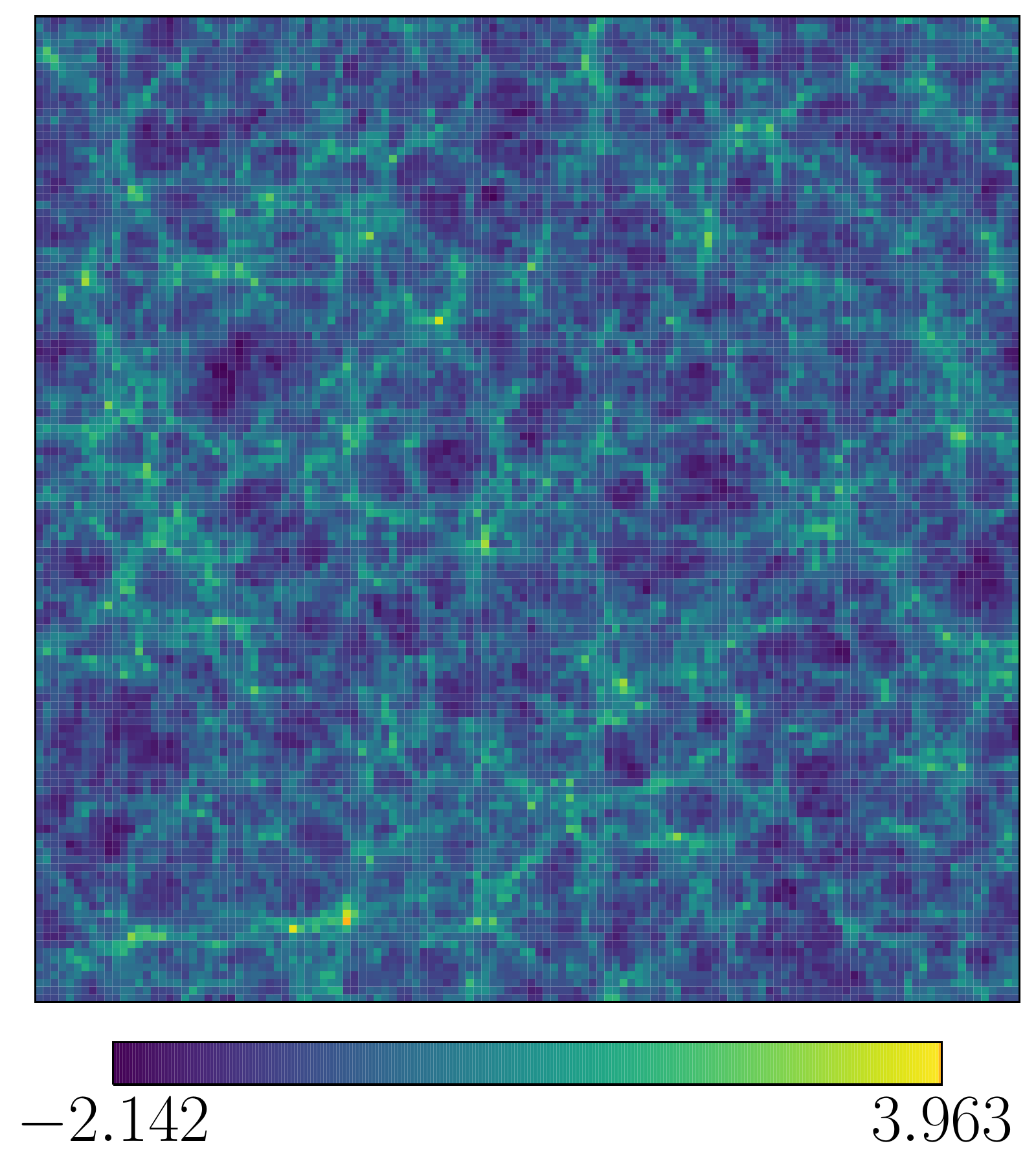}
\includegraphics[scale=0.36,valign=t]{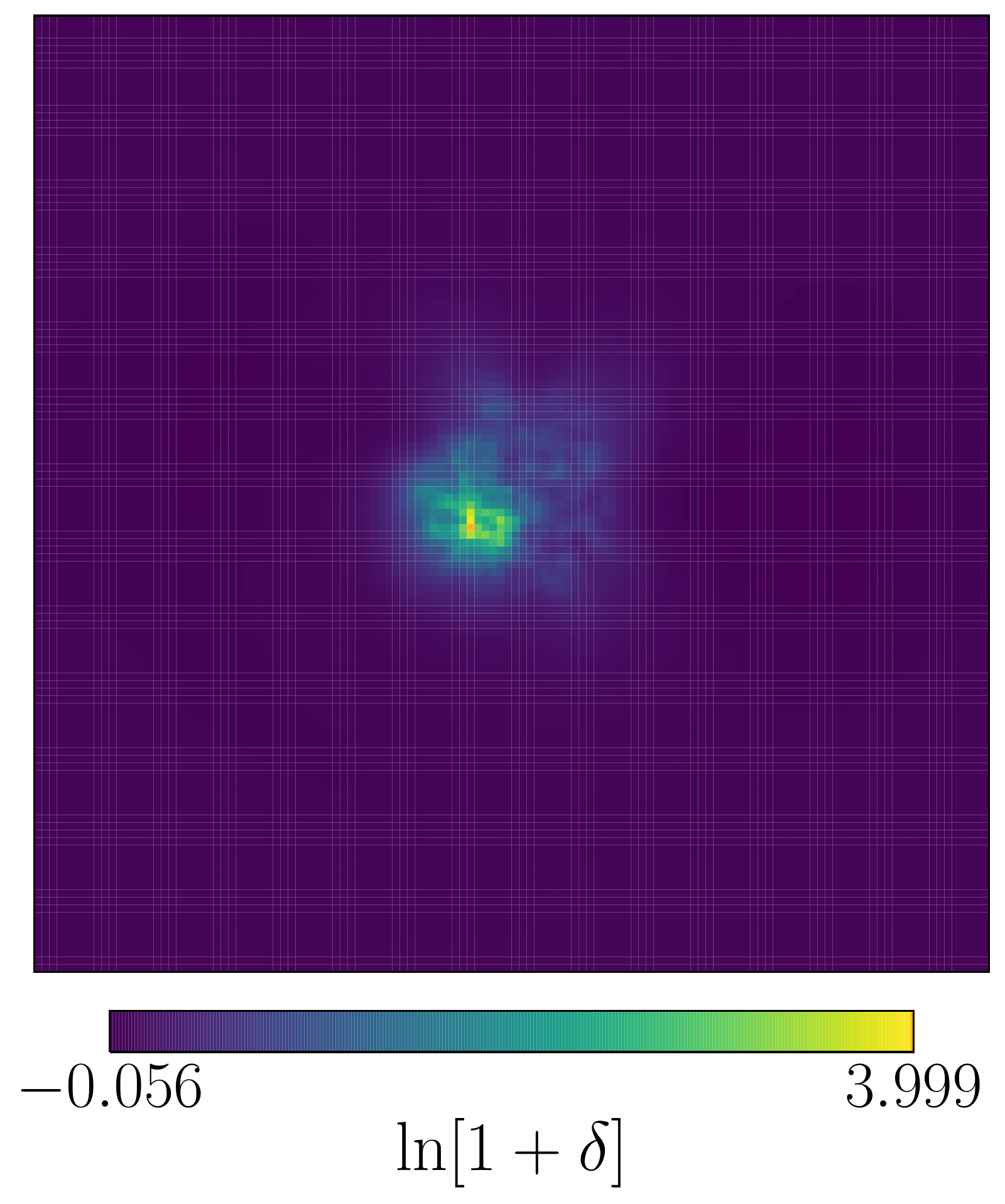}
\includegraphics[scale=0.36,valign=t]{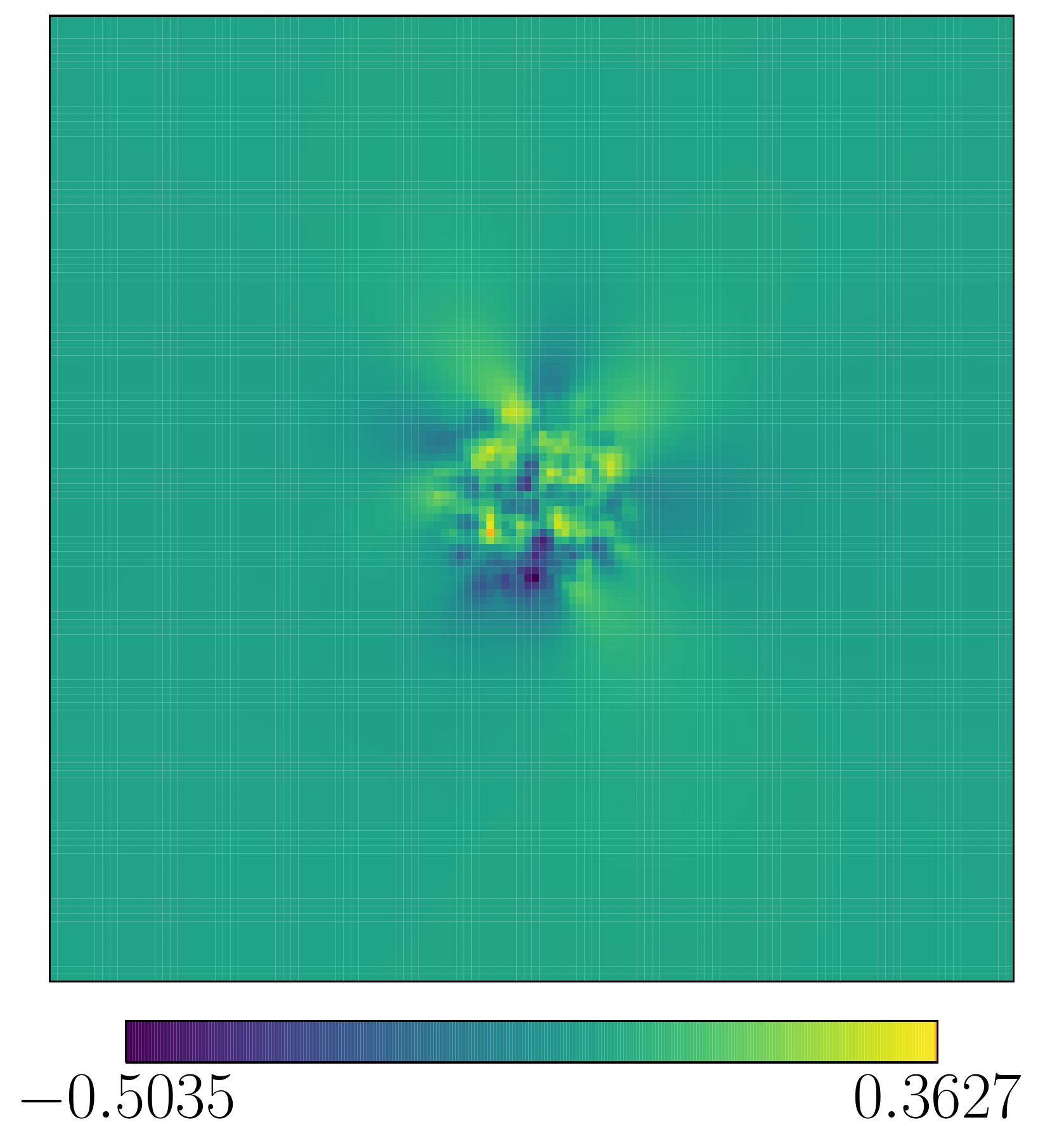}
\end{center}
\caption{\label{fig:TestC3} Results of test C: reconstructions from CFHTLenS-like mock data with realistic shape noise. In the left column we show slices through the underlying density field, in the central and middle columns the corresponding slices through the lognormal and WF reconstructions, respectively. In the upper row, we show a central slice through the $y$-axis. Both reconstructions suffer from a strong line of sight degeneracy. The Wiener filter seems to slightly better reconstruct underdensities, while the lognormal model seems more sensitive to overdensities. The latter even overestimates the densities in certain slices normal to the observational direction. We interpret this as a consequence of mass sheet degeneracy. To verify that the algorithm identifies real structures in these overdense slices, we show one of them (the lower one) in the lower row.}
\end{figure}

In the last test of our test series, we add realistic shape noise with $\sigma_\epsilon=0.25$ to the mock data on the light cone. This value is close to the value in CFHTLenS, $\sigma_\epsilon=0.279$~\cite{CFHTLens_Tomo+IA}. Compared to test B, we use the same opening angle of $7.15^o$ but a slightly higher galaxy density of $15\, \mathrm{gal/arcmin^2}$. We show results of this final test in Figures~\ref{fig:TestC3PDFs} and \ref{fig:TestC3}. The left panel of Figure~\ref{fig:TestC3PDFs} shows that the lognormal model is more likely to reconstruct the correct density in a pixel. However, the PDF in the right panel seems to have a spurious tail of high density values. This tail does not significantly deviate from the high density tail of the underlying reconstruction and is therefore hardly suppressed by the prior. Figure ~\ref{fig:TestC3PDFs} reveals that these high density values are due to an overestimation of densities in slices normal to the direction of observation. We have verified that these overdense sheets are not due to a wrong minimum or saddle-point by starting the algorithm at either the WF solution or the true input field and changing convergence criteria. We interpret these features as follows: Adding a constant density offset to a slice normal to the observational direction does not significantly change the likelihood, because of the well known mass-sheet degeneracy - the fact that adding a homogeneous mass sheet does not change the lensing signal. The lognormal prior does not punish or even encourages these overdensities since it expects a skewed density distribution. The Gaussian prior suppresses strong deviations from a symmetric density distribution, which is why these features are not present in the Gaussian case.

\section{Dicussion and Future Improvements}
\label{sec:conclusions}

We have presented a novel, fully Bayesian, lensing tomography algorithm that reconstructs the three-dimensional matter distribution from measurements of individual galaxy shapes. Given a data vector of measured galaxy shears, the galaxies' angular positions and their redshifts, this algorithm finds the maximum a posteriori estimate of the density contrast between the lensed galaxies and the observer. Since galaxy shapes provide only limited information about the intervening structure, this reconstruction problem is in general ill-constrained and requires regularization. In case of a maximum a posteriori estimate this regularization is provided by the prior probability distribution - the distribution which ideally contains all available information about the underlying field before the lensing measurement.

This ideal distribution, however, is in reality unaccessible. Similar previous works have used a Gaussian prior in this context. A Gaussian prior (together with a Gaussian noise model) results in a Gaussian posterior distribution which has a distinct minimum and no saddle points. Maximizing this posterior is comparably straightforward. The algorithm presented in this work was especially designed to also handle non-Gaussian posteriors. We show this ability by using a lognormal prior which should be an improvement over the Gaussian approximation in two ways: First, it enforces the strict positivity of the density while a Gaussian prior allows unphysical, negative densities. Second, it incorporates an a priori knowledge of the presence of odd moments in the matter distribution, which arise as a consequence of non-linear structure formation. These corrections are relevant since the cosmic shear signal probes structures down to scales that lie well within the non-linear regime.

We note that neither of these priors describe the exact distribution of the non-linearly evolved density contrast. In regions with high signal-to-noise this is not problematic since the prior is updated by the information contained in the data. In less well constrained regions the prior will paint in missing information. This can lead to  systematic deviations from the true distribution. However, non optimal priors can be sufficient depending on the quantity and scales one is interested in reconstructing. For example a Gaussian prior will be a good choice for large scales, where the field is still Gaussian. On small scales the lognormal model will yield better reconstructions since it is informed about the skewness of the distribution and avoids negative densities.

To test the algorithm and assess the differences between a reconstruction with a lognormal and a Gaussian prior (in the latter case the reconstruction simplifies to a Wiener filter), we have applied it in both configurations to increasingly realistic mock data sets. Those data sets were produced by applying the algorithm's inherent data model to non-linearly evolved densities from N-body simulations. 
In tests with negligible shape noise, but a realistic CFHTLenS-like source distribution, we find that the reconstruction with a lognormal prior avoids negative densities that are abundantly present in the Wiener filter reconstruction. It is also better in reconstructing the highest local peaks in the density field, and it leads to a higher point-wise correlation between the true underlying density and its reconstruction. The Gaussian model seems better in tracing low density regions and capturing the largest scales. In regions with low signal-to-noise both priors tend towards the mean density. 

For realistic shape noise, the data contains too little information to break the line-of-sight degeneracy of the tomographic problem with either prior. This is expected for current data sets, which are commonly analyzed in a limited number of broad redshift bins in order to yield sufficient signal to noise per bin. In addition, the lognormal model is prone to mass-sheet degeneracy: since the prior encourages a skewed distribution with more high density pixels, the reconstruction tends towards inserting sheets of higher density normal to the observational direction.
The situation might slightly improve for future surveys with a higher source density. But since the shape noise scales only with the square root of the source density, even future data is unlikely to contain sufficient information for a 3D reconstruction without this degeneracy.

The fidelity of the reconstructions in this work might also be overestimated since we assumed precisely known source redshifts and point spread function, and ignored intrinsic alignments. In this work we show how photometric redshift uncertainties can be included, but leave the implementation for future work. Other sources of uncertainty could be included in a similar fashion.
Another approximation we made concerns the source distribution. In reality the source distribution is correlated with the underlying density field, which we ignore in our test cases. A more clustered distribution of sources would render the fidelity of the reconstruction more inhomogeneous since structures in front of high density regions would be detected with higher signal to noise than structures in front of low density regions. Since the volume fraction of high density regions is smaller than the volume fraction of low density regions, this could  decrease the overall fidelity of the reconstruction. To get an estimate of the effect of the clustering, future tests of the algorithm should account for the correlation between galaxies and density field by using the halo positions identified in the simulation by a halo finder.

Possible improvements of the algorithm could go into two directions, 1) improving the prior or data model, 2) including additional sources of information.
An improved data model could be obtained by forward modeling structure formation numerically for an initially Gaussian distributed field. While this is conceptually straightforward to integrate in the algorithm, it is numerically very expensive. In general, we expect any prior model that is informed about the relative orientation of structures (as e.g. the Zeldovich approximation) to partly break the line-of-sight degeneracy or inhibit spurious mass sheets.
Additional information could come from the clustering of galaxies. This would require the inference of at least one additional bias parameter (depending on the bias model), but would add information about the relative orientation of structure, also possibly breaking the line-of-sight and mass-sheet degeneracy.

In the current implementation the algorithm assumes a fixed set of cosmological parameters for both models. For future applications the parameter estimation could be made part of the algorithm by analyzing the joint posterior of the density distribution and cosmological parameters.
The cosmology enters the density estimate through the theoretical matter power spectrum, the growth factor and the background evolution (that determines e.g. the redshift-distance relation). Lensing measurements are mostly sensitive to a combination of $\sigma_8$ and $\Omega_m$. For two cosmological parameters only, sampling of the joint posterior would be feasible already with the current serialized implementation, for one could sample the $\sigma_8$-$\Omega_m$ plane (e.g. on a grid) and evaluate the minimum of $P(\delta|\sigma_8,\Omega_m,d)$ for each parameter combination in parallel.
Since the total information content of the data does not depend on the analysis method, a joint reconstruction of the 3D density field and cosmological parameters should yield competitive constraints to traditional analyses that make use of redshift binning.

\begin{acknowledgments}
We thank Jakob Knollm\"uller for helpful insights on covariance probing and Reimar Leike and Steffen Hagstotz for useful comments on the draft. 
\end{acknowledgments}
%

\end{document}